\documentclass[letterpaper, numberedappendix]{emulateapj} 

\usepackage{graphicx}
\usepackage{epsfig}
\usepackage{amsmath}
\usepackage{subfigure} 
\usepackage{amssymb}
\usepackage{lscape} 
\usepackage{apjfonts}
\usepackage{latexsym}
\usepackage{enumitem}

\shorttitle{Vertical perturbations in nearby disk-type galaxies}
\shortauthors{Urrejola-Mora et al.}

\begin{document}

\title[]{WiNDS: An $\mathrm {H_{\alpha}}$ kinematics survey of nearby spiral galaxies - Vertical perturbations in nearby disk-type  galaxies}

\author{Catalina Urrejola-Mora\altaffilmark{1}}
\author{Facundo A. G{\'o}mez\altaffilmark{1,2}}
\author{Sergio Torres-Flores\altaffilmark{1}}
\author{Philippe Amram\altaffilmark{3}}
\author{Beno{\^i}t Epinat\altaffilmark{3,4}}
\author{Antonela Monachesi\altaffilmark{1,2}} 
\author{Federico Marinacci\altaffilmark{5}}
\author{Claudia Mendes de Oliveira\altaffilmark{6}}

\altaffiltext{1}{Departamento de Astronom{\'i}a, Universidad de La Serena, Av. Juan Cisternas 1200 Norte, La Serena, Chile}
\altaffiltext{2}{Instituto de Investigaci{\'o}n Multidisciplinar en Ciencia y Tecnolog{\'i}a, Universidad de La Serena, Ra{\'u}l Bitr{\'a}n 1305, La Serena, Chile}
\altaffiltext{3}{Aix Marseille Univ, CNRS, CNES, LAM, Marseille, France}
\altaffiltext{4}{Canada-France-Hawaii Telescope, CNRS, 96743 Kamuela, Hawaii, USA}
\altaffiltext{5}{Department of Physics \& Astronomy, University of Bologna, via Gobetti 93/2, 40129 Bologna, Italy}
\altaffiltext{6}{Departamento de Astronomia, Instituto de Astronomia, Geof{\'i}sica e Ci{\^e}ncias Atmosf{\'e}ricas da USP, Cidade Universitaria, CEP: 05508-900, S{\~a}o Paulo, SP, Brazil}
 
\begin{abstract}

We present the Waves in Nearby Disk galaxies Survey (WiNDS) consisting of 40 nearby low inclination disk galaxies observed through $\mathrm {H_{\alpha}}$ high-resolution Fabry Perot interferometry. WiNDS consists of 12 new galaxy observations and 28 data archived observations obtained from different galaxy surveys. We derive two-dimensional line-of-sight velocity fields that are analyzed to identify the possible presence of vertical velocity flows in the galactic disks of these low-inclination late-type galaxies using velocity residual maps, derived from the subtraction of an axisymmetric rotation model to rotational velocity map. Large and globally coherent flows in the line-of-sight velocity of nearly face-on galaxies can be associated with large vertical displacement of the disk with respect to its mid-plane. Our goal is to characterize how frequent vertical perturbations, such as those observed in the Milky Way, arise in the Local Universe. Our currently available data have allowed us to identify 20$\%$ of WiNDS galaxies with strong velocity perturbations that are consistent with vertically perturbed galactic disks.

\end{abstract}


\section{Introduction}
\label{sec:intro}

Over the last ten years, oscillatory perturbations towards the outskirts of the Solar Neighborhood in the Galactic disk have been reported several times \citep[e.g.][]{2002A&A...394..883L,2006A&A...451..515M,2012ApJ...750L..41W,2014ApJ...791....9S,2015MNRAS.452..676P,2015ApJ...801..105X,2018Natur.561..360A,2021A&A...649A...8G}. In particular, using  11k main-sequence stars from Sloan Digital Sky Survey \citep[SDSS,][]{2000AJ....120.1579Y}, \citet{2012ApJ...750L..41W} detected asymmetries in both, \textit{i)}  the bulk velocity of Solar Neighborhood stars associated to a breathing pattern, i.e. compression and rarefaction motions; and \textit{ii)} in the vertical stellar number density distribution. The latter was related to a bending pattern, i.e. local displacements of the disk from the midplane \citep{2014MNRAS.440.1971W}. This study was followed up by \citet{2015ApJ...801..105X} who analyzed measurements of stellar number counts of main-sequence stars located at Galactic latitudes $110^{\circ}$ < l <  $229^{\circ}$, as a function of Galactocentric radius. They showed that the  amplitude of  the perturbations in the vertical stellar number counts, i.e. the displacement of the disk with respect to its midplane, increases towards the outskirts of the Galaxy. This type of oscillatory perturbation on the stellar and/or gaseous component of the disk is known as a corrugation pattern.  Thanks to the analysis of the full 6D phase-space information for more than six million stars from \citet{2018A&A...616A..11G}, \citet{2018Natur.561..360A} showed that, indeed, our own Galactic disk is undergoing phase mixing of a non-equilibrium configuration, a perturbation likely associated with the interaction of the Milky Way with Sagittarius dwarf galaxy in the past \citep[e.g.][]{2013MNRAS.429..159G,2018MNRAS.481..286L,2019MNRAS.484..245L,2020MNRAS.492L..61L}. More recently \citet{2021A&A...649A...8G} confirmed that the Milky Way disk has very complex dynamics, with vertical velocity perturbations that can be partially described as a bending wave.

Interestingly, evidence of such complex vertical oscillatory patterns on external galaxies, such as those observed in the Milky Way, are still extremely limited. Vertically perturbed disks have been extensively observed, especially in edge-on galaxies. Studies of 21 cm neutral hydrogen line observations show  that most galaxies that have extended HI disks, with respect to the optical, are warped. Using a sample of 26 edge-on galaxies located in different environments with  inclinations  $ i > 75^{\circ}$ and blue diameters larger than 1.5$'$, from Westerbork HI Survey of Spiral and Irregular Galaxies \citep[WHISP,][]{2001ASPC..240..451V}, \citet{2002A&A...394..769G} showed that all galaxies in their sample with an extended HI disk ($\sim76\%$) are warped with respect to the optical disk. They also found that warped disk galaxies are more often found in less dense environments, indicating that tidal interactions are not the only mechanism producing Galactic warps. However, in denser environments, the disk vertical perturbations typically show larger
amplitudes than those in less dense environments. This is likely due to the fact that, in denser environments, galaxies are more likely to undergo violent tidal interactions with companion galaxies.

For observations in optical bands, \citet{2006NewA...11..293A} analyzed a sample of 325 galaxies from the SDSS, which consists of a  majority of late-type spiral galaxies with inclinations $ i > 84^{\circ}$ and isophotal major axis length  $D_{25} > 1'$ at $ \mu_{B} = 25\ \mathrm{mag\ arcsec^{-2}}$. The study showed that 73$\%$ of the sample contained a warped disk: 51$\%$ corresponds to S-shaped and 22$\%$ to U-shaped perturbations. For the analysis of the environment-dependence on the frequency of warps, \citet{2006NewA...11..293A} analyzed 75 galaxies with redshift data from the SDSS DR3, of which 56 have companion galaxies. For the subset of galaxies that have companions, 64$\%$ show warped disks. For galaxies without companions, 81$\%$ show warped disks. This result is consistent with those presented in \citet{2002A&A...394..769G}, i.e, in poor environments a higher frequency of warped disks is found than in rich environments. Vertical perturbations have also been studied in edge-on galaxies using dust lines. \citet{2020MNRAS.495.3705N} analyzed a sample of five nearby objects with inclinations $ i > 80^{\circ}$ and also showed that such perturbations are common.

S-shaped warps are expected to be a very common feature of both interacting and isolated late-type galaxies. \citet{2016MNRAS.456.2779G,2017MNRAS.465.3446G} studied in detail the origin and evolution of vertical perturbation in the disk of late-type galaxies  using high-resolution cosmological magneto-hydrodynamical simulations of the Auriga Project \citep{2017MNRAS.467..179G}. By means of the  construction of mass-weighted mean height, $\langle Z \rangle$, and mean vertical velocities maps, $\langle V_{\rm Z} \rangle$, for both the stellar and cold gas components, these  studies found that $70\%$ of the Milky Way-like models showed strongly vertically perturbed disks. Interestingly, while only half of the vertically perturbed disks showed integral-sign (S-shaped) warps, the remaining half showed a more complex vertical structure, i.e. corrugation patterns, indicating that such structures are expected to be common. 

A possible reason behind the lack of detected vertical patterns relates to the complexity behind their observation. On edge-on galaxies, such pattern can be easily confused with S-shaped warps due to projection effects. However, as discussed in \citet{2017MNRAS.465.3446G}, such a corrugation pattern should be detectable along line-of-sight velocity (V$_\mathrm{los}$) of nearly face-on galaxies. Due to the oscillating nature of a corrugation, global patterns on the (V$_\mathrm{los}$) field can be directly linked to global patterns in a $\langle Z \rangle$ map. Until recently, previous efforts to detect corrugations in the cold  gas disk component of nearby galaxies were based  on $\mathrm {H_{\alpha}}$ emission through long-slit spectroscopy.  \citet{2001ApJ...550..253A} analyzed NGC 5427, a galaxy with an inclination angle of 30$^{\circ}$, and found wavy-like variations in the velocity profile of the ionized gas. \citet{2015MNRAS.454.3376S} analyzed a sample of four spiral galaxies and showed that two presented corrugation patterns. However, due to the limited coverage associated with the long-lit spectroscopic observations, such patterns could be confused with the effects of local perturbations such as fountain flows. In a recent work, \citet{2021ApJ...908...27G} presented a full 2D velocity map of the low inclination galaxy, VV304a. Their study was based on  Fabry-Perot interferometer $\mathrm {H_{\alpha}}$ observations. This technique has the great advantage of providing high spectral resolution in a narrow frequency range, along with good spatial resolution (in this work $\sim 2''$).  As such, it allows resolving local velocity perturbations with amplitudes of the order of $10\ \mathrm{km\ s^{-1}}$. The study showed, for the first time, that VV304a presents global and coherent velocity perturbations that are consistent with a corrugation pattern. They also showed that these velocity perturbations cannot be described by the effects associated with the presence of axisymmetric features on its disk, such as the bar and spiral structure. Thus, they conclude that these perturbations are a direct consequence of the gravitational interaction with its similar mass companion galaxy, VV304b.
 
The characterization of  vertical perturbations in nearby disk galaxies can provide very valuable information about their recent interaction with their environment. Several mechanisms can be behind the formation of warps and corrugation patterns \citep{2013pss5.book..923S,2016MNRAS.456.2779G,2017MNRAS.465.3446G}. One of them is the tidal distortion of a pre-existing disk by an external perturber, such as the cases of our own Milky Way \citep{2018Natur.561..360A,2019MNRAS.484..245L} and VV304a \citep{2021ApJ...908...27G}. The strong tidal torques being exerted on a pre-existing disk as relatively massive satellites fly by or merge can induce strong vertical perturbations \citep{1989MNRAS.237..785O,1993ApJ...403...74Q,1999MNRAS.304..254V,2003ApJ...583L..79B,2009ApJ...700.1896K,2013MNRAS.429..159G,2016ApJ...823....4D,2017MNRAS.465.3446G,2018MNRAS.473.1218L,2018MNRAS.481..286L}. In addition, the torques associated with a non-spherical mass distribution of dark matter can also trigger the formation of vertical patterns \citep{1999ApJ...513L.107D,1999MNRAS.303L...7J,2006MNRAS.370....2S, 2012MNRAS.426..983D,2015MNRAS.452.2367Y,2016MNRAS.456.2779G,2018MNRAS.473.1218L}. Other possibilities may include misaligned accretion of cold gas due to the cooling of hot gas halo, infalling from the cosmic web, or being left by gas-rich host-satellite interactions \citep{2008A&ARv..15..189S,1999MNRAS.303L...7J,2010MNRAS.408..783R,2013MNRAS.434.3142A,2014ApJ...780..105R,2017MNRAS.465.3446G} and ram-pressure of the surrounding intergalactic material \citep{2014MNRAS.443..186H}. 
The objective of this work is to search for velocity perturbations in the disks of nearby late-type galaxies. To achieve this goal we use high-resolution observations of near face-on galaxies obtained with a Fabry-Perot interferometer. For this purpose, we present the Waves in Nearby Disk galaxies Survey (WiNDS), which consists of a data set with high spectral resolving power (R\ $\sim 10000$) and large spatial coverage, making them ideal for investigating the kinematics of the ionized gas in galactic disks. This study is the first step in an effort that aims to characterize the history of interactions of nearby galaxies, and estimate the frequency with which vertical disturbances such as those observed in the Milky Way arise in the Local Universe.  

The paper is organized as follows. In Section 2 we introduce the WiNDS data sample and we describe our new observations. In Section 3, we discuss the data reduction procedure. Section 4 and Section 5 describes the data analysis and selection and quantification criteria of bending modes. Finally, we present our results in Section 6 and in Section 7 we provide a discussion and our conclusions. 
In Appendix A the comments for each individual galaxy and maps of the new observations made in this work are given.
The residual velocity maps are shown in Appendix B. The rotation curves are displayed in Appendix C. In Appendix D the image processing method is described.

\section{WiNDS Data Sample} \label{sec:data_sample}

In this Section, we present our survey WiNDS, an ongoing observational campaign currently comprising 40 nearby nearly-face-on spiral galaxies. WiNDS consist of 12 new observational data and we complement it with additional archival data of 28 late-type galaxies from different surveys. The entire data sample contains 3D data cubes obtained using Fabry-Perot interferometer with a resolving power at the $\mathrm {H_{\alpha}}$ rest wavelength of R $\approx$ 10000.
The galaxies in the WiNDS sample were selected according to three main criteria:
\begin{itemize}
    \item Distance: nearby spiral galaxies were selected to have systemic velocities $ v\ \leq 3000\ \mathrm{km\ s^{-1}}$, corresponding to less than 45 Mpc.
    \item Inclination: all galaxies have previously estimated low inclination angles $ i \leq 40\ \mathrm{degrees}$. Both, morphologically and kinematically estimated inclination were considered in this step.
    \item Size: all galaxies have projected diameters, $a_{p}$, between $2\ \mathrm{arcmin} \leq  $ $a_{p}$ $\leq 4\ \mathrm{arcmin}$.
    
\end{itemize}

The goal of this selection criteria is to allow us to resolve velocity perturbations in the observed disks with amplitudes as small as $10\ \mathrm{km\ s^{-1}}$. In what follows we discuss WiNDS new observations in detail. We also briefly describe the additional surveys utilized to complement WiNDS. In Table \ref{tab:table1} we list some of the main properties of the WiNDS sample.
The blue histograms in Figure \ref{fig:hist_distribution} show the distribution of the most relevant parameters for the final WINDS sample. The solid curves depict the smooth continuous approximation of the underlying histograms. These smooth histograms were obtained using a Gaussian Kernel Density Estimator. The green histograms in the same figure show the parameter distribution of the vertically perturbed galaxy candidates within the WiNDS sample (see Section \ref{sec:quantification}). Although the WiNDS sample is not complete in terms of, e.g., mass and magnitude, it allows performing a first systematic search of vertically perturbed disks in the nearby Universe. The 40 selected galaxies span different environments such as clusters, groups, and field galaxies, and a wide variety of morphologies (Sa to Sc). Note however that this is an evolving project, and more objects will be added in the future to complement the available sample. Figure \ref{fig:sample} shows the optical images of the WiNDS sample, obtained from SDSS DR9 using the \textit{g, r} and {\it z} bands.

\subsection{WiNDS: New data cubes} \label{subsec:winds}

The new observations for WiNDS were obtained using the Fabry-Perot interferometer at the Observatoire de Haute Provence Observatory (OHP, France). 
In addition to the selection criteria mentioned in Section \ref{sec:data_sample}, we focus on galaxies with previously reported $\mathrm {H_{\alpha}}$ observations \citep{2004A&A...414...23J, 2010PASP..122.1397S}. The new data cubes include galaxies with inclination angles ranging from 6$^{\circ}$ to 33$^{\circ}$, except for NGC 2500 which has a $i \approx 40^{\circ}$. 
The sample contains galaxies with a range of morphological types (between Sa to Sc) and a wide absolute $B$-band magnitude range ($-17.9 \ \leq $ $M_{B}$  $ \leq -21.9\ \mathrm{mag}$). It also contains both isolated and interacting galaxies belonging to different environments. The 12 new galaxies observed as part of WiNDS are mainly located in the Northern Hemisphere, the list of 12 galaxies is reported in Table \ref{tab:table1}. In particular, NGC 2763 is the only galaxy in our sample observed from the Southern Hemisphere using the SOAR Adaptive Module Fabry–Perot \citep[SAM-FP,][]{2017MNRAS.469.3424M}. 
This galaxy was added to the sample at a later time, based on the availability of the SAM-FP.

\subsubsection{WiNDS: Observation of new data cubes}
 \label{sec:observations}

Most WiNDS observations for the new galaxies were performed in February 2019 using the 1.93-m telescope at the OHP. 
The observations were made with a mean integration time of 2 hours per galaxy. We note that due to poor weather conditions three galaxies in the sample were observed with a seeing  slightly over $3\ \mathrm{arcsec}$.

The new datacubes were obtained through the Gassendi HAlpha survey of Spirals (GHASP) instrument, a focal reducer containing a Fabry-Perot scanning interferometer which has a large field-of-view (FoV), of $\simeq 5.9 \times 5.9\ \mathrm{arcmin^{2}}$, an aperture ratio of $f/3.9$, and a pixel scale $0.68\ \mathrm{arcsec\ pixel^{-1}}$. The Free Spectral Range (FSR) for the two used interferometers with interference orders of $798$ and $2600$ are $375.9\ \mathrm{km\ s^{-1}}$ and $115.4\ \mathrm{km\ s^{-1}}$, respectively.
The resolving power reached in our survey was $R = 10000 - 28600$, which translates into a velocity resolution of $11.6 - 3.6\ \mathrm{km\ s^{-1}}$, respectively. As discussed in \citet{2019A&A...631A..71G} the detector, Image Photon Counting System (IPCS), has the advantage of a zero-readout noise, which allows a very fast scan of the interferometer through the entire FSR.
The use of an IPCS makes it possible to carry out short exposures (10 seconds) and to make a large number of cycles (typically 20) to average the variations in observation conditions.
Within new observation of WiNDS, the $\mathrm {H_{\alpha}}$ line can be found within the range going from 6563 {\AA} to 6627 {\AA}. A total of seven filters, each with a Full Width at Half Maximum (FWHM) of 15 {\AA}, are available to isolate the $\mathrm {H_{\alpha}}$ emission line. 

For the wavelength calibration of the data, a cube is obtained before and after the observation of each galaxy to account for possible changes in observing conditions. Each calibration cube is obtained using the Neon emission line (6598.95 {\AA}), which is isolated using a narrow-band interference filter ($\sim 15$ {\AA}) centered on this wavelength. This line is used because it is intrinsically narrow and close to the wavelength of the redshifted $\mathrm {H_{\alpha}}$ line.

An additional galaxy, NGC 2763, was observed in February 2016 using the 4.1-meter telescope at the Southern Astrophysical Research (SOAR), Cerro Pach{\'o}n, in Chile. 
The instrument used is called SOAR Adaptive Module Fabry-Perot (SAM-FP). It is a restricted-use instrument providing a new mode of operation of SAM for spectroscopy using a Fabry-Perot (FP) and SAM-Imager (SAMI).
The SAM-FP provides a FoV of $\simeq 3 \times 3\ \mathrm{arcmin^{2}}$, a CCD pixel scale of $0.045\ \mathrm{arcsec\ pixel^{-1}}$ \ (after electronic binning on the detector) and 48 wavelength channels, in the third dimension of the datacube.
For the SAM-FP observations, the interference order was 609 with a FSR of $492.6\ \mathrm{km\ s^{-1}}$ and resolving power of $R = 10000$ which translates into velocity sampling of $10.3\ \mathrm{km\ s^{-1}}$ per channel (for a resolution of $\sim 20.6\ \mathrm{km\  s^{-1}}$), see \citet{2017MNRAS.469.3424M} for more details.

\begin{table*}[ht]
\centering
    \begin{tabular}{@{}clcccccccccccc@{}}
    \hline
NGC & Morph & t$_{morph}$ & $R_{opt}$ & v$_{sys}$              & $d$    & $M_{B}$ & v$_{rot}^{max}$         & log(M$_{star}$) & log(M$_{HI}$) & log(SFR) & $i$          & $PA$            \\

    &  (Type) &           &  (kpc)    & ($\mathrm{km\ s^{-1}}$) & (Mpc) &  (mag)  & ($\mathrm{km\ s^{-1}}$) & ($M_{\odot}$)    & ($M_{\odot}$) &        & ($^{\circ}$) &  ($^{\circ}$) &  \\ 
    (1) &  (2) &  (3) &  (4) &  (5) &  (6) &  (7) &  (8) &  (9) &  (10) &  (11) &  (12) &  (13) \\

    \hline \hline

         New observations & & & & & & & & & & & & & \\

         628             & SAc   & 5.2 & 16.4 &  657.2 & 10.8 & -20.7 $\pm$ 0.3 &  64 & 10.3$^{a}$ & 9.7$^{b}$  &           &  7$^{m}$ & 252 \\
         1058            & SAc   & 5.1 &  4.3 &  518.0 &  9.7 & -18.9 $\pm$ 0.4 &  13 &            & 8.8$^{d}$  &           &  6$^{m}$  & 152 \\ 
         2500            & SBd   & 7.0 &  4.5 &  503.9 & 10.5 & -17.9 $\pm$ 1.7 & 131 &  9.4$^{a}$ & 9.0$^{c}$  &           & 40$^{k}$   &  86 \\
         2763 (SAM-FP)   & SBcd  & 9.9 & 10.0 & 1891.6 & 29.6 & -19.8 $\pm$ 0.3 & 117 &            &            &           & 30$^{k}$  &  50 \\
         3147            & SAbc  & 3.9 & 26.1 & 2801.9 & 46.1 & -21.9 $\pm$ 0.2 & 335 & 11.4$^{a}$ &            &           & 32$^{m}$  & 147 \\
         3184            & SABcd & 5.9 & 11.4 &  592.1 & 10.9 & -19.1 $\pm$ 0.4 & 208 & 10.4$^{a}$ & 9.3$^{c}$  &           & 16$^{k}$  &   1 \\
         3423            & SAcd  & 6.0 &  8.5 & 1004.0 & 15.2 & -20.0 $\pm$ 0.2 & 128 &  9.7$^{a}$ & 9.1$^{b}$  &           & 19$^{k}$  &  45 \\
         3485            & SBb   & 3.2 &  9.4 & 1436.0 & 28.2 & -19.1 $\pm$ 0.8 & 188 &  9.9$^{a}$ & 9.3$^{b}$  & 0.6$^{e}$ & 20$^{k}$  & 116 \\
         3642            & SAbc  & 4.0 & 23.3 & 1588.0 & 30.0 & -20.6 $\pm$ 0.6 &  48 & 10.4$^{a}$ &            &           & 25$^{k}$  & 123 \\
         4136            & SABc  & 5.2 &  4.2 &  608.9 &  7.2 & -18.5 $\pm$ 1.5 & 102 &  9.5$^{a}$ & 8.9$^{b}$  &           & 22$^{k}$  &  72 \\
         4900            & SBc   & 5.1 &  3.2 &  959.6 &  9.8 & -19.2 $\pm$ 1.1 & 112 & 10.4$^{a}$ & 9.1$^{b}$  &           &  5$^{k}$  &   0 \\
         5194            & SAbc  & 4.0 & 15.8 &  462.9 &  9.8 & -21.3 $\pm$ 0.3 & 134 & 10.9$^{a}$ & 9.3$^{c}$  &           &  20$^{k}$  & 13 \\

\hline

         Data Archive & & & & & & & & & & & & & \\
         
         864 (G)         & SABc  & 5.1 & 15.3  & 1561.9 & 22.6 & -19.9 $\pm$ 0.4 & 134 & 10.2$^{a}$ & 9.6$^{b}$ &           & 35$^{k}$  &  27 \\
         2775 (G)        & SAab  & 1.6 & 14.3  & 1350.0 & 23.0 & -20.6 $\pm$ 0.8 & 296 & 10.9$^{a}$ & 8.6$^{b}$ &           & 38$^{k}$  & 157 \\
         3344 (G)        & SABbc & 4.0 &  7.7  &  580.1 &  7.4 & -19.6 $\pm$ 0.3 & 217 &  9.7$^{a}$ & 9.6$^{b}$ &           & 18$^{k}$  & 156 \\
         3346 (G)        & SBcd  & 5.9 & 10.3  & 1274.1 & 24.5 & -19.1 $\pm$ 0.9 & 126 & 10.2$^{a}$ & 9.1$^{b}$ & 0.6$^{e}$ & 34$^{m}$  & 113 \\
         3351 (S)        & SBb   & 3.1 & 10.5  &  777.9 &  9.7 & -19.8 $\pm$ 0.1 & 151 & 10.5$^{a}$ & 9.0$^{b}$ &           & 40$^{k}$  &  11 \\   
         3504 (G)        & SABab & 2.1 & 11.7  & 1525.0 & 29.8 & -20.5 $\pm$ 0.7 & 194 & 10.4$^{a}$ & 8.8$^{b}$ & 3.6$^{e}$ & 39$^{k}$  & 164 \\
         3596 (G)        & SABc  & 5.2 & 14.0  & 1192.9 & 24.2 & -19.7 $\pm$ 0.9 & 157 & 10.0$^{a}$ & 9.0$^{b}$ &           & 17$^{k}$  &  77 \\
         3631 (H)        & SAc   & 5.1 & 17.3  & 1155.9 & 23.8 & -21.0 $\pm$ 0.8 &  79 & 10.2$^{a}$ &           & 2.7$^{e}$ & 24$^{k}$  & 171 \\
         3938 (S)        & SAc	 & 5.1 & 12.9  &  809.1 & 16.6 & -20.1 $\pm$ 1.1 & 128 & 10.5$^{a}$ & 9.3$^{d}$ &           &  8$^{k}$  &  17 \\
         4037 (H)        & SBb   & 3.3 &  5.5  &  932.1 & 15.1 & -18.0 $\pm$ 0.1 & 101 &  9.7$^{a}$ & 8.5$^{b}$ & 0.2$^{e}$ & 32$^{k}$  & 151 \\
         4189 (V)        & SABcd & 5.9 &  5.3  & 2115.0 & 15.1 & -19.7 $\pm$ 0.5 & 196 & 10.5$^{a}$ & 9.4$^{b}$ &           & 31$^{k}$  &  70 \\
         4321 (V)        & SABbc & 4.0 & 16.3  & 1570.9 & 15.1 & -21.2 $\pm$ 0.1 & 279 & 10.9$^{a}$ & 9.4$^{b}$ & 6.2$^{e}$ & 38$^{k}$  & 149 \\ 
         4411B (H)       & SABcd & 6.2 &  5.5  & 1272.0 & 14.9 & -19.4 $\pm$ 0.2 &  76 &  9.5$^{a}$ & 9.0$^{b}$ &           & 18$^{k}$  &  52 \\
         4430 (H)        & SBb   & 3.5 &  4.9  & 1451.0 & 14.7 & -20.3 $\pm$ 0.5 &  87 &  9.8$^{a}$ & 8.6$^{b}$ & 0.7$^{e}$ & 28$^{m}$  &  60 \\
         4519 (V)        & SBd   & 6.9 &  6.9  & 1218.1 & 14.8 & -19.3 $\pm$ 0.2 & 117 & 10.2$^{a}$ & 9.5$^{b}$ & 0.9$^{e}$ & 40$^{k}$  & 180 \\
         4625 (S)        & SABm  & 8.7 &  3.6  &  620.9 & 11.3 & -17.9 $\pm$ 1.3 &  39 &  9.1$^{a}$ & 8.5$^{c}$ &           & 36$^{k}$  & 126 \\
         4725 (S)        & SABab & 2.2 & 45.0  & 1206.1 & 28.8 & -20.7 $\pm$ 0.2 & 257 & 10.9$^{a}$ & 9.6$^{b}$ & 1.6$^{e}$ & 30$^{m}$  &  28 \\   
         4736 (S)        & SAab  & 2.3 &  8.4  &  307.9 &  5.2 & -19.7 $\pm$ 0.3 & 182 & 10.5$^{a}$ & 8.3$^{c}$ &           & 36$^{k}$  & 118 \\
         5204 (G)        & SAm   & 8.9 &  4.4  &  200.9 &  5.9 & -17.0 $\pm$ 0.1 &  56 &  8.7$^{a}$ & 8.6$^{c}$ &           & 40$^{k}$  &  17 \\
         5334 (H)        & SBc   & 5.3 & 15.9  & 1385.9 & 26.1 & -19.1 $\pm$ 1.1 & 133 & 10.4$^{a}$ & 9.1$^{c}$ & 0.5$^{e}$ & 38$^{k}$  &   6 \\
         5430 (G)        & SBb	 & 3.1 & 16.1  & 2961.1 & 50.5 & -20.8 $\pm$ 0.4 & 188 & 10.7$^{a}$ &           &           & 32$^{k}$  &   8 \\
         5585 (G)        & SABd  & 6.9 &  7.1  &  292.9 &  8.2 & -18.7 $\pm$ 0.2 &  79 &  9.3$^{a}$ & 8.8$^{c}$ &           & 36$^{k}$  &  50 \\
         5668 (G)        & SAd   & 6.9 & 14.3  &  218.4 & 29.7 & -20.1 $\pm$ 0.6 &  73 & 10.2$^{a}$ & 9.9$^{b}$ & 1.6$^{e}$ & 18$^{k}$  & 146 \\
         5669 (H)        & SABcd & 6.0 & 15.8  & 1367.9 & 27.3 & -18.6 $\pm$ 0.7 & 102 & 10.2$^{a}$ & 9.3$^{b}$ & 1.0$^{e}$ & 36$^{m}$  &  71 \\
         5713 (S)        & SABbc & 4.0 & 13.5  & 1898.9	& 33.8 & -21.2 $\pm$ 0.5 & 108 & 10.6$^{a}$ &           &           & 33$^{k}$  &  23 \\
         6217 (G)        & SBbc  & 4.0 & 11.4  & 1362.0 & 26.1 & -20.5 $\pm$ 0.7 & 113 & 10.5$^{a}$ & 9.6$^{c}$ &           & 34$^{k}$  & 105 \\
         6946 (G)        & SABcd & 5.9 & 10.0  &   39.9 &  5.9 & -20.9 $\pm$ 0.2 & 315 &            & 9.9$^{c}$ &           & 17$^{m}$  &  61 \\
         UGC 3685 (G)    & SBb   & 3.0 & 14.6  & 1797.0 & 30.3 & -19.9 $\pm$ 0.6 &  41 &            & 9.8$^{c}$ &           & 12$^{k}$  & 134 \\
    \end{tabular}
    \caption{WiNDS galaxy targets and Data Archive parameters. 
    (1) Galaxy name in the NGC catalog except for UGC 3685 that do not have NGC name, the letter in parenthesis corresponds to the survey name: (G) for GHASP \citep{2008MNRAS.390..466E,2008MNRAS.388..500E}, (S) for SINGS \citep{2006MNRAS.368.1016D,2008MNRAS.385..553D}, (V) for Virgo \citep{2006MNRAS.366..812C} and (H) for HRS \citep{2019A&A...631A..71G};
    (2) and (3) Morphological Type found in the HyperLeda database;  
    (4) Optical radius considering the isophotal radius at the limiting surface brightness of 25 B mag arcsec$^{-2}$ from Third Reference Catalog of Bright Galaxies \citep[RC3,][]{1991rc3..book.....D} corrected for the Virgo and Great Attractor (GA) infall, assuming $H_{0} = 67.8\ \mathrm{ km s^{-1} Mpc^{-1}}$; 
    (5) Systemic velocity from NED;
    (6) Distance in Mpc from NED, corrected for the Virgo and GA infall, assuming $H_{0} = 67.8\ \mathrm{ km s^{-1} Mpc^{-1}}$; 
    (7) Absolute B-band magnitude (HyperLeda);
    (8) Maximum rotation velocity corrected for inclination from HyperLeda; 
    (9) Stellar mass ${a}$ from \citet{2010PASP..122.1397S};
    (10) HI mass ${b}$ from \citet{2018ApJ...861...49H}, ${c}$ from \citet{2013AJ....145..101K}, ${d}$ from radio survey Westerbork survey of HI in SPirals (WHISP)
galaxies \textit{https://www.astro.rug.nl/~whisp/};
    (11) Star formation rate. ${e}$ from \citet{2015A&A...579A.102B};
    (12) Inclinations ${k}$ correspond to kinematic values and ${m}$ correspond to morphologycal values from literature;
    (13) Position angle}.
    \label{tab:table1}
\end{table*}
\begin{table}[ht]
\centering
    \begin{tabular}{@{}ccccc@{6cm}}
    \hline
    Instrument              & Telescope         &  Galaxies         &     Survey            \\
    (1)                     & (2)               &       (3)         &        (4)            \\
    \hline

    GHASP                   & OHP               &     30            &  WiNDS, GHASP, HRS    \\
    Cigale                  & ESO               &     1             &        Virgo          \\
    FANTOMM                 & OMM/ESO/CFHT      &     8             & SINGS, Virgo          \\
    SAM-FP                  & SOAR              &     1             & WiNDS                 \\
\end{tabular}
    \caption{Setup. (1) Instrument; (2) Telescope; (3) Number of galaxies observed for each instrument; (4) Survey name.
    \newline OMM: Observatoire du mont M{\'e}gantic, Qu{\'e}bec, Canada, 1.6-m telescope; ESO: European Southern Observatory, La Silla, Chile, 3.6-m telescope; CFHT: 3.6-m Canada–France–Hawaii Telescope, Hawaii, USA, 3.6-m telescope}
    \label{tab:table2}
    
\end{table}

\subsection{Additional Data Archive} \label{subsec:add_data}

The new data cubes described in Section \ref{subsec:winds} were complemented with 28 galaxies previously observed, available at the Fabry-Perot interferometer database\footnote{\label{note1}\url{https://cesam.lam.fr/fabryperot/}} and Herschel database\footnote{\label{note2}\url{https://hedam.lam.fr/}} at the Laboratoire d'Astrophisique de Marseille.
The galaxies were selected from the GHASP (G), SINGS--$\mathrm {H_{\alpha}}$ (S), VIRGO--$\mathrm {H_{\alpha}}$ (V) and HRS--$\mathrm {H_{\alpha}}$ (H) surveys, each of them briefly discussed below. Some basic parameters, describing the corresponding observations, are listed in Table \ref{tab:table1}.

The data cubes available from the previously mentioned surveys were reprocessed in order to homogenize the galaxy analysis. This was performed considering the same signal-to-noise and smoothing method for all data cubes. Data cubes from the GHASP, VIRGO--$\mathrm {H_{\alpha}}$, and SINGS--$\mathrm {H_{\alpha}}$ surveys are available in the Fabry-Perot database website. On the other hand, HRS--$\mathrm {H_{\alpha}}$ data cubes are available in the Herschel Database.

\begin{figure*}
  \centering
  \includegraphics[width=150mm,clip,angle=0]{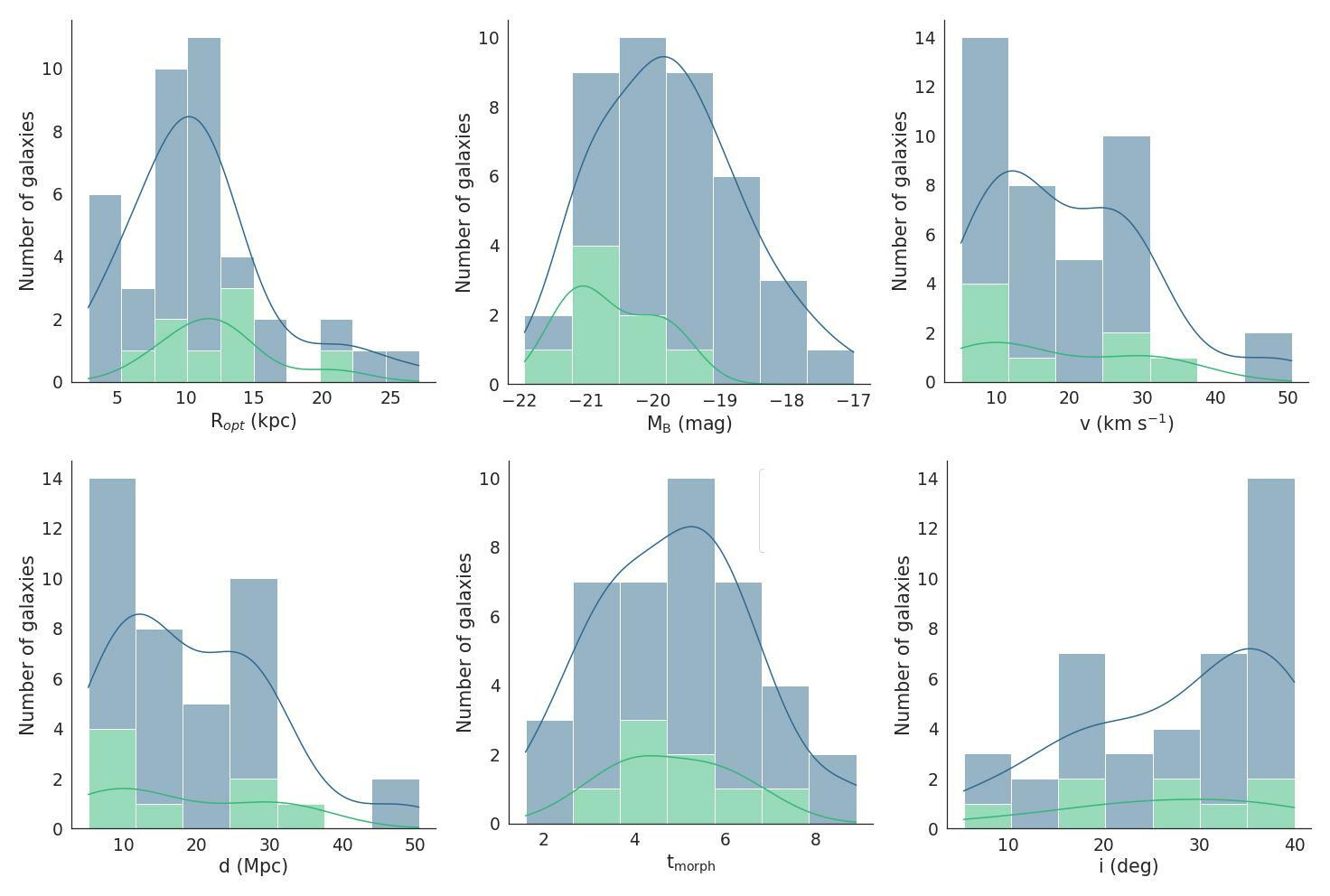}  

   \caption{Basic parameter distributions for our sample of 40 WiNDS galaxies. Top left: Optical radius distribution at the 25 mag $\mathrm{arcsec^{-2}}$ corrected for the effects of projection and extinction. 
   Top middle: Absolute B-band magnitudes. 
   Top right: Radial velocity distribution. 
   Bottom left: Distribution of distances in Mpc (corrected for the Virgo, Great Attractor and Shapley supercluster infall). 
   Bottom middle: Galaxy type distribution. 
   Bottom right: Inclination angles distribution.  
   The solid line corresponds to the smooth continuous approximation of the underlying distribution. The blue distributions correspond to the total sample of WiNDS galaxies, while the green distributions  to the vertically perturbed galaxy candidates.}
   \label{fig:hist_distribution}
\end{figure*}
\vspace{0.5cm}

\subsubsection{GHASP}
GHASP \citep{2008MNRAS.390..466E, 2008MNRAS.388..500E} survey consists of 3D $\mathrm {H_{\alpha}}$ data cubes of 203 late-type galaxies (spiral and irregular). This sample contains a wide range of morphological types, from Sa to Irr, with absolute magnitudes in the range of $-16\ \leq $ $M_{B}$  $ \leq -22\ \mathrm{mag}$, stellar masses ranging from $ 10^{9}$ M$_{\odot}$ to  5 $ \times 10^{11}$\  M$_{\odot}$ and are located in nearby low-density environments. From the total sample of 203 galaxies, 83 are strongly barred galaxies (SB or IB) and 53 moderately barred galaxies (SAB or IAB).
According to our selection criteria described in Section \ref{sec:data_sample}, thirteen galaxies were selected and are reported in Table \ref{tab:table1}. 

\subsubsection{SINGS--$\mathrm {H_{\alpha}}$}

The Legacy survey $Spitzer$ Infrared Nearby Galaxies Survey \citep[SINGS,][]{2003PASP..115..928K} consists of 75 nearby galaxies selected with a wide coverage of morphological type (E to Im), luminosity types (IR quiescent to luminous IR galaxies) and CO/HI ratio covering over 3 orders of magnitudes. These galaxies are located in different environments such as galaxy groups, clusters, and low-density fields. The SINGS survey was created to characterize the infrared emission in a wide range of galaxy properties and star formation environments. The SINGS--$\mathrm {H_{\alpha}}$ data is a $\mathrm {H_{\alpha}}$ kinematic follow-up survey of SINGS \citep{2006MNRAS.367..469D,2008MNRAS.385..553D} and consists of 65 late-type galaxies which present HII regions. According to the criteria described in Section \ref{sec:data_sample}, the selected galaxies are reported in Table \ref{tab:table1}.

\subsubsection{VIRGO--$\mathrm {H_{\alpha}}$}

The  3D $\mathrm {H_{\alpha}}$ data cubes from the Virgo survey, VIRGO--$\mathrm {H_{\alpha}}$, \citep[][]{2006MNRAS.366..812C} consists of 30 spiral and irregular galaxies. This survey is a subsample of the Virgo Cluster Catalog \citep{1985AJ.....90.1681B,1993A&AS...98..275B}, which is the nearest cluster to the Milky Way and counts  $\sim$ 1400 members, mainly dwarf-type galaxies. The galaxies  considered in VIRGO--$\mathrm {H_{\alpha}}$ have an apparent magnitude greater than $B^{0}_{t}$ = 12 mag, wide morphological types (S0/a to Im), and  inclination angles between 25$^{\circ}$ and  89$^{\circ}$. All galaxies are located in the cluster's core and its extension towards M49. 
According to our criteria selection described in Section \ref{sec:data_sample}, the three selected galaxies are reported in Table \ref{tab:table1}.

\subsubsection{HRS--$\mathrm {H_{\alpha}}$}
The Herschel Reference Survey, HRS--$\mathrm {H_{\alpha}}$, \citep[][]{2019A&A...631A..71G} consists of 152 star-forming galaxies observed using the 1.93-m telescope at the OHP. This survey aimed to study the relationship between the baryonic and dynamical mass of galaxies. The HRS--$\mathrm {H_{\alpha}}$ galaxies sample spans a wide range of morphologies (from Sa to Sm-types, including Blue Compact Dwarfs) and stellar masses 
($10^{8}\ M_{\rm \odot} \leq M_{\rm star} \leq 10^{11}\  M_{\rm \odot}$).

From the HRS--$\mathrm {H_{\alpha}}$ sample, and excluding overlap with previous surveys (GHASP, SINGS--$\mathrm {H_{\alpha}}$ and VIRGO--$\mathrm {H_{\alpha}}$), we found seven galaxies that follow our selection criterion. The names of the six targets are listed in Table \ref{tab:table1}.

\begin{figure*}
  \centering
  \includegraphics[width=150mm,clip,angle=0]{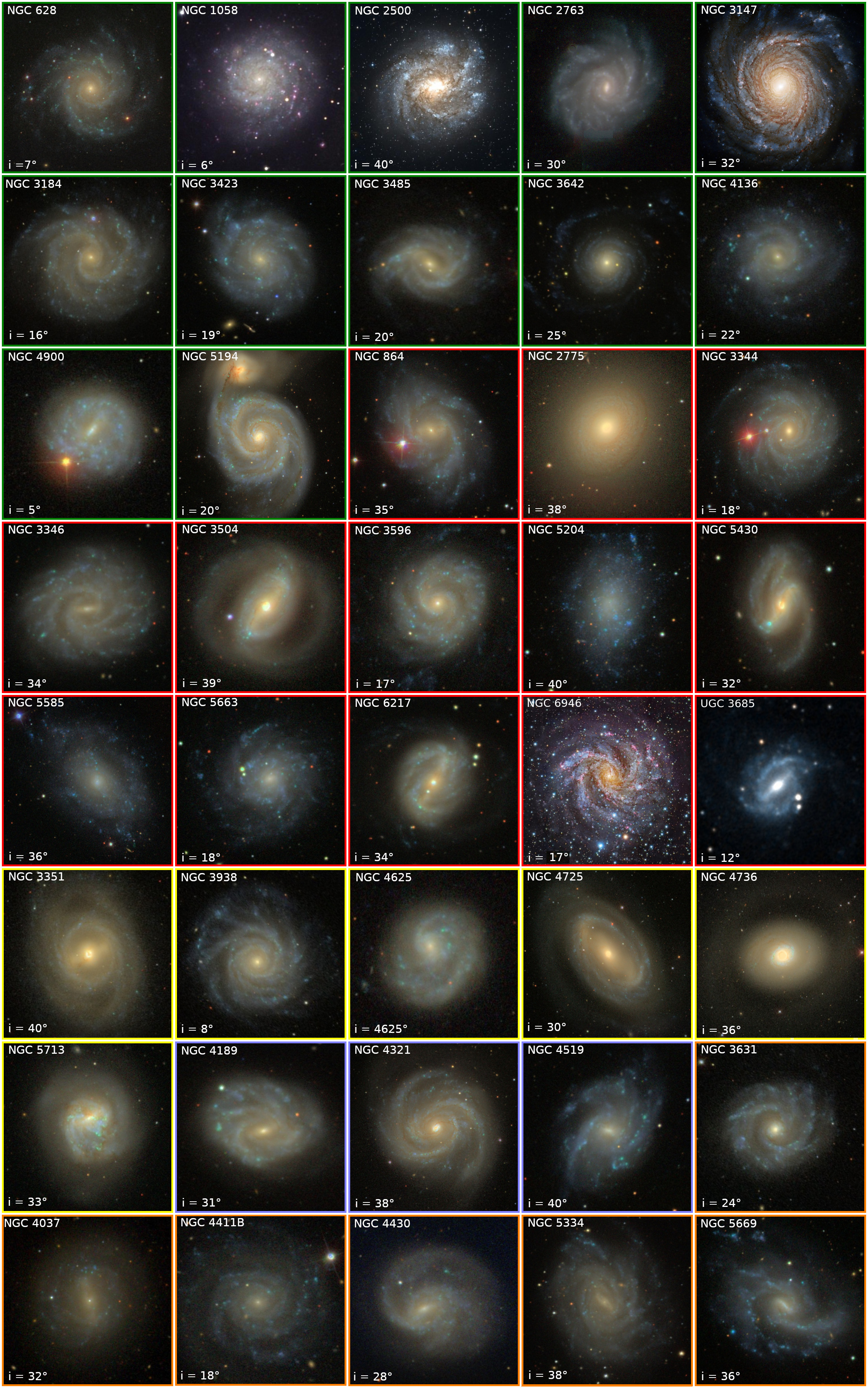}  
  
  \caption{Optical images of WiNDS galaxies from SDSS:  New observations (green outline); GHASP (red outline); SINGS (yellow outline); VIRGO survey (blue outline) and HRS (orange outline)}
  \label{fig:sample}
\end{figure*}

Note that not all of the galaxies come from the same source. Table \ref{tab:table2} shows our observations broken-down by instrument, telescope, survey of origin, and number of galaxies observed. The most relevant observations parameters are listed in Table \ref{tab:table3}. For completeness, in the same table we also list the properties of the additional observations extracted from the data archive.

\section{WiNDS: Data Reduction}
\label{sec:reduction}

The WiNDS data sample  was reduced using the pipeline based program \textsc{computeeverything}\footnote{/https://projets.lam.fr/projects/computeeverything} and the \textsc{reducWizard} \footnote{https://projets.lam.fr/projects/fpreducwizard} interface, following the steps described in \citet{2006MNRAS.368.1016D} and \citet{2008MNRAS.388..500E}.

The data reduction process includes {\it a)} the wavelength calibration, {\it b)} the night sky lines subtraction, {\it c)} the astrometry process, {\it d)} the adaptive spatial binning using the 2D Voronoi tessellation, whose implementation finally produces a smoothing specially adapted to a given SNR, {\it e)} the generation of the 2D momentum maps, {\it f)} a semi-automatic cleaning of the velocity fields and {\it g)} velocity dispersion correction. 
In the following we briefly describe these processes:

\begin{enumerate}[label={\it \alph*)}]
\item The calibration process consists of applying a phase map to provide the same wavelength/velocity origin to each profile, for each pixel of the observation datacube. The 2D phase map is computed from the two calibration datacubes. During the phase correction process, the individual exposures are re-centered with respect to each other, using field stars, to minimize telescope drifts and instrument bend.

\item The OH emission lines are the main foreground sky contamination. The sky subtraction is done by dividing the collapsed datacube into a galaxy-dominant and a sky-dominant region. The night-sky emission is interpolated from this galaxy-free region using a two-degree polynomial, and subsequently removed from the whole datacube.

\item The next step is astrometry which is done through the KOORDS task in the KARMA\footnote{KARMA tools package is available on website https://www.atnf.csiro.au/computing/software/karma/} package \citep{1996ASPC..101...80G}, comparing star fields between XDSS\footnote{ESO Online Digitized Sky Survey https://archive.eso.org/dss/dss} $R$-band images and our continuum images of each galaxy.

\item For the estimation of The Signal-to-Noise ratio ($S/N$), which is related to the flux in the line, to the spectral resolution and the r.m.s. in the continuum. We use adapted spatial binning through Voronoi tessellation, as described in \citet{2006MNRAS.368.1016D}. This allows us to obtain the highest spatial resolution for a given $S/N$, which is the main advantage of this binning technique. The Voronoi bins are constructed iteratively from a given pixel, by accreting adjacent pixels until the resulting spectrum reaches the desired $S/N$. Pixels where the $S/N$ is greater than the chosen threshold value will not be affected by the binning process, while regions of low $S/N$ will emerge from noise without being contaminated by adjacent regions of higher $S/N$.
For the WiNDS sample, the objective is to obtain a $S/N$ $\geq 7$ per spatial bin, where $S/N$ is considered as the square root of the flux in the line. We implemented a Hanning smoothing of a one-spectral-element of the spectrum which preserves the flux.

\item From this Voronoi binned datacube, the different momentum maps are computed as explained in \citet{2006MNRAS.368.1016D}: after identifying the $\mathrm {H_{\alpha}}$ line boundaries, the continuum emission around the line  is estimated and subtracted from the spectrum. The moments of the line are then estimated. Namely, the line or monochromatic flux is the zeroth-order moment (intensity integrated within the line boundaries), the line-of-sight velocity ($V_{los}$) is the first-order moment (intensity-weighted velocity sum within line boundary), and the LoS velocity squared velocity dispersion (variance) as the second-order moment (difference between the intensity-weighted squared velocity and squared of the intensity-weighted velocity within the line boundaries). 

\item In some cases, the lowest $S/N$ regions are strongly affected by night sky line contamination.  Thus a semi-automatic cleaning is performed. Indeed, to
achieve the desired signal-to-noise ratio in the outer- most galactic regions very large Voronoi bins are produced. These are mainly associated with sky subtraction residuals and background emission. Those bins are first semi-automatically erased and then manually deleted when the semi-automatic process is not sufficient.

\item Finally, velocity dispersion maps were corrected by instrumental width. This correction was done by subtracting the mean dispersion of the instrument contribution to the observed velocity dispersion map, as follows:

\begin{equation}
\label{eq:sigma_corr}
\sigma_{corr} = \sqrt{\sigma_{obs} ^{2} - \sigma_{inst} ^{2}}
\end{equation}

\noindent where $\sigma_{obs}$ corresponds to the observed velocity dispersion and $\sigma_{inst}$ is the mean dispersion of the instrument contribution. Moreover, $\sigma_{inst}$ is a function of the $\mathrm{FSR}$, the  interference order $p$, and the resolving power $R$:

\begin{equation}
\label{eq:sigma_inst}
\sigma_{inst} = \frac{\mathrm{FSR}\ p}{R}
\end{equation}
\end{enumerate}

The results of the data reduction process are presented below for a WINDS subsample where panels a), b) and c) show the XDSS B-band image, the $\mathrm {H_{\alpha}}$ line-of-sight velocity map, and the $\mathrm {H_{\alpha}}$ velocity dispersion map, respectively. In addition, the results of the data reduction for the remaining new observations made in this work, where XDSS blue image (top left panel), the $\mathrm {H_{\alpha}}$ monochromatic image (top right panel), $\mathrm {H_{\alpha}}$ velocity field (bottom left panel), and the $\mathrm {H_{\alpha}}$ residual velocity field (bottom right panel) are shown in Appendix \ref{sssec:new_obs_maps} and \ref{sssec:monochromatic_maps}.

\begin{table*}[ht]
\centering
    \begin{tabular}{@{}ccccccccccc@{}}
    \hline
    NGC  & UGC & RA         & DEC                    & Cube        & Pixel scale & Int Order & Date         & $\lambda_{scan}$ & R    & seeing \\
         &     & (hh mm ss) & ($^{\circ}$ $'$ $''$)  & $x y z$     & (arc sec)   &           & (yyyy-mm-dd) & (\AA)            &      & (arc sec) \\ 

    (1)  & (2) & (3)        & (4)                    & (5)         & (6)         & (7)       & (8)          & (9)              & (10) & (11) \\

    \hline \hline

    New observations &   &            &           &              &       &      &            &        &       &        \\
    628           & 1149 & 01 36 41.8 & 15 47 00  & 512 512 32   & 0.68  & 2600 & 2019-02-05 & 6577.2 & 28600 & $\star$\\
    1058          & 2193 & 02 43 30.1 & 37 20 28  & 512 512 32   & 0.68  & 798  & 2019-02-03 & 6574.2 & 10000 & $\star$ $\star$\\ 
    2500          & 4165 & 08 01 53.2 & 50 44 14  & 512 512 32   & 0.68  & 798  & 2019-02-04 & 6573.8 & 10000 & $\star$ $\star$\\
    2763 (SAM-FP) &      & 09 06 49.1 & -15 29 59 & 1024 1028 48 & 0.045 & 609  & 2016-02-18 & 6604.2 & 10000 & $\star$\\
    3147          & 5332 & 10 16 53.7 & 73 24 03  & 512 512 32   & 0.68  & 798  & 2019-02-08 & 6624.1 & 10000 & $\star$ $\star$\\
    3184          & 5557 & 10 18 16.9 & 41 25 28  & 512 512 32   & 0.68  & 2600 & 2019-02-05 & 6575.8 & 28600 & $\star$ $\star$\\
    3423          & 5962 & 10 51 14.3 & 05 50 24  & 512 512 32   & 0.68  & 2600 & 2019-02-07 & 6584.8 & 28600 & $\star$ $\star$\\
    3485          & 6077 & 11 00 02.4 & 14 50 29  & 512 512 32   & 0.68  & 798  & 2019-02-03 & 6574.2 & 10000 & $\star$ $\star$\\
    3642          & 6385 & 11 22 17.9 & 59 04 28  & 512 512 32   & 0.68  & 2600 & 2019-02-06 & 6597.6 & 28600 & $\star$ $\star$\\
    4136          & 7134 & 12 09 17.7 & 29 55 39  & 512 512 32   & 0.68  & 2600 & 2019-02-06 & 6576.1 & 28600 & $\star$ $\star$\\
    4900          & 8116 & 13 00 39.2 & 02 30 03  & 512 512 32   & 0.68  & 798  & 2019-02-04 & 6583.8 & 10000 & $\star$ $\star$\\
    5194          & 8493 & 13 29 52.7 & 47 11 42  & 512 512 32   & 0.68  & 2600 & 2019-02-06 & 6572.9 & 28600 & $\star$\\

\hline

    Data Archive  &      &            &           &              &       &      &            &        &                & \\
    864 (G)       & 1736  & 02 15 27.6 & 06 00 09 & 512 512 24  & 0.68  & 793 & 2000-10-23 & 6597.0 & $\simeq$ 10000 & $\star$ $\star$\\
    2775 (G)      & 4820  & 09 10 20.1 & 07 02 17 & 512 512 24  & 0.68  & 793 & 2003-03-06 & 6592.4 & $\simeq$ 10000 & $\star$ $\star$\\
    3344 (G)      & 5840  & 10 43 31.2 & 24 55 20 & 512 512 24  & 0.68  & 793 & 2002-03-20 & 6575.5 & $\simeq$ 10000 & $\star$ $\star$\\
    3346 (G)      & 5842  & 10 43 38.9 & 14 52 19 & 512 512 24  & 0.68  & 793 & 2004-03-20 & 6590.7 & $\simeq$ 10000 & $\star$ $\star$\\
    3351 (S)      & 5850 & 10 43 57.7 & 11 42 13  & 512 512 48   & 1.61  & 765  & 2005-02-03 & 6579.8 & 13750          & $\star$ $\star$ $\star$\\
    3504 (G)      & 6118  & 11 03 11.2 & 27 58 21 & 512 512 24  & 0.68  & 793 & 2002-03-18 & 6596.2 & $\simeq$ 10000 & $\star$ $\star$\\
    3596 (G)      & 6277  & 11 15 06.2 & 14 47 13 & 512 512 24  & 0.68  & 793 & 2004-03-19 & 6588.9 & $\simeq$ 10000 & $\star$ $\star$\\
    3631 (H)      & 6360 & 11 21 02.9 & 53 10 10  & 512 512 32   & 0.68  & 798  & 2016-02-03 & 6587.3 & $\simeq$ 10000 & $\star$ $\star$ $\star$\\
    3938 (S)      & 6856 & 11 52 49.5 & 44 07 15  & 512 512 48   & 1.61  & 765  & 2004-03-11 & 6579.7 & 12852          & $\star$ $\star$ $\star$\\
    4037 (H)      & 7002 & 12 01 23.7 & 13 24 04  & 512 512 32   & 0.68  & 798  & 2016-02-15 & 6582.4 & $\simeq$ 10000 & $\star$ $\star$ $\star$\\
    4189 (V)      & 7235 & 12 13 47.3 & 13 25 29  & 512 512 24   & 0.68  & 793  & 2003-03-07 & 6609.1 & 7950           & $\star$ $\star$\\
    4321 (V)      & 7450 & 12 22 54.8 & 15 49 19  & 512 512 52   & 1.61   & 899  & 2003-02-25 & 6597.2 & 21000          & $\star$ $\star$\\
    4411B (H)     & 7546  & 12 26 47.2 & 08 53 05 & 512 512 32  & 0.68  & 798 & 2016-02-04 & 6589.0 & $\simeq$ 10000 & $\star$ $\star$ $\star$\\
    4430 (H)      & 7566 & 12 27 26.4 & 06 15 46  & 512 512 32   & 0.68  & 798  & 2016-02-15 & 6593.7 & $\simeq$ 10000 & $\star$ $\star$ $\star$\\
    4519 (V)      & 7709 & 12 33 30.3 & 08 39 17  & 512 512 48   & 1.61  & 899  & 2003-04-04 & 6589.5 & 21000          & $\star$\\
    4625 (S)      & 7861 & 12 41 52.7 & 41 16 26  & 512 512 48   & 0.48  & 899  & 2003-04-06 & 6576.4 & 14294          & $\star$ $\star$ $\star$\\
    4725 (S)      & 7989 & 12 50 26.6 & 25 30 03  & 512 512 48   & 1.61  & 765  & 2004-02-19 & 6589.2 & 14305          & $\star$ $\star$ $\star$\\
    4736 (S)      & 7996 & 12 50 53.1 & 41 07 14  & 512 512 48   & 1.61  & 765  & 2005-05-11 & 6569.6 & 12745          & $\star$ $\star$ $\star$\\
    5204 (G)      & 8490  & 13 29 36.5 & 58 25 07 & 256 256 24  & 0.68  & 793 &            & 6567.2 & $\simeq$ 10000 & $\star$ $\star$\\
    5334 (H)      & 8790 & 13 52 54.5 & -01 06 53 & 512 512 32   & 0.68  & 798  & 2018-12-07 & 6592.1 & $\simeq$ 10000 & $\star$ $\star$ $\star$\\
    5430 (G)      & 8937  & 14 00 45.7 & 59 19 42 & 512 512 24  & 0.96  & 897 & 2003-03-03 & 6627.6 & $\simeq$ 10000 & $\star$ $\star$\\
    5585 (G)      & 9179  & 14 19 48.2 & 56 43 45 & 512 512 24  & 0.68  & 793 & 2004-03-20 & 6569.2 & $\simeq$ 10000 & $\star$ $\star$\\
    5668 (G)      & 9363  & 14 33 24.3 & 04 27 02 & 512 512 24  & 0.68  & 793 & 2003-04-26 & 6597.3 & $\simeq$ 10000 & $\star$ $\star$\\
    5669 (H)      & 9353 & 14 32 43.5 & 09 53 26  & 512 512 32   & 0.68  & 798  & 2016-02-16 & 6591.9 & $\simeq$ 10000 & $\star$ $\star$ $\star$\\
    5713 (S)      & 9451 & 14 40 11.5 & -00 17 20 & 512 512 24   & 0.42  & 609  & 2004-04-13 & 6604.4 & 5785           & $\star$ $\star$ $\star$\\
    6217 (G)      & 10470 & 16 32 39.2 & 78 11 53 & 512 512 24  & 0.68  & 793 &           & 6592.6  & $\simeq$ 10000 & $\star$ $\star$\\
    6946 (G)      & 11597 & 20 34 52.3 & 60 09 14 & 512 512 24  & 0.68  & 793 & 2002-06-14 & 6563.7 & $\simeq$ 10000 & $\star$ $\star$\\ 
                  & 3685  & 07 09 05.9 & 61 35 44 & 512 512 24  & 0.68  & 793 & 2002-03-17 & 6602.1 & $\simeq$ 10000 & $\star$ $\star$\\

    \end{tabular}
    \caption{Log of the WiNDS. 
    (1) Galaxy name in the NGC catalog except for UGC 3685 that do not have NGC name, the letter corresponds to the survey name; 
    (2) Name of the galaxy in the UGC catalog when available 
    (3-4) Right Ascension and Declination (J2000); 
    (5) Cube dimension: \textit{x} (Right ascension $\alpha$), \textit{y} (Declination $\delta$) and \textit{z} (number of channels);  
    (6) Pixel scale; 
    (7)Interference order (at the $\mathrm {H_{\alpha}}$ rest wavelength); 
    (8) Date of observation; 
    (9) Scanning wavelength; 
    (10) Spectral resolution ($\Delta \lambda / \lambda $)  according to the computed Finesse; 
    (11) $\star$ is for a seeing < 2 arc sec, $\star$ $\star$ is for a seeing between 2 and 4 arc sec, $\star$ $\star$ $\star$ see papers (H) \citep{2019A&A...631A..71G} and (S) \citep{2006MNRAS.367..469D,2008MNRAS.385..553D}}
    \label{tab:table3}.
\end{table*}

\section{Residual Velocity Fields}
\label{sec:residual_field}

Our goal is to analyze the previously derived line-of-sight velocity map, \textit{V}$_{\rm los}$, to search for evidence of large, global and, coherent kinematic perturbations on our sample of late-type, low-inclination galaxies. In particular, we seek for perturbations that are consistent with bending modes produced by warps and corrugation patterns. In this Section, we discuss the procedure followed to  analyze our resolved velocity fields. 

As discussed in \citet{2021ApJ...908...27G}, the observed \textit{V}$_{\rm los}$ field of each galaxy is bound to contain contributions from three different velocity components: the distributions within the disk plane, radial \textit{V}$_{\rm R}$ and  rotational, \textit{V}$_{\rm rot}$; and the perpendicular velocity  to the disk plane, \textit{V}$_{\rm Z}$. More precisely, each \textit{V}$_{\rm los}$ can be described as:

\begin{align}
\label{eq:vlos_general}
\begin{split}
\textit{V}_\mathrm{los} = & \textit{V}_\mathrm{sys}(r) + \textit{V}_\mathrm{\rm rot}(r) \cos \theta \sin i \\
                 & + \textit{V}_\mathrm{R}(r) \sin \theta \sin i                 \\
                 & + \textit{V}_\mathrm{z}(r) \cos i,
\end{split}
\end{align}

\noindent where \textit{V}$_\mathrm{sys}$ is the systemic velocity of the galaxy and it is considered as a fixed value. The polar coordinates (r, $\theta$) in the plane of the galaxy are measured from the position angle (PA), inclination (i) and sky position ($\alpha$, $\delta$) of the rotation center.
The first step to identify global perturbations in the data is to subtract from the \textit{V}$_{\rm los}$ map an axisymmetric model of \textit{V}$_{\rm rot}$. To obtain such an axy-symmetric velocity model we assume that, at first order, the contributions from the radial and the vertical velocities to \textit{V}$_{\rm los}$ are negligible. Therefore, the observed velocity Equation \ref{eq:vlos_general} reduces to:

\begin{equation}
\label{vlos_first_order}
\textit{V}_\mathrm{los}  = \textit{V}_\mathrm{sys}(r) + \textit{V}_\mathrm{rot}(r) \cos \theta \sin i
\end{equation}

To obtain \textit{V}$_\mathrm{rot}(R)$ we use the tilted-ring method \citep{1987PhDT.......209B}, which assumes that the galaxy can be analyzed using concentric rings along the major axis, which are described by the parameters \textit{V}${_{\rm rot}^r,i^r,PA^r}$. Here the supra index $r$ indicates that we are referring to the rings. 
In this work, we use an improved tilted-ring method, described in detail in \citet{2008MNRAS.390..466E}, and derive \textit{V}$_\mathrm{rot}(R)$ for each galaxy bin is modeled using the modified Zhao function \citep{2008MNRAS.390..466E},

\begin{equation}
\label{eq:vrot}
\textit{V}_\mathrm{rot}(r)=\textit{v}_\mathrm{t}\dfrac{(r/r_\mathrm{t})^{g}}{1+(r/r_\mathrm{t})^{a}}.
\end{equation}

\begin{table}[ht]
\centering
    \begin{tabular}{@{}cccccc@{}}
    \hline
    ID        & i$_{k}$           & PA$_{k}$          & $V_{sys}$     & v$_{rot}^{max}$\\
              & ($^{\circ}$)      & ($^{\circ}$)      & ($km s^{-1}$) & ($km s^{-1}$)  \\
    (1)       & (2)               & (3)               & (4)           & (5) \\
    \hline \hline

    New observations \\

      NGC 628           &  7$^{*}$    & 155 $\pm$ 2  &  658 $\pm$  1  & 147 \\ 
      NGC 1058          &  6$^{*}$    &  34 $\pm$ 2  &  521 $\pm$  0  & 140 \\ 
      NGC 2500          & 41 $\pm$ 11 &  85 $\pm$ 2  &  512 $\pm$  1  &  49 \\
      NGC 2763  & 30 $\pm$  8 &  47 $\pm$ 1  & 1884 $\pm$  1  & 123 \\ 
      NGC 3147          & 32 $\pm$ 20 &  37 $\pm$ 8  & 2894 $\pm$  3  & 164 \\ 
      NGC 3184          & 16 $\pm$  9 &   5 $\pm$ 1  &  592 $\pm$  0  & 155 \\ 
      NGC 3423          & 40 $\pm$  8 & 134 $\pm$ 1  & 1007 $\pm$  1  &  76 \\ 
      NGC 3485          & 30 $\pm$  7 & 156 $\pm$ 2  & 1428 $\pm$  1  & 146 \\ 
      NGC 3642          & 36 $\pm$  8 &  70 $\pm$ 8  & 1581 $\pm$  1  &  72 \\
      NGC 4136          & 20 $\pm$  6 & 108 $\pm$ 1  &  606 $\pm$  1  &  82 \\ 
      NGC 4900          &  5$^{*}$    &  80 $\pm$ 2  &  959 $\pm$  1  & 297 \\ 
      NGC 5194          & 20 $\pm$ 11 &  13 $\pm$ 2  &  464 $\pm$  2  & 189 \\ 

      \hline 
      Data Archive \\
      
      NGC 864 (G)       & 35 $\pm$ 18 &  30 $\pm$ 3  & 1525 $\pm$  3  &  52 \\
      NGC 2775 (G)      & 38 $\pm$ 3  & 157 $\pm$ 0  & 1350 $\pm$  1  & 262 \\
      NGC 3344 (G)      & 15 $\pm$ 11 & 153 $\pm$ 2  &  580 $\pm$  1  & 262 \\
      NGC 3346 (G)      & 29 $\pm$  8 & 113 $\pm$ 2  & 1245 $\pm$  1  &  95 \\
      NGC 3351 (S)      & 21 $\pm$  5 & 169 $\pm$ 1  &  781 $\pm$  1  & 289 \\
      NGC 3504 (G)      & 39 $\pm$ 13 & 160 $\pm$ 3  & 1523 $\pm$  2  & 100 \\
      NGC 3596 (G)      & 17 $\pm$ 12 &  78 $\pm$ 2  & 1190 $\pm$  2  &  87 \\
      NGC 3631 (H)      & 24 $\pm$ 14 & 151 $\pm$ 2  & 1150 $\pm$  2  &  59 \\
      NGC 3938 (S)      & 14 $\pm$  8 & 164 $\pm$ 1  &  809 $\pm$  0  & 143 \\      
      NGC 4037 (H)      & 32 $\pm$ 20 & 151 $\pm$ 3  &  924 $\pm$  1  &  50 \\
      NGC 4189 (V)      & 31 $\pm$ 11 & 169 $\pm$ 1  & 2102 $\pm$  1  & 128 \\
      NGC 4321 (V)      & 25 $\pm$  9 &  29 $\pm$ 1  &  172 $\pm$  1  & 190 \\ % 1.57 & 25.61 \\
      NGC 4411B (H)     & 18 $\pm$ 15 &  49 $\pm$ 3  & 1267 $\pm$  1  &  37 \\
      NGC 4430 (H)      & 28 $\pm$ 13 &  75 $\pm$ 2  & 1424 $\pm$  1  & 180 \\
      NGC 4519 (V)      & 40 $\pm$ 16 & 180 $\pm$ 4  & 1216 $\pm$  2  &  124 \\
      NGC 4625 (S)      &  8 $\pm$ 12 & 124 $\pm$ 3  &  626 $\pm$  1  & 105 \\
      NGC 4725 (S)      & 43 $\pm$  4 & 156 $\pm$ 1  & 1198 $\pm$  2  & 227 \\
      NGC 4736 (S)      & 25 $\pm$  7 & 122 $\pm$ 1  &  318 $\pm$  1  & 245 \\
      NGC 5204 (G)      & 40 $\pm$ 11 & 166 $\pm$ 1  &  186 $\pm$  1  &  87 \\
      NGC 5334 (H)      & 38 $\pm$  4 &   6 $\pm$ 1  & 1402 $\pm$  7  & 105 \\
      NGC 5430 (G)      & 32 $\pm$  8 &   2 $\pm$ 1  & 2963 $\pm$  3  & 255 \\
      NGC 5585 (G)      & 36 $\pm$ 20 &  53 $\pm$ 3  &  298 $\pm$  2  &  59 \\
      NGC 5668 (G)      & 18 $\pm$ 14 & 146 $\pm$ 2  & 1575 $\pm$  1  &  94 \\ 
      NGC 5669 (H)      & 36 $\pm$  9 &  65 $\pm$ 1  & 1377 $\pm$  1  & 116 \\
      NGC 5713 (S)      & 12 $\pm$ 12 & 154 $\pm$ 3  & 1896 $\pm$  1  &  84 \\ 
      NGC 6217 (G)      & 18 $\pm$ 10 & 104 $\pm$ 1  & 1356 $\pm$  1  & 207 \\
      NGC 6946 (G)      & 17 $\pm$ 14 &  61 $\pm$ 2  &   38 $\pm$  2  &  185 \\ 
      UGC 3685 (G)      & 12 $\pm$ 19 & 118 $\pm$ 10 & 1815 $\pm$  3  &  25 \\

\end{tabular}
    \caption{Model Parameters. (1) Galaxy name, the letter corresponds to the survey name; 
    (2) Kinematic inclination, those marked with an asterisk (*) have been fixed equal to morphological value; 
    (3) Kinematic Position Angle; 
    (4) Systemic Velocity;
    (5) Maximum rotation $\mathrm {H_{\alpha}}$ velocity corrected for inclination derived in this work}
    \label{tab:table4}
    
\end{table}

This four-parameter model was especially chosen for its versatility to adjust to a very diverse set of different rotation curve shapes. The four parameters involved in the \textit{V}$_\mathrm{rot}$ model are \textit{v}$_\mathrm{t}$ and \textit{r}$_\mathrm{t}$ which correspond to the velocity and radius when the rotation curve changes from an increasing velocity to a flat regime, respectively; $a$ and $g$ which are related with sharpness of the turnover. As discussed in \citet{2008MNRAS.390..466E}, for each galaxy the fitting procedure requires a set of initial values for the Zhao function parameters as well as for ($\alpha$, $\delta$), $i$ and $PA$.  In our work, the values of ($\alpha$, $\delta$) are set to the peak of the emission in the continuum image and set as fixed values. For the  systemic velocity, $PA$, and $i$, initial values were extracted from the literature. 
The resulting values of $i$ and  $PA$, obtained as a result of the fitting procedure, are referred to as kinematically inferred values. Note that, in those cases  where disks are projected nearly perfectly face-on ($ i < 10^{\circ}$), the value of $i$ is not allowed to vary and, thus, is kept fixed to their estimated morphological value. Instead, \textit{v}$_\mathrm{t}$, \textit{r}$_\mathrm{t}$, $a$, $g$ parameters are always allowed to vary. Finally, \textit{V}$_\mathrm{rot}$ for each Voronoi binning is estimated through the minimization of $\chi^{2}$ based on the Levernberg-Marquardt method \citep{1992nrca.book.....P}, computing an iterative 3.5$\sigma$ clipping on the observed bin-centroid velocity field. 
All kinematical parameters ($i_{k}, PA_{k}, V_{sys}$) estimated for WiNDS galaxies are listed in Table \ref{tab:table4}. The estimation of errors in the determination of kinematic parameters for the WiNDS sample was calculated using the power spectrum of the residual velocity field and the application of a Monte Carlo method, as described in more detail in \citet[][]{2008MNRAS.390..466E}. It is worth noting that the degeneracy between $V_{\rm max}$ and $i$ when fitting our models \citep{1989A&A...223...47B}, especially for very low inclined disk, does not affect our results as we are focusing our analysis on residual velocity fields, $V_{\rm res}$. The $\mathrm {H_{\alpha}}$ $V_{\rm max}$ values  derived in this work are listed in Table \ref{tab:table4}

Once an unperturbed axisymmetric model of \textit{V}$_{\rm los}$ is obtained for each galaxy, we generate their corresponding \textit{V}$_{\rm res}$ fields. This is done by subtracting the properly \textit{V}$_{\rm los}$ model from the observed \textit{V}$_{\rm los}$ maps. It is worth recalling that global and coherent features in a \textit{V}$_{\rm res}$ field of a late-type galaxy could be the result of kinematic perturbations induced by features such as a bar or spiral structures, but also from improper model parameters (such as center, systemic velocity, position angle of major axis, inclination or rotation curve model). The multi-polar signatures expected for ill-defined parameters \citep{1973MNRAS.163..163W} are not observed in our data.

It is also worth noticing that small variations in the magnitude of the \textit{V}$_{\rm res}$ could be attributed to uncertainties in the determination of the kinematic parameters (PA and center). In order to verify the sensitivity of the method used to the initially chosen and fixed inclination value, the model is re-run considering the extreme values of the inclination that are within the error range shown in Table \ref{tab:table4}.

The contributions from in-plane flows are expected to be small in low inclination galaxies such as those studied in this work. Using a suite of numerical simulations, \citet{2021ApJ...908...27G} characterized the contribution from in plane flows to the \textit{V}$_{\rm los}$ driven by non-axisymmetric structures. Their models consisted of a disk galaxy, projected into an inclination of $35^{\circ}$, considering a bar with similar characteristics to that of the Milky Way and spiral overdensities that ranged from 100\% to 1000\% density contrast ($\Delta \rho / \rho$) with respect to the disk background density. Even for $\Delta \rho / \rho = 1000$ the resulting velocity perturbations in the corresponding \textit{V}$_{\rm res}$ field were $\lesssim 10\ \mathrm{km\ s^{-1}}$.

\section{Quantification and Selection Criteria of Bending Modes}
\label{sec:quantification}

\begin{figure}
  \centering
  \includegraphics[width=65mm,clip,angle=0]{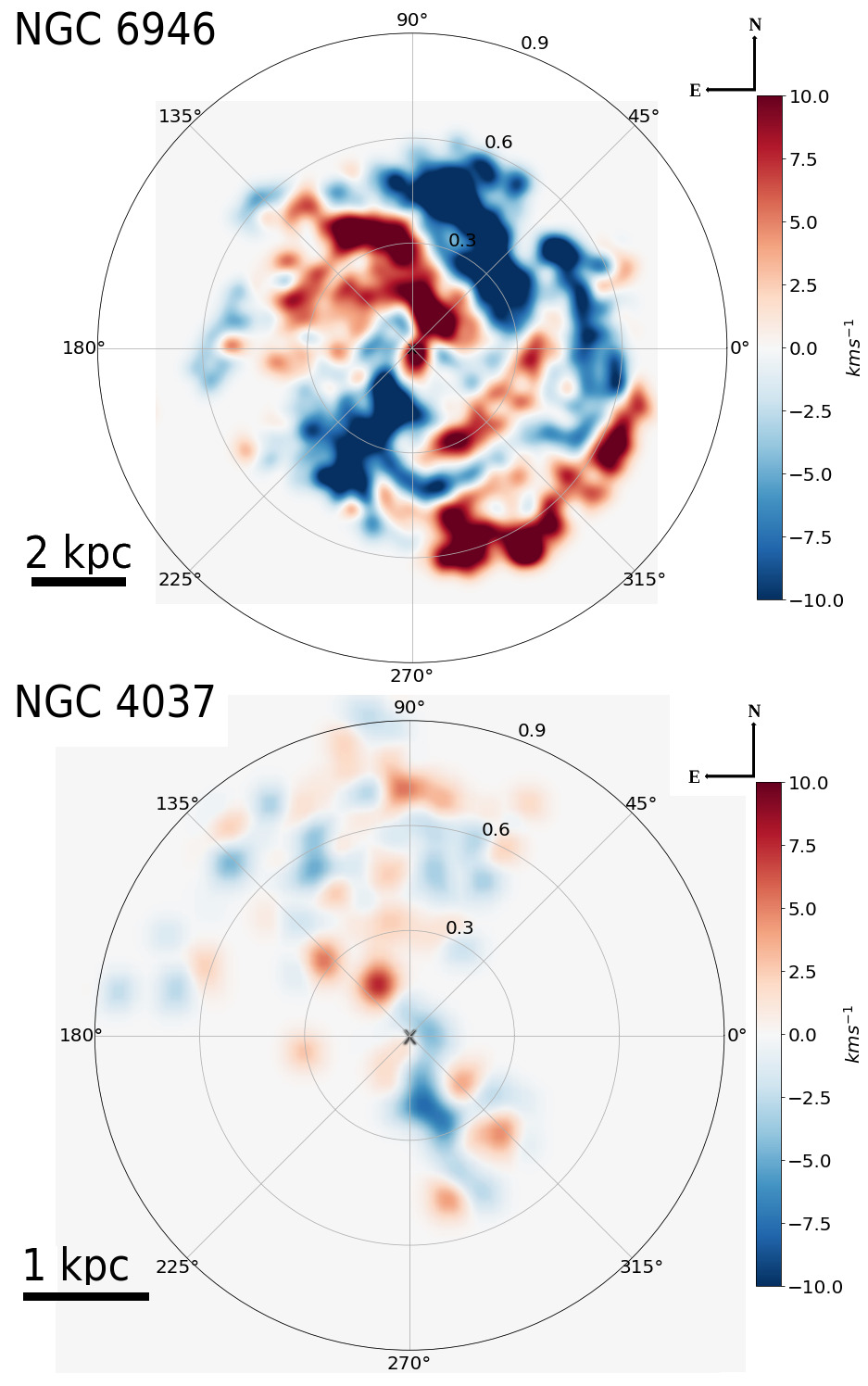}  
  
  \caption{Filtered \textit{V}$_\mathrm{res}$ images. Top panel: NGC 6946 galaxy with well-defined velocity perturbations covering azimuthal extensions > $60^{\circ}$ and radial extension $\leq 0.3\ R_{opt}$ at a given azimuthal radius. Bottom panel: NGC 4037 galaxy without signs of global perturbation. The contours represent 0.3 $R_{opt}$ for each galaxy.}
  \label{fig:example_filter_image}
\end{figure}

In this Section we describe the selection criteria applied to select those galaxies within the WiNDS sample that show velocity perturbations consistent with a bending mode, such as a warp or corrugation pattern. 
To avoid selecting discrete and local vertical perturbations, such as those associated, e.g., fountain flows we have established the main criteria to select strong candidates are the following:
\begin{itemize}
\item We focus on galaxies that present an extended $\mathrm {H_{\alpha}}$ coverage, extending for at least 0.7 $R_{\rm opt}$. This criterion allows us to globally explore the kinematics of the disks, especially on the  outer regions where bending modes typically show their stronger amplitudes.

\item We focus on galaxies where global perturbations show amplitudes that are $> 10\ \mathrm{km\ s^{-1}}$. This enhances the chances that the observed perturbations in this low inclination sample of disks are not mainly driven by the axisymmetric components of the galaxies.

\item {We applied a Gaussian low-pass filter to the \textit{V}$_\mathrm{res}$ images in Fourier space with cutoff frequency equivalent to a spatial distance of $\approx 500$ pc to highlight relevant and large perturbations} (see Appendix \ref{sssec:low_pass_filter}). 
This allows us to smooth out local and discrete perturbations, and focus on large-scale and coherent features. Based on these images, using a circular grid centered at the center of the galaxy, evenly spaced by 0.3 $R_{\rm opt}$ as shown in Fig \ref{fig:example_filter_image}, we select as potential candidates all those galaxies where the perturbations cover at least an azimuthal extension of 60 degrees. We consider  perturbations that at a given azimuthal angle cover a radial extension  $\lesssim 0.3$ $R_{\rm opt}$ as corrugation patterns, whereas the remaining as typical warps.

\end{itemize}

Examples of the resulting filtered images are shown in Figure \ref{fig:example_filter_image}.
The upper panel shows an example of well-defined global and coherent velocity perturbations. The bottom panel shows the case where the map was consistent with an unperturbed velocity field.  According to the these criteria eight galaxies show \textit{V}$_\mathrm{res}$ maps consistent with a vertical perturbed disk.

 We note that previous studies have analyzed velocity dispersion maps, \textit{V}$_{\rm disp}$, to identify vertically perturbed disk galaxies. For example, \citet{2000A&A...358..812J} used \textit{V}$_{\rm disp}$ to detect plausible shells or chimneys in the galactic disk of NGC 5668. This was done by comparing the geometry of the \textit{V}$_{\rm res}$ and \textit{V}$_{\rm disp}$ maps and associating local perturbations in both maps to vertical motions in the galaxy. In our work, \textit{V}$_{\rm disp}$ maps are obtained as a sub product of the data reduction process (see Section \ref{sec:reduction}) and, thus, we will present them together with the \textit{V}$_{\rm res}$ maps. However, as opposed to \citet{2000A&A...358..812J}, our goal here is to identify global rather than localized and discrete velocity perturbations. As a result, \textit{V}$_{\rm disp}$ maps were not explored in detail. We will further analyze these maps in follow-up work.

We emphasize that our selection criteria cannot confirm or rule out the presence of vertical perturbations in our disks. Although the objective is to detect global and coherent perturbations, it should be noted that our method does not exclude the presence of fountain flows, which are shown as local and discrete perturbations but rather, both types of perturbations could coexist. Nonetheless, note that we have carefully checked the $\mathrm {H_{\alpha}}$ intensity profiles in all galaxies that were selected as potential vertically-perturbed objects. We found that all candidates within the WiNDS sample show, in general, well-behaved Gaussian profiles. That is, there is no presence of strong and intense multiple components in the $\mathrm {H_{\alpha}}$ profiles due to the presence of other velocity components  \citep[such as those observed in merging systems such as HCG31, for example,][]{2007A&A...471..753A}.

\section{Results}
\label{sec:results_candidates}

In this Section, we analyze and discuss the resulting \textit{V}$_\mathrm{res}$ fields of the eight WiNDS galaxies that fulfilled our quantification criteria described in Section \ref{sec:quantification}. However, we discuss the remaining new observed galaxies in the Appendix \ref{sssec:comments_new_obs} and we present in Appendix \ref{sssec:maps_without_perturbations} the \textit{V}$_\mathrm{res}$ maps of galaxies that do not accomplish the criteria described in Section \ref{sec:quantification}.

\begin{figure*}
  \centering
  \includegraphics[width=180mm,clip,angle=0]{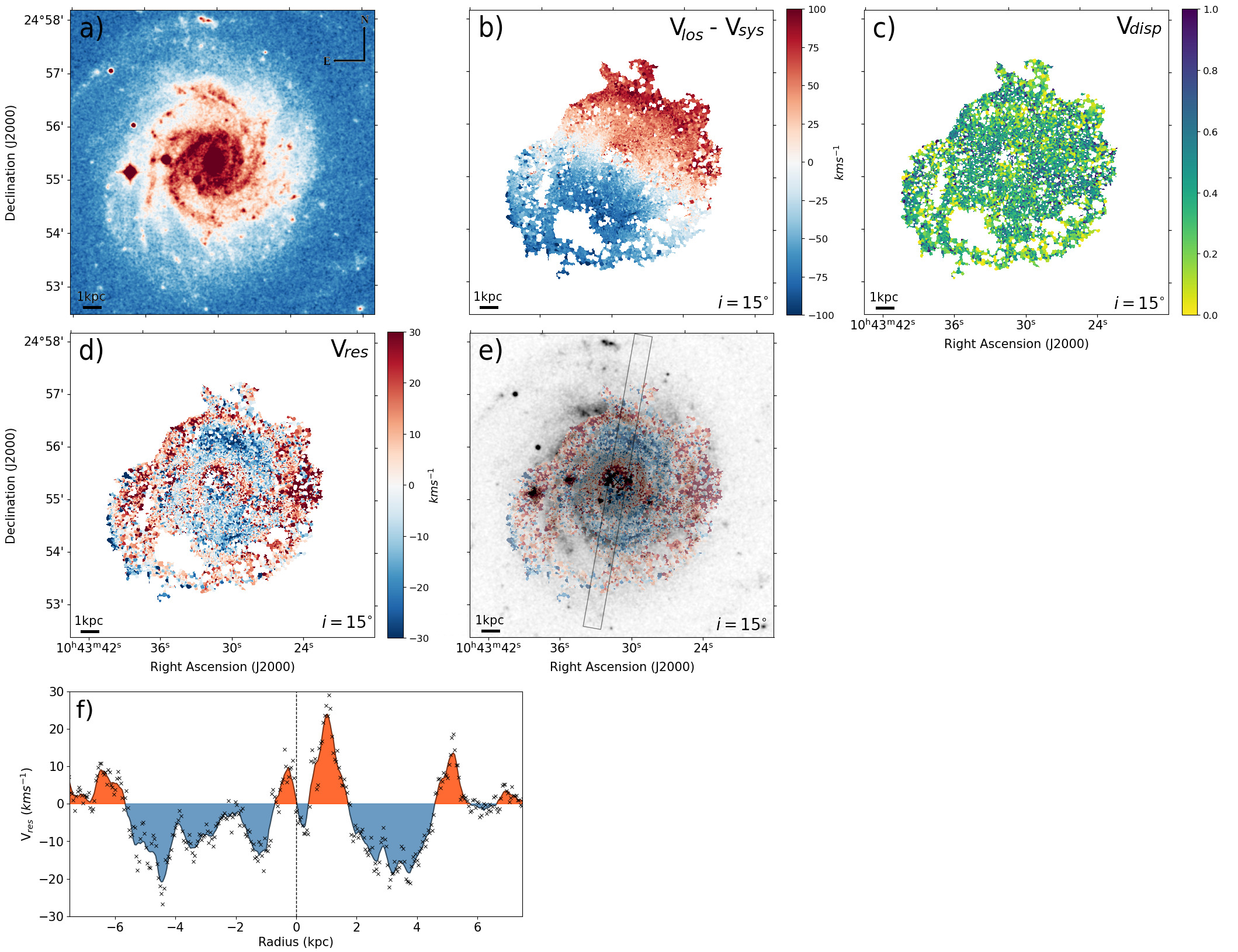}  

  \caption{\textbf{NGC 3344}. Panel (a): XDSS blue-band image.
  Panel (b): $\mathrm {H_{\alpha}}$velocity field.
  Panel (c):  $\mathrm {H_{\alpha}}$velocity dispersion map normalized to $44.2 \ \mathrm{km\ s^{-1}}$ corresponding to the 95th percentile corrected for instrumental broadening.
  Panel (d): $\mathrm {H_{\alpha}}$residual velocity field.
  Panel (e): Optical band image with $\mathrm {H_{\alpha}}$residual velocity field.
  Panel (f): Residual velocity radial profile considering a slit indicated with black lines with position angle of 10$^{\circ}$ in Panel (e). The black line corresponds to smoothed curve and the dashed line fits center of galaxy indicated in the panel (e). The red (blue) area corresponds to velocities above (below) the midplane of the galaxy. The derived kinematic inclination is indicated in the lower right of each panel.}
  \label{fig:candidate_ngc3344}
\end{figure*}

 In Figure~\ref{fig:candidate_ngc3344}, we show the results obtained for the galaxy NGC 3344. This galaxy has a very low morphologycal inclination angle \citep[$i\sim18$\degr,][]{1997A&AS..124..109P} and, thus, represents an ideal candidate to search for vertical perturbations on its galactic disk. In panel a) we show a $B$-band image of the galaxy, obtained from the XDSS. The galaxy shows a well-defined flocculent spiral structure and a weak bar inside the inner ring \citep{2000A&A...356..827V}. It has an optical radius, $R_{\rm opt}$, of 7.7 $\mathrm{kpc}$ \citep[RC3,][]{1991rc3..book.....D}, i.e. approximately half the size of the Milky Way. In panel b) we show the resulting \textit{V}$_{\rm los}$ map, obtained following the procedure outlined in Section \ref{sec:reduction}. The map shows a good $\mathrm {H_{\alpha}}$ coverage of the overall disk ($1\ R_{\rm opt}$), despite being limited by the FoV of the GHASP instrument. The amplitude of \textit{V}$_{\rm los}$ reaches $\approx 100$ km/s. Panel c)  shows the $\mathrm {H_{\alpha}}$ velocity dispersion map, normalized to  $\sigma_{95th} = 44.2\ \mathrm{km\ s^{-1}}$ 
\footnote{A normalization of the velocity dispersion to the 95th percentile was used for all galaxies in order to avoid outlying dispersion measurements and better show possible correlations between residual velocity fields and velocity dispersion.}

 Panel d) shows the residual velocity field \textit{V}$_{\rm res}$, obtained after subtracting the axisymmetric rotational velocity model from \textit{V}$_{\rm los}$. Interestingly the $V_{\rm res}$ map reveals a global, strong, and coherent oscillating-like pattern throughout the entire disk of the galaxy, with an amplitude of the order of $\pm 30\ \mathrm{km\ s^{-1}}$. To compare the disk morphology to the structure observed on its residual velocity field, on panel e) we show the galaxy $B$-band image with the contours obtained from the $V_{\rm res}$ field.  
 Notice that no clear correlation between axisymmetric patterns and velocity perturbations can be observed.
 The extended black box or slit, placed across the disk on panel e), highlights the multiple transitions from positive to negative values experienced by \textit{V}$_{\rm res}$ as a function of galactocentric distance. This is better shown on panel g) where we highlight the behaviour of \textit{V}$_{\rm res}$ across this particular slit (black crosses). Here the black line represents the corresponding smoothed data using a moving average function with a 7 data points window, while the dashed vertical line indicates the galaxy centre. The direction of the slit was chosen to highlight  transitions of the residual velocity values.
 NGC 3344 is a galaxy that is cataloged as isolated, with no nearby galaxies within a projected radial distance of 150 kpc and radial velocity difference $\Delta \leq 500\ \mathrm{km\ s^{-1}}$. 
 We recall that NGC 3344 has a very mildly inclined disk, $i \sim 18^{\circ}$. Thus, the contribution from in-plane flows to the resulting \textit{V}$_{\rm res}$, associated with e.g. bars and/or spiral structure, is expected to be small.
 As previously discussed, a plausible interpretation of these \textit{V}$_{\rm res}$ perturbations is associated with the contribution from a vertical corrugation pattern. Such corrugation pattern could be the result of several mechanisms, such as close interactions with satellites \citep{2008A&ARv..15..189S,1999MNRAS.303L...7J,2010MNRAS.408..783R,2013MNRAS.434.3142A,2014ApJ...780..105R,2017MNRAS.465.3446G}, torques associated with either misaligned triaxial DM halos or even DM overdensity wakes \citep{1989MNRAS.237..785O,1993ApJ...403...74Q,1999MNRAS.304..254V,2003ApJ...583L..79B,2009ApJ...700.1896K,2013MNRAS.429..159G,2016ApJ...823....4D}. Vertical perturbations on this galaxy have been previously reported by other authors. Indeed \citet{1990ApJ...352...15B} reports the presence of a warp in the outer regions of NGC 3344 after analysing observations of the HI 21 cm line. \citet{2000A&A...356..827V} presented a detailed study of this galaxy, also using HI 21 cm lines. The authors suggest a possible relationship between the inner and outer star-forming rings with a strong spiral structure and the warp in the HI layer. However, they were not able to clearly link the warp structure with the perturbation in the more internal regions. Our analysis of NGC 3344, with a spectral resolution $\sim 10$ km s$^{-1}$, reveals a very complex structure of NGC 3344. NGC 3344 is a very interesting candidate of the vertically perturbed galaxy that will be analyzed in greater detail on a follow-up study.

\begin{figure*}
  \centering
  \includegraphics[width=180mm,clip,angle=0]{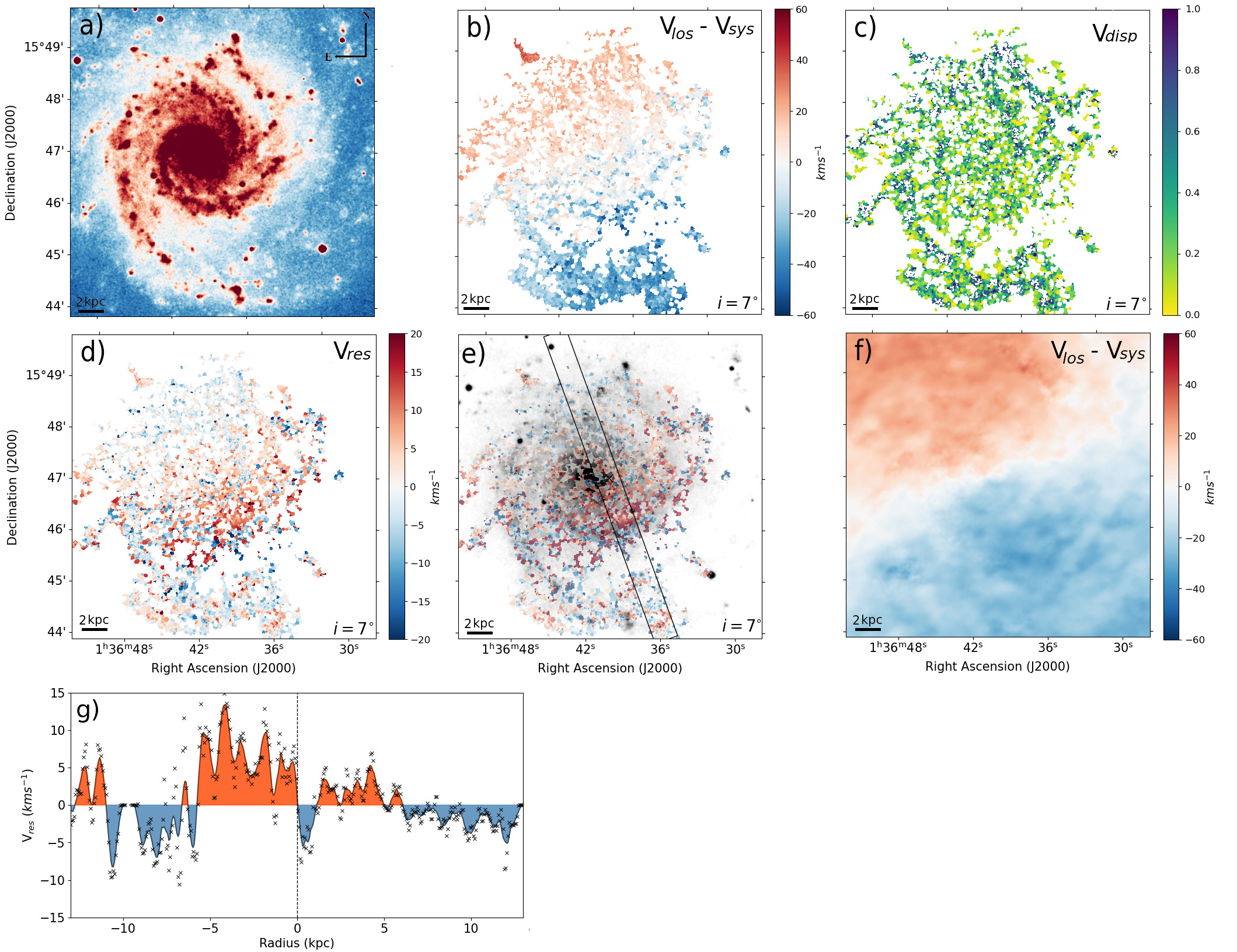}  

  \caption{\textbf{NGC 628}. Panel (a): XDSS blue-band image. 
  Panel (b): $\mathrm {H_{\alpha}}$ velocity field. 
  Panel (c): $\mathrm {H_{\alpha}}$ velocity dispersion map normalized to $14\ \mathrm{km s^{-1}}$ corresponding to the 95th percentile.
  Panel (d): $\mathrm {H_{\alpha}}$residual velocity field. 
  Panel (e): Optical band image with $\mathrm {H_{\alpha}}$ residual velocity field. 
  Panel (f): HI velocity field from THINGS survey \citet{2008AJ....136.2563W}. 
  Panel (g): Residual velocity radial profile considering a slit indicated with black lines with position angle of 160$^{\circ}$ in Panel (e). 
  The black line correspond to smoothed curve and the dashed line fits center of galaxy indicated in the panel (e). The red (blue) area corresponds to velocities above (below) the midplane of the galaxy. The derived kinematic inclination is indicated in the lower right of each panel.}
   \label{fig:candidate_ngc628}
\end{figure*}

In Figure~\ref{fig:candidate_ngc628} we now show the results obtained for NGC 628 (M74). This is a nearly face-on galaxy with an inclination angle of $6.5^{\circ}$; thus, an ideal candidate to search for possible vertical perturbations. On panel a) we display a $B$-band image of the galaxy, obtained from the XDSS. NGC 628 is a non-barred galaxy with two main spiral arms emerging from its bulge.
The $\mathrm {H_{\alpha}}$ observation coverage of NGC 628 corresponds to a mean radius of 12.5 $\mathrm{kpc}$, which represents 0.8 $R_{\rm opt}$. The $\mathrm {H_{\alpha}}$ velocity map (panel b), shows an amplitude of $\sim 60$ km/s, as expected from a low inclination galaxy. For comparison we also show on panel f) the HI velocity map from The HI Nearby Galaxy Survey (THINGS) \citep[THINGS]{2008AJ....136.2563W}, observed with Very Large Array (VLA), which is consistent with the $\mathrm {H_{\alpha}}$ velocity map. Interestingly,  studies based on neutral hydrogen \citep{1992A&A...253..335K} have detected the presence of an elongated warp structure at around 12$'$ projected distance from the nucleus. On panel c) we show the derived $\mathrm {H_{\alpha}}$ velocity dispersion map, normalized to $\sigma_{95th} = 14\ \mathrm{km\ s^{-1}}$. 
The \textit{V}$_{\rm res}$ field of NGC 628 is shown on  panel d). Interestingly, global and coherent velocity perturbations, of the order of $20\ \mathrm{km\ s^{-1}}$, can be clearly observed. Panel e) allows us to correlate NGC 628-disk morphology with its residual velocity map. The velocity perturbations on this galaxy are more clearly highlighted on panel g) where we follow the mean \textit{V}$_{\rm res}$ along the highlighted slit, suggesting a warp-like structure.

As shown on its $B$-band image (panel a), an extended tail is observed in the southwest direction of the disk. This substructure  has been previously reported in \citet{1992A&A...253..335K}. Although NGC 628 has currently no close companions, a dwarf galaxy system (UGC 1176 and UGC 1171) is located at 140 kpc. The overall mass of the satellite system has been reported to lie on the order of magnitude of the total NGC 628 mass \citep{2020A&A...638A..47M}. This system may have played an important role in the evolution of NGC 628. As discussed by \citet{1986ApJ...300..613B}, if the dwarf system is at rest  at its current location concerning for to NGC 628, it would fall into its host in about 1 - 2 Gyr. However, a more recent study by \citet{2020A&A...638A..47M} suggests that a tidal origin for the asymmetric HI tail located on the south-western outskirt is an unlikely scenario. They argue that the tail does not resemble recent tidal features, whereas an older feature would wind almost symmetrically around the galaxy. In addition, recently induced tidal tails typically show $m=2$ patterns, with two nearly symmetrical arms. Instead, NGC 628 shows a strong $m=1$ like spiral structure. Nonetheless, they do not rule out a possible previous interaction with the neighboring dwarf pair. It is interesting to compare with the results presented in \citet{2017MNRAS.465.3446G}, based on fully cosmological simulations. In particular, their model S13 shows similar morphology to NGC 628, with a strong $m=1$ arm extending from the inner galactic regions. 
This model displays a vertically perturbed stellar and cold gas disk, but no recent interaction with massive satellites ($M_{\rm sat}/M_{\rm host }> 0.03$). Instead, S13 experienced a massive gas-rich merger 5 Gyr in the past. As a result, its pre-existing disk is destroyed but a new thin disk quickly forms thereafter due to the re-accretion of misaligned gas. This gas accretion gives rises to both the vertical perturbations observed in this modeled galaxy and its strong $m=1$ morphology. 
This suggests that the velocity perturbations observed in  NGC 628, as well as its $m=1$ spiral morphology, could be linked to the recent accretion of misaligned cold gas.

\begin{figure*}
  \centering
  \includegraphics[width=180mm,clip,angle=0]{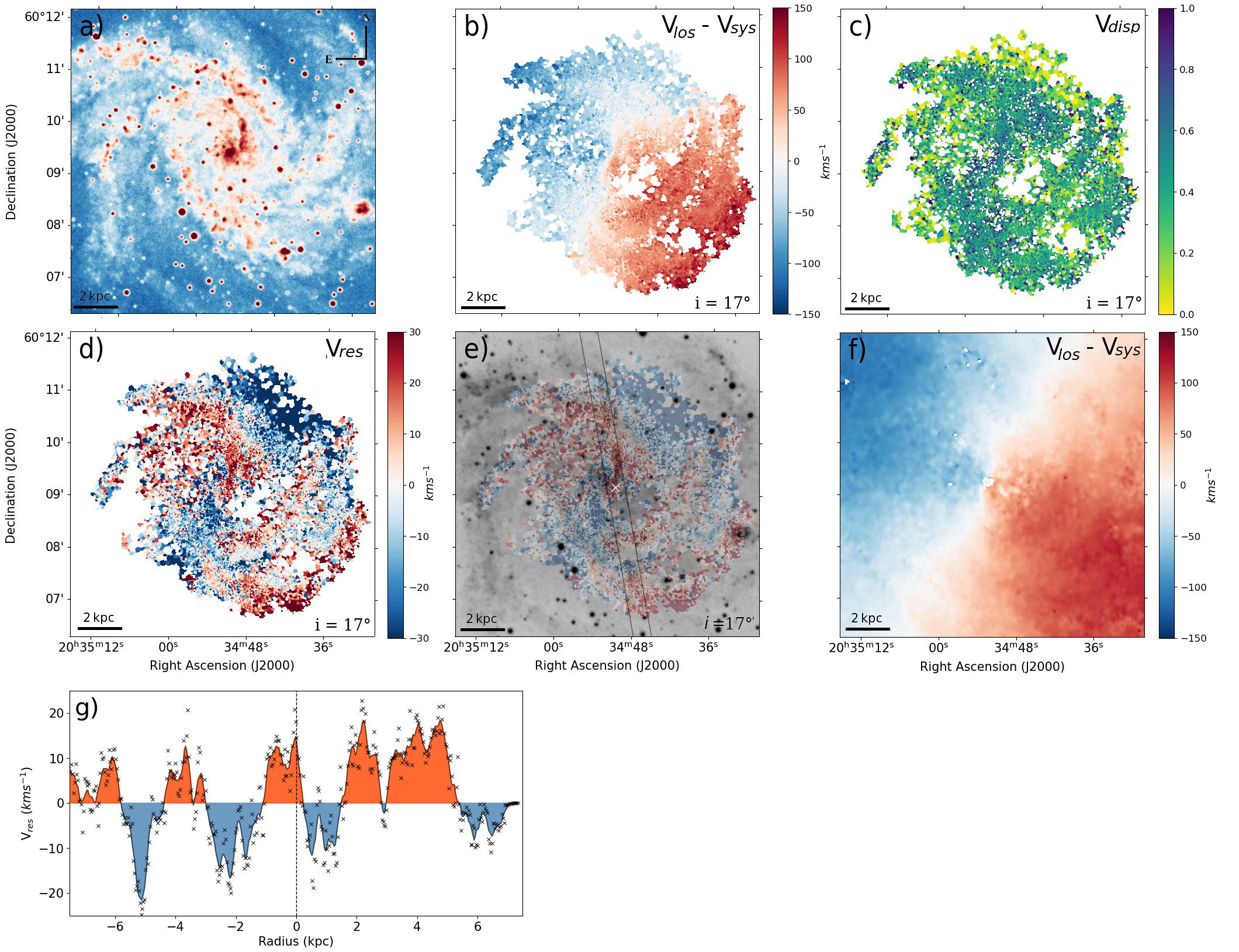}  

  \caption{\textbf{NGC 6946}. Panel (a): XDSS blue-band image. 
  Panel (b): $\mathrm {H_{\alpha}}$velocity field. 
  Panel (c): $\mathrm {H_{\alpha}}$velocity dispersion map normalized to $55.1\ \mathrm{km s^{-1}}$ corresponding to the 95th percentile.
  Panel (d): $\mathrm {H_{\alpha}}$residual velocity field. 
  Panel (e): Optical band image with $\mathrm {H_{\alpha}}$residual velocity field.  
  Panel (f): HI velocity field from THINGS survey \citet{2008AJ....136.2563W}. 
  Panel (g): Residual velocity radial profile considering a slit indicated with black lines with position angle of 170$^{\circ}$ in Panel (e). 
  The black line corresponds to smoothed curve and the dashed line fits center of galaxy indicated in the panel (e). The red (blue) area corresponds to velocities above (below) the midplane of the galaxy. The derived kinematic inclination is indicated in the lower right of each panel.}
   \label{fig:candidate_ngc6946}
\end{figure*}

In Figure~\ref{fig:candidate_ngc6946} we focus on NGC 6946. This is a  spiral galaxy classified as SABcd  with a kinematical inclination of $i \sim 17^{\circ}$, an optical radius of $\approx 10$ kpc, and an estimated total mass of $1.2 \times 10^{11}$ M$_{\odot}$ \citep{2000MNRAS.319..821P}. Optical and neutral hydrogen observations of NGC 6946 have revealed its strong spiral arms, a particularly high star formation rate, a nuclear starburst, and a weak bar \citep{2000MNRAS.319..821P,2006ApJ...649..181S}. Panel a) shows its XDSS $B$-band image, where we can clearly see its spiral nature. NGC 6946 is considered to be isolated \citep{1988JBAA...98..316T}. Using HI observations, previous studies have detected a dozen of low mass irregular dwarf galaxies companions \citep{2000MNRAS.319..821P, 2000A&A...362..544K, 2005A&A...434..935K}. However, at the present day NGC 6946 does not show any sign of undergoing strong direct gravitational interactions with its satellite population. Its two most massive companions, UGC 11583 and L149 have estimated total masses of $1.5 \times  10^9$ and $3.0 \times 10^8$ M$_{\odot}$ and projected distances of 83 and 75 kpc, respectively \citep{2000MNRAS.319..821P}. The $\mathrm {H_{\alpha}}$ \textit{V}$_{\rm los}$ maps of NGC 6946, shown in panel b), has an amplitude of $\sim 150\ \mathrm{km\ s^{-1}}$. Following our selection criteria, the map reveals a significant $\mathrm {H_{\alpha}}$ coverage, reaching at least a radius of 7 kpc, corresponding to 0.7 $R_{\rm opt}$. As in the case of NGC 3344, the coverage for NGC 6946 is limited by the FoV of the GHASP instrument.
For comparison, we show (on panel f) the HI velocity map, extracted from \citet{2008AJ....136.2563W}. Both maps are consistent, even though they are considering different components of the ISM and come from different instruments and techniques. Note, however, that our observations have a better spatial resolution, of 3 arcsec against the 6 - 12 arcsec resolution in the HI observations. The $\mathrm {H_{\alpha}}$ velocity dispersion map is shown on panel c), normalized to $\sigma_{95th} = 55.2\ \mathrm{km\ s^{-1}}$. On panels d) and e) we show the galaxy \textit{V}$_{\rm res}$ map, which reveals  global and coherent perturbations, reaching amplitudes $\lesssim 35\ \mathrm{km\ s^{-1}}$. The  $V_{\rm res}$ velocity contours, overlaid on the $B$-band image (panel e), allow us to directly contrast this velocity structure with the morphology of NGC 6946. The oscillating nature of this velocity field is better  highlighted on panel g), where we show the mean \textit{V}$_{\rm res}$ along the highlighted slit on panel e). 
 It is worth recalling NGC 6946 spiral nature, and  its low but not negligible inclination. As a result, a contribution from in-plane flow to the $V_{\rm res}$ field can be expected. However, as shown by \citet{2021ApJ...908...27G}, for disks as inclined as NGC 6946 ($i \sim 40^{\circ}$), even an spiral structure 1000 times denser than its mean background disk density cannot generate velocity perturbations with amplitudes $\gtrsim 10$ km/s. This indicates that the observed global velocity perturbation could be partially linked to vertical velocity flows. 
 
It is interesting to consider what mechanisms could be driving the perturbations in this galaxy. As previously discussed, NGC 6946 does not have nearby massive satellites that could be directly tidally interacting with its disk. However, even low mass satellites at relatively large galactocentric distances could significantly perturb an embedded disk through the excitation of dark matter overdensity wakes \citep{1998MNRAS.299..499W, 2000ApJ...534..598V}. For example, using cosmological simulations of Milky Way-mass galaxies, \citet{2016MNRAS.456.2779G} studied the onset and evolution of a strong vertical pattern in the disk. The vertical pattern in this model, with an amplitude of $\sim 50\ \mathrm{km\ s^{-1}}$, is the result of a satellite-host halo-disk interaction. Interestingly, the satellite had a total mass of $\sim 5$ per cent of the host and a pericentre distance of 80 kpc. The satellite was not massive enough to directly perturb the galactic disk but the density field of the host dark matter halo responded to the satellite passage and strongly amplified its  perturbative effects. A similar scenario could be taking place in the case of NGC 6946, considering its low mass companions at a relatively large projected galactocentric distance. Other possible mechanisms, in addition to the previously discussed contribution from in-plane flows,  could be related to smooth accretion of misaligned cold gas.

\begin{figure*}
  \centering
  \includegraphics[width=180mm,clip,angle=0]{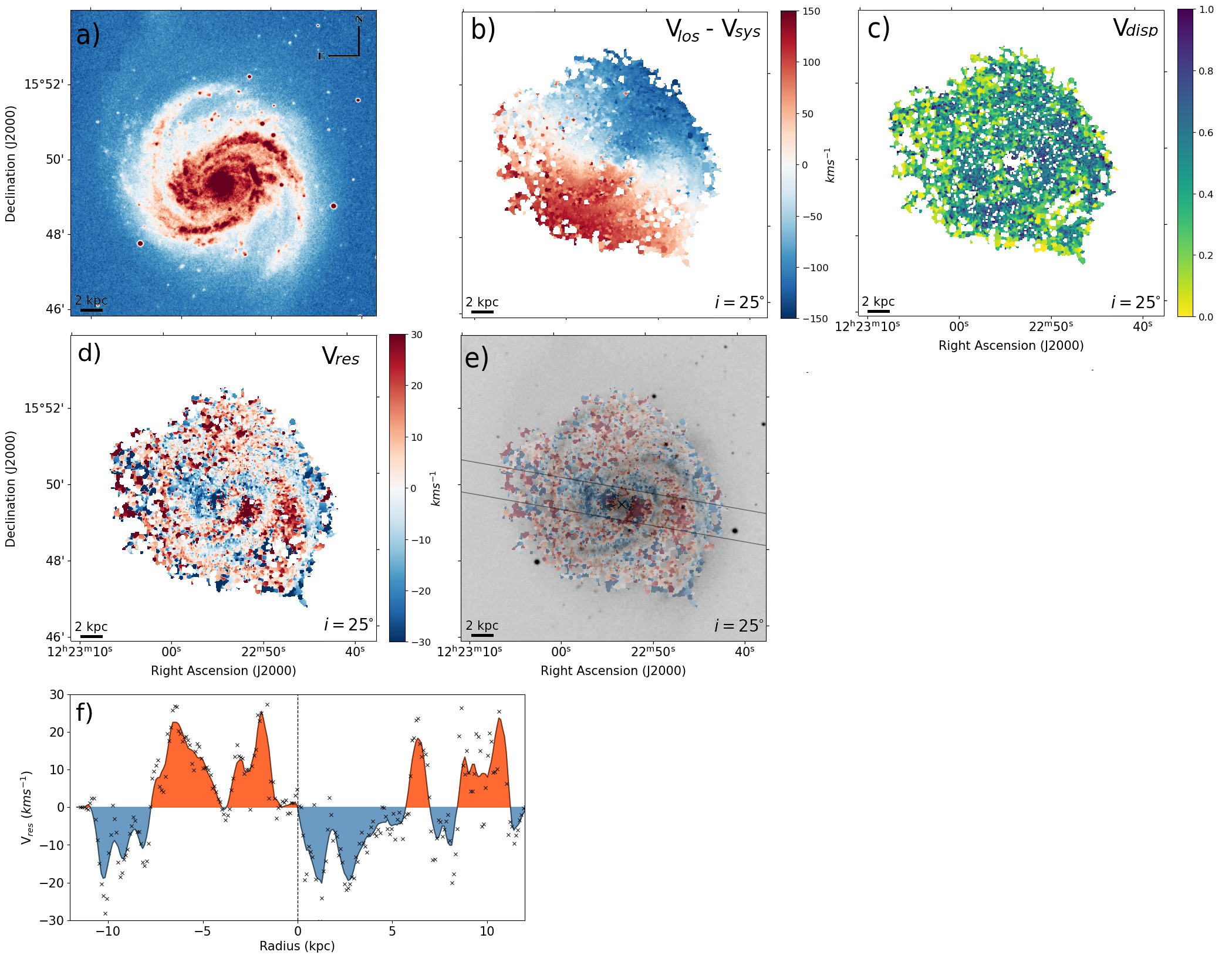}  

  \caption{\textbf{NGC 4321}. Panel (a): XDSS blue-band image.
  Panel (b): $\mathrm {H_{\alpha}}$velocity field.
  Panel (c): $\mathrm {H_{\alpha}}$velocity dispersion map normalized to $33.1 \ \mathrm{km s^{-1}}$ corresponding to the 95th percentile.
  Panel (d): $\mathrm {H_{\alpha}}$residual velocity field.
  Panel (e): Optical band image with $\mathrm {H_{\alpha}}$residual velocity field.
  Panel (f): Residual velocity radial profile considering a slit indicated with black lines with position angle of 100$^{\circ}$ in Panel (e). The black line corresponds to smoothed curve and the dashed line fits center of galaxy indicated in the panel (e). The red (blue) area corresponds to velocities above (below) the midplane of the galaxy. The derived kinematic inclination is indicated in the lower right of each panel.}
   \label{fig:candidate_ngc4321}
\end{figure*}

In Figure \ref{fig:candidate_ngc4321} we show NGC 4321 (M100), a grand-design nearby spiral galaxy belonging to the Virgo Cluster, classified as SABbc (RC3). The galaxy has a small bulge and two well-defined, prominent and, symmetric spiral arms, as shown on its $B$-band image (panel a). The nucleus of the galaxy is compact and bright. NGC 4321 is a very low-inclination galaxy, with a kinematical inclination of $i \approx 25^{\circ}$ and an optical radius of ${R_{\rm opt}} = 16.3$ kpc from RC3. Panel b) shows that the coverage of the \textit{V}$_{\rm los}$ distribution, obtained from the $\mathrm {H_{\alpha}}$ observations,  is extensive and reaches nearly to its optical radius; i.e. 13 kpc $ \sim 0.8 {R_{\rm opt}}$. The \textit{V}$_{\rm los}$ map shows an amplitude of $\approx 150\ \mathrm{km\ s^{-1}}$. For completeness, we show on panel c) the $\mathrm {H_{\alpha}}$  velocity dispersion map, normalized to $\sigma_{95th} = 33.1\ \mathrm{km\ s^{-1}}$. 

The derived residual velocity map, obtained after subtracting the axisymmetric velocity model is shown on panels d) and e). As in previous examples, this map reveals  global and coherent velocity flows with amplitudes that can reach $ \sim 30\ \mathrm{km\ s^{-1}}$.  Panel f) more clearly shows the  radial variations of the residual velocity along the highlighted slit  in panel e). Note the oscillatory behavior of \textit{V}$_{\rm res}$ across the disk. 
These perturbations, observed in the ionized gas of NGC 4321, could be the result of a recent interaction with one of its satellite galaxies. Indeed, NGC 4321 has two dwarf companion galaxies, NGC 4328 and NGC 4323, located at projected distances of 24 kpc and 28 kpc from their host, respectively \citep{2005ApJ...632..253H}. \citet{1993ApJ...416..563K} showed that the HI component of NGC 4321 is mostly contained within the optical disk. However, they also identify a large HI extension that could be the result of the tidal interaction with NGC 4323. Indeed, \citet{1993ApJ...416..563K} find important differences in the behaviour of its rotation between the approaching and the receding sides. They suggest these differences could be caused by deviations from circular motions in the outer disk probably due to a close passage of its companion galaxy. They also suggest this interaction could be the cause of the observed asymmetry in the total HI distribution. NGC 4321 also shows diffuse stellar extension in two directions, one of them towards NGC 4323, likely the result of ongoing interaction with this satellite \citep{2005ApJ...632..253H}.

\begin{figure*}
  \centering
  \includegraphics[width=180mm,clip,angle=0]{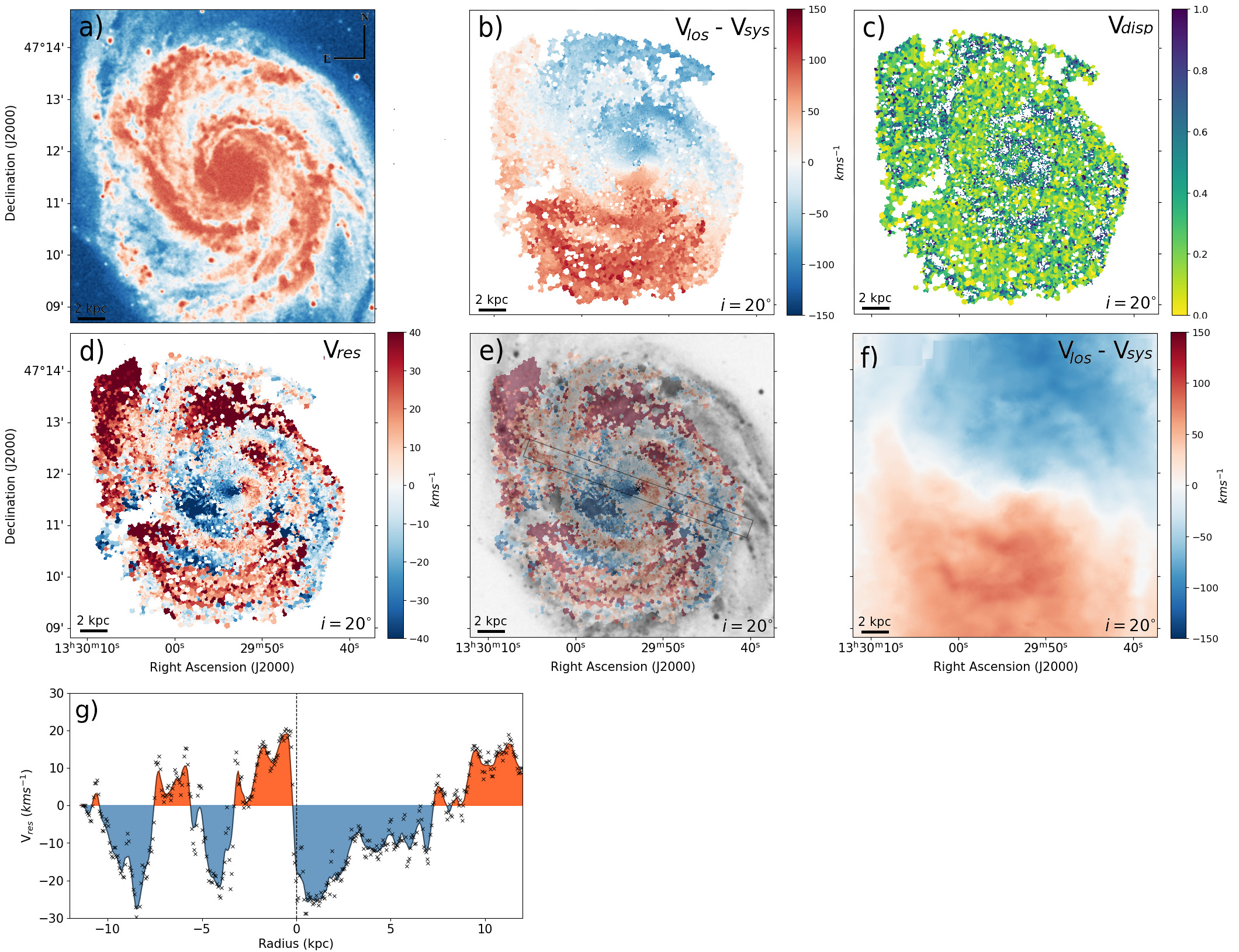}  

  \caption{\textbf{NGC 5194}. Panel (a): XDSS blue-band image.
  Panel (b): $\mathrm {H_{\alpha}}$velocity field.
  Panel (c):  $\mathrm {H_{\alpha}}$velocity dispersion map normalized to $15\ \mathrm{km s^{-1}}$ corresponding to the 95th percentile.
  Panel (d): $\mathrm {H_{\alpha}}$residual velocity field.
  Panel (e): Optical band image with $\mathrm {H_{\alpha}}$residual velocity field.
  Panel (f): HI velocity field from THINGS survey \citet{2008AJ....136.2563W}.
  Panel (g): Residual velocity radial profile considering a slit indicated with black lines with position angle of 110$^{\circ}$ in Panel (e). The black line corresponds to smoothed curve and the dashed line fits center of galaxy indicated in the panel (e). The red (blue) area corresponds to velocities above (below) the midplane of the galaxy. The derived kinematic inclination is indicated in the lower right of each panel.}
   \label{fig:candidate_ngc5194}
\end{figure*}

NGC 5194 (M51a) is a well-known nearby grand design spiral galaxy classified as non-barred by RC3, with a very low  inclination angle $i = 20^{\circ}$. NGC 5194 is tidally interacting with its companion NGC 5195 (M51b), an early-type SB0 galaxy. NGC 5194 shows an intense star formation activity at its center and along its spiral arms. Panel a), on Figure \ref{fig:candidate_ngc5194}, clearly shows its strong spiral structure with signs of being lopsided, likely due to the interaction with its companion. Panel b) shows that the $\mathrm {H_{\alpha}}$ observations provide wide coverage of the disk, which allows us to obtain its velocity field up to distances of 12 kpc $\sim 0.8\ R_{\rm opt}$. The resulting $V_{\rm los}$ map shows a  velocity range of $\approx 150\ \mathrm{km\ s^{-1}}$. It is interesting to compare our $\mathrm {H_{\alpha}}$ velocity map with the one derived from HI observations, obtained from THINGS using VLA \citep{2008AJ....136.2563W},  shown in panel f). As seen in NGC 628 and NGC 6946, the HI velocity map of NGC 5194, constrained to the same FoV as the $\mathrm {H_{\alpha}}$ data, shows a very good agreement with the ionaized gas velocity map, despite differences in both the spatial and spectral resolutions.
In panel c), we show the velocity dispersion map normalized to $\sigma_{95th} = 15\ \mathrm{km\ s^{-1}}$.

The residual velocity map is shown in panels d) and e). It is interesting to note the large velocity perturbations throughout the disk, reaching amplitudes of $40\ \mathrm{km\ s^{-1}}$.  We recall the very low inclination angle of this galaxy, thus rendering it unlikely that these velocity perturbations are mainly due to in-plane velocity flow. Panel e) allows comparing the substructure of the  $V_{\rm res}$ map with the galaxy morphology as seen in the $B$-band image. In panel g) we follow the smoothed residual velocity profile along the highlighted slit. As before, this allows appreciating the radial oscillatory behavior of \textit{V}$_{\rm res}$, consistent with what is expected for a corrugated disk. \citet{2007ApJ...665.1138S} presented a deep kinematic study of NGC 5194 using full 2D velocity distributions from interferometric CO and Fabry-Perot $\mathrm {H_{\alpha}}$ observations. They report a complex velocity field, with variations of the disk's PA and inclination along the galactic radius that strongly suggest a vertically perturbed disk and very significant out-of-plane motions. An HI warp has also been previously reported in NGC 5194 \citep{1990AJ....100..387R,2014PASJ...66...77O,2014MNRAS.443..186H}, which has been attributed to the tidal interaction with NGC 5195. The presence of vertical velocity flows in this very interesting galaxy will be further analyzed in a follow-up study, following the procedure described in \citet{2021ApJ...908...27G}.

\begin{figure*}
  \centering
  \includegraphics[width=180mm,clip,angle=0]{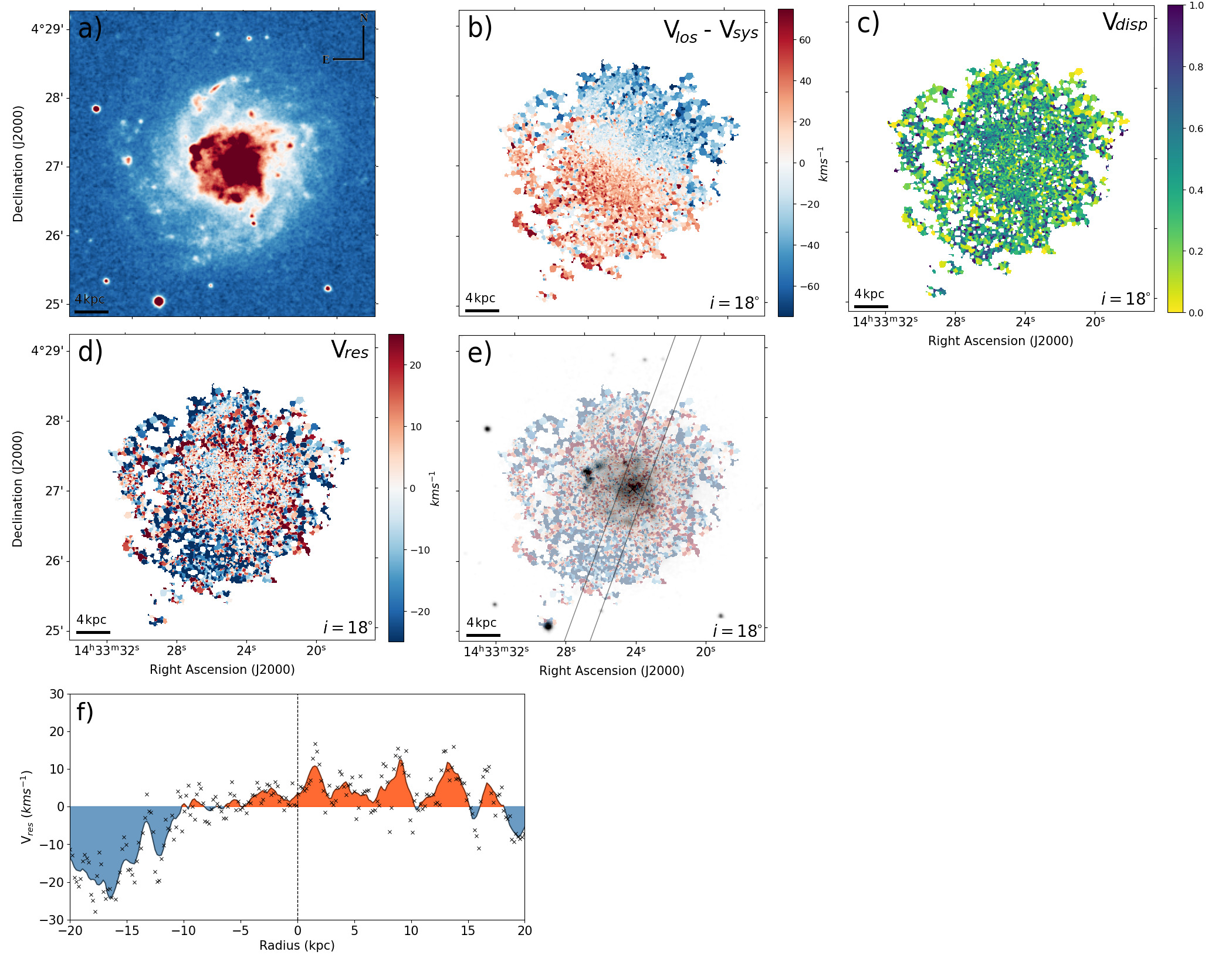}  

  \caption{\textbf{NGC 5668}. Panel (a): XDSS blue-band image. 
  Panel (b): $\mathrm {H_{\alpha}}$velocity field. 
  Panel (c):  $\mathrm {H_{\alpha}}$velocity dispersion map normalized to $57.5 \ \mathrm{km s^{-1}}$ corresponding to the 95th percentile, corrected for instrumental broadening.
  Panel (d): $\mathrm {H_{\alpha}}$residual velocity field. 
  Panel (e): Optical band image with $\mathrm {H_{\alpha}}$residual velocity field. 
  Panel (f): Residual velocity radial profile considering a slit indicated with black lines with position angle of 20$^{\circ}$ in Panel (e). 
  The black line corresponds to smoothed curve and the dashed line fits center of galaxy indicated in the panel (e). The red (blue) area corresponds to velocities above (below) the midplane of the galaxy. The derived kinematic inclination is indicated in the lower right of each panel.}
   \label{fig:candidate_ngc5668}
\end{figure*}

In Figure~\ref{fig:candidate_ngc5668} we present NGC 5668, a nearly face-on late-type spiral galaxy with an estimated inclination angle $18^{\circ}$ \citep{1996AJ....112..960S} and  R$_{\rm opt}$ of $\sim 14.3$ kpc, located at $27.6\ \mathrm{Mpc}$. The galaxy is classified as a SA(s)d by RC3 and, on its optical image (panel a), it shows a weak bar or oval structure on its inner $12''$ region. Observations of NGC 5668 were extracted from the GHASP sample \citep{2008MNRAS.390..466E}. The $\mathrm {H_{\alpha}}$ coverage reaches out to 20 kpc corresponding to $\sim 1.4$ R$_{\rm opt}$. On panel b) we show the resulting $V_{\rm los}$ field, with a velocity amplitude reaching $\approx 70\ \mathrm{km\ s^{-1}}$.  The velocity dispersion map, normalized to $\sigma_{95th} = 57.5\ \mathrm{km\ s^{-1}}$, is shown on panel c). 

The \textit{V}$_{\rm res}$ map, shown on panel d), reveals global perturbations with amplitudes $> 20\  \mathrm{km\  s^{-1}}$ km/s which are also consistent with a  warped $\mathrm {H_{\alpha}}$ disk.
In panel e) we present the optical image with the overlay of the residual velocity map. The warp is better shown in panel f), where we can see the mean velocity profile along the highlighted slit on panel e). Note the large amplitude of this kinematical perturbation reaching velocities $> 20\  \mathrm{km\ s^{-1}}$ at the disk outskirts.
High-resolution observations in the 21-cm line of the neutral hydrogen velocity field, from the Very Large Array (VLA),  have detected a kinematic warp that begins, at least, at the end of the optical radius \citep{1996AJ....112..960S}. Note that the $\mathrm {H_{\alpha}}$ warp is very well aligned with the previously reported HI warp, indicating that the perturbation extends even within the optical radius of the galaxy. In addition to the HI kinematic warp, \citet{1996AJ....112..960S} detected high-velocity wings beyond the double-horned 21 cm profile, related to High-Velocity Clouds (HVCs) in the disk and halo of the galaxy. As discussed by \citet{1996AJ....112..960S}, the high-velocity wings located outside the optical disk may be the result of infalling material, comparable to the Magellanic Stream observed in the Milky Way. 

The origin of the kinematic warp, and the infalling material, could be related to a past interaction with its neighbor galaxy UGC 9380, currently located at a projected distance of $\sim 200$ kpc of NGC 5668. It is worth mentioning that \citet{2000A&A...358..812J} also analyzed the 2D velocity field of this galaxy through its $\mathrm {H_{\alpha}}$ emission using Fabry-Perot interferometer. However, they did not cover the region where we detect the kinematic warp. Instead, they reported the detection of HVCs and localized high residual velocity dispersion regions, associating them to shell/chimney regions. Even though we present a velocity dispersion map in panel c), in this work we have not attempted to recover these regions, typically associated to vertical motions of ionized gas due to star-forming processes.

\begin{figure*}
  \centering
  \includegraphics[width=180mm,clip,angle=0]{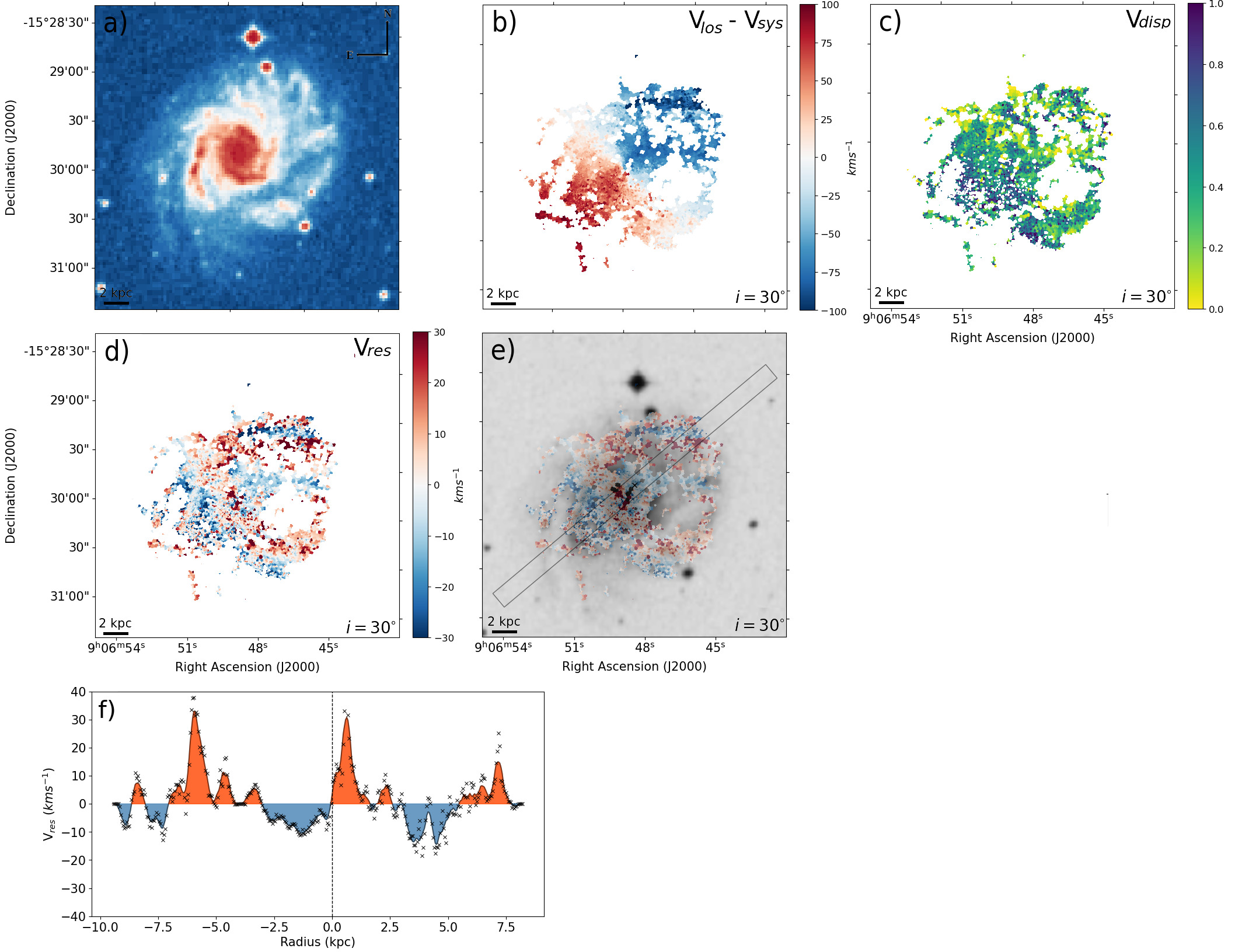}  

  \caption{\textbf{NGC 2763}. 
  Panel (a): XDSS blue-band image. 
  Panel (b): $\mathrm {H_{\alpha}}$velocity field. 
  Panel (c): $\mathrm {H_{\alpha}}$velocity dispersion map normalized to $37.2 \ \mathrm{km s^{-1}}$ corresponding to the 95th percentile.
  Panel (d): $\mathrm {H_{\alpha}}$residual velocity field.
  Panel (e): Optical band image with $\mathrm {H_{\alpha}}$residual velocity field.
  Panel (f): Residual velocity radial profile considering a slit indicated with black lines with position angle of 40$^{\circ}$ in Panel (e).
  The black line corresponds to smoothed curve and the dashed line fits center of galaxy indicated in the panel (e). The red (blue) area corresponds to velocities above (below) the midplane of the galaxy. The derived kinematic inclination is indicated in the lower right of each panel.}
   \label{fig:candidate_ngc2763}
\end{figure*}

In Figure \ref{fig:candidate_ngc2763} we show NGC 2763. This low inclination disk galaxy ($i\approx 29.5^{\circ}$) is the only object in WiNDS observed in the Southern Hemisphere, and one of the first galaxies observed with the SAM-FP instrument \citep{2017MNRAS.469.3424M}. The $B$-band image on panel a) reveals two main spiral arms, each with multiple sub-arms, and a small bar. In addition, it shows a significant lopsided structure. The galaxy shows an extended $\mathrm {H_{\alpha}}$ emission across the disk, reaching up to  0.9 ${R_{\rm opt}}$, with ${R_{\rm opt}} \approx 9.9$ kpc. Panel b) shows  its \textit{V}$_{\rm los}$ maps, with an amplitude of $\approx 100\ \mathrm{km\ s^{-1}}$. The velocity dispersion map, normalized to $\sigma_{95th} = 37.2\ \mathrm{km\ s^{-1}} $ is shown on panel c). 

The resulting $V_{\rm res}$ maps, shown on panel d), reveals a very complex structure with global residual velocity flows reaching amplitudes $\gtrsim 20\ \mathrm{km\  s^{-1}}$.
In panel e) we can observe the $B$-band image of NGC 2763 with the residual velocity map overlapped, highlighting the region where  complex residual velocity perturbations are observed.
Interestingly, no detailed kinematical studies of NGC 2763 velocity field, nor companion galaxies within 150 kpc,  have been reported. We have confirmed the lack of massive nearby companions by performing a systematic search within a radial projected distance of 250 kpc and radial velocity difference $\Delta \leq 1000\ \mathrm{km\  s^{-1}}$ using NASA/IPAC Extragalactic Data base. Thus, NGC 2763 is considered as an isolated galaxy. As such, it is not clear whether its significantly perturbed morphology and kinematics are the results of a previous minor merger event or due to significant misaligned smooth gas accretion.

\begin{figure*}
  \centering
  \includegraphics[width=180mm,clip,angle=0]{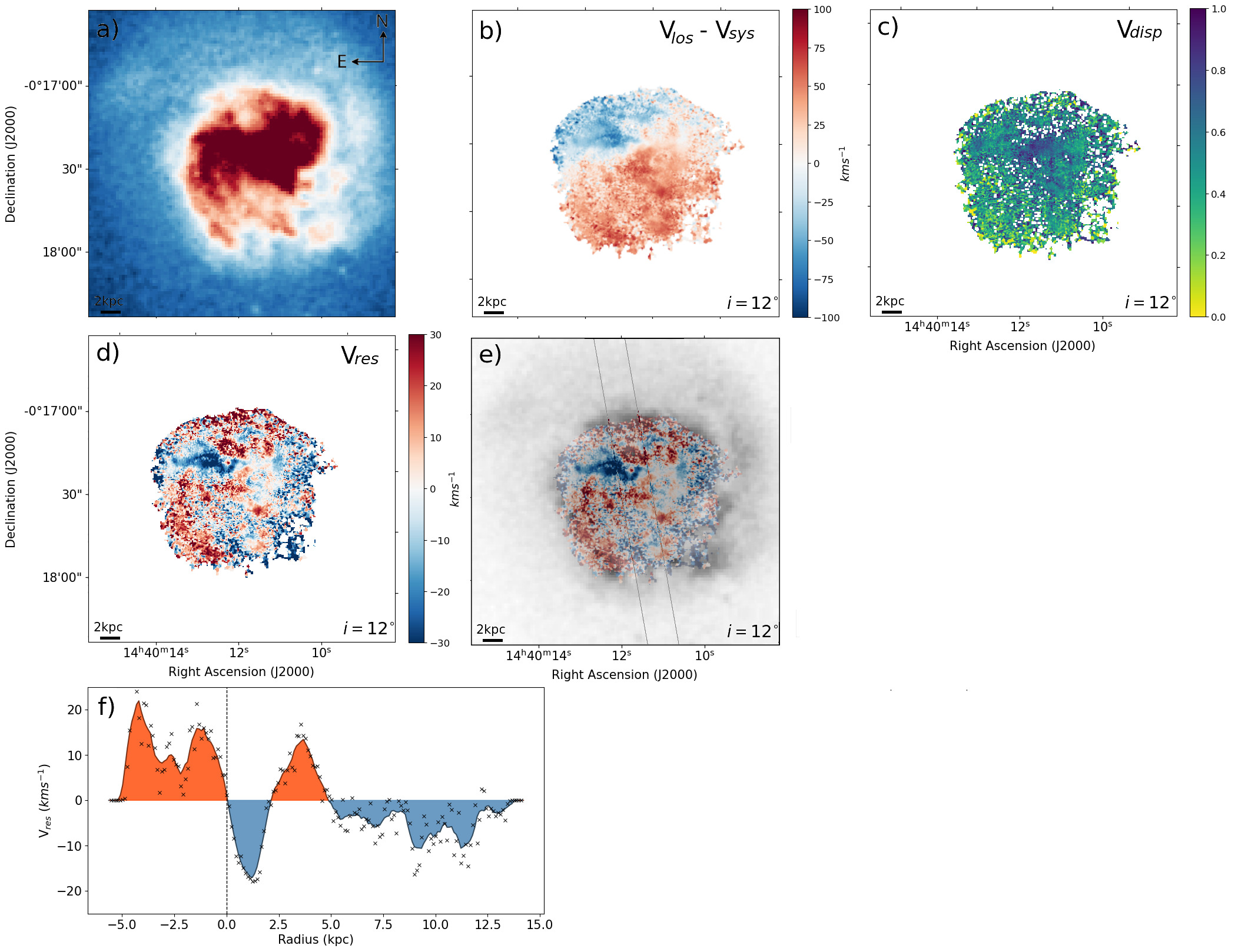}  

  \caption{\textbf{NGC 5713} Panel (a): XDSS blue-band image. 
  Panel (b): $\mathrm {H_{\alpha}}$velocity field. 
  Panel (c): $\mathrm {H_{\alpha}}$velocity dispersion map normalized to $54.2 \ \mathrm{km s^{-1}}$ corresponding to the 95th percentile. 
  Panel (d): Optical band image with $\mathrm {H_{\alpha}}$residual velocity field. 
  Panel (e): $\mathrm {H_{\alpha}}$residual velocity field. 
  Panel (f): Residual velocity radial profile considering a slit indicated with black lines with position angle of 170$^{\circ}$ in Panel (e). 
  The black line corresponds to smoothed curve and the dashed line fits center of galaxy indicated in the panel (e). The red (blue) area corresponds to velocities above (below) the midplane of the galaxy. The derived kinematic inclination is indicated in the lower right of each panel.}
   \label{fig:candidate_ngc5713}
\end{figure*}

NGC 5713 is oriented nearly face-on, with inclination $i \approx 10^{\circ}$, and shows a significant lopsided morphology. This can be seen in panel a) of Figure \ref{fig:candidate_ngc5713}, where we show its XDSS $B$-band image. NGC5713 has an $R_{\rm opt} \sim 13.5$ kpc, it is a barred and multiple arm spiral galaxy   \citep[SABb-type][]{1991rc3..book.....D}, and is located at 33.4 Mpc. The $\mathrm {H_{\alpha}}$ \textit{V}$_{\rm los}$ map, shown on panel b) covers approximately a region of 0.9 ${R_{\rm opt}}$, and shows an amplitude of $\sim 100\ \mathrm{km\  s^{-1}} $. The velocity dispersion map, normalized to $\sigma_{95th} = 54.2\ \mathrm{km\  s^{-1}}$, is shown on panel c).

The \textit{V}$_{\rm res}$ (panel d) reveals a very peculiar structure, with several global and coherent perturbations reaching peak velocities $\geq 30\ \mathrm{km\  s^{-1}} $. The galaxy is currently undergoing a very strong tidal interaction with its similar mass Sab-type companion, NGC 5719 \citep{2007A&A...463..883V}, and shows two HI tidal tails. Both galaxies (NGC 5713 and NGC 5719) show an optical radius at a similar galactocentric distance and are connected by two HI tidal bridges. The projected distance between these two galaxies is $\sim 77$ kpc. NGC 5719 shows an ionized gas disk that counter rotates with respect to its neutral gas counterpart. According to \citet{2007A&A...463..883V}, this counter-rotating disk is the result of the accretion of NGC 5713 material onto NGC 5719 during their closest approach. Interestingly NGC 5719 shows a nearly edge-on configuration with a clearly visible warped disk. In addition, NGC 5713 is interacting with the Sm-type dwarf galaxy, PGC 135857, which is located at a projected distance of $\sim 64$ kpc \citep{2007A&A...463..883V}. These interactions are likely behind the observed NGC 5713 strongly perturbed velocity field. The interacting system will be compared against numerical models in a follow-up study.

\section{Discussion and Conclusions} 
\label{sec:conclusions}

In this study, we have presented the Waves in Nearby Disk galaxies Survey (WiNDS), which currently consists of 40 nearby low-inclination spiral galaxies, including objects with diverse morphological features. The WiNDS sample contains galaxies within a distance limit of 45 Mpc and absolute B magnitude between -17.0 and -21.9 $\mathrm{mag}$, all observed in $\mathrm {H_{\alpha}}$ with Fabry-Perot interferometer. These observations allow us to obtain very detailed velocity maps of the ionized gas distribution, with velocity sampling precision as low as $3\ \mathrm{km\  s^{-1}}$, which translates into  a resolution of $6\ \mathrm{km\ s^{-1}}$. Our  sample   was selected such that galaxies have an inclination angle $i \lesssim 40^{\circ}$. WiNDS consists of 12 new data observed, and includes archival data for 28 low-inclination late-type galaxies, extracted from the GHASP, SINGS--$\mathrm {H_{\alpha}}$, VIRGO--$\mathrm {H_{\alpha}}$, HRS--$\mathrm {H_{\alpha}}$ surveys. For each galaxy in the final sample, we derived their line-of-sight velocity field, an unperturbed axy-symmetric velocity model, kinematical parameters, the corresponding residual velocity fields, \textit{V}$_{\rm res}$ and rotation curves (see Appendix \ref{sssec:rotation_curves}).

Our main goal in this work was to examine the resulting V$_\mathrm{res}$ fields to search for evidence of large, global, and coherent kinematic perturbations in our sample of low-inclined late-type galaxies. In particular, we seek perturbations that are consistent with those produced by warps and corrugation patterns. Through three criteria we identify clear signatures of perturbations. First, we are interested in objects that present a wide $\mathrm {H_{\alpha}}$ coverage ($ \geq 0.7 R_{\rm opt}$) so that the kinematics of the disks could be globally explored. Second, we focused on galaxies with V$_\mathrm{res}$ amplitude that are $> 10\ \mathrm{km\ s^{-1}}$, which allows us to be more confident in observing perturbations in the low-inclination disks that are not the result of the axisymmetric components of the galaxies. Finally, we also searched for perturbations in the \textit{V}$_\mathrm{res}$ fields that show global and coherent velocity flows, avoiding local and discrete perturbations that could be linked to, e.g., fountain flows.

We emphasize that our selection criteria cannot confirm nor rule out the presence of vertical perturbations in our disks. Instead, our goal is to select potentially strong candidates for follow-up studies. Note however that, as shown by \citet{2016MNRAS.461.3835M} and \citet{2021ApJ...908...27G}, the amplitude of the observed velocity flows in these low-inclined disks are, typically, much too large to be driven either by spiral structure or by a bar.
Within the subset of galaxies with vertical perturbations considered as candidates, those displaying discrete and local perturbations that could be associated to fountain flows are not excluded since such signatures can coexist with the global and coherent perturbations that we aim at detecting in this study.

A vertical perturbation such as a warp or a corrugation pattern is manifested by an extended and oscillatory vertical displacement of the disk with respect to the overall mid-plane. Its characterization can allow us to constrain the recent interaction history of a galaxy.
Previous studies based on cosmological simulations \citep[e.g.][]{2017MNRAS.465.3446G} have identified four main mechanisms behind such perturbations: distant fly-by massive companions, close satellite encounters, re-accretion of cold gas from progenitors of a gas-rich major merger and accretion of misaligned cold gas. 
Out of 40 observed low inclination galaxies, we find that approximately 20 percent of the sample shows clear signatures of global and coherent perturbations. 
Adding to this statistic the results of our pilot study based on VV304a \citep{2021ApJ...908...27G} brings this number to 22 percent. We note however that, as further discussed below, some of our observations were either not sufficient or not adequate to identify velocity perturbation on the disk outskirts and, thus, this percentage could be higher. In fact, out of the 40 galaxies in the WiNDS sample, only 70$\%$ show good $\mathrm {H_{\alpha}}$ coverage. If we only consider this subset, the fraction of vertically perturbed  galaxy candidates rises to $32\%$.

From the galaxies that show signs of potential vertical perturbation, six of them are interacting with a satellite galaxy and/or belong to a group. Thus, the observed kinematic perturbations are likely directly linked with recent environmental interactions. The rest of the perturbed subsample of galaxies are considered to be isolated. Therefore, their perturbations could be the result of previous minor mergers or misaligned cold gas accretion. A deeper and more detailed study must be performed for each galaxy to constrain the origin of their perturbed velocity field.
As shown in Fig. \ref{fig:hist_distribution}, the vertically perturbed candidates within WiNDS sample (green distribution) shows no preferential distribution of the main parameters with respect to the overall WiNDS sample.

Previous studies, either in HI \citep{2002A&A...394..769G} or in the optical \citep[][for the edge-on galaxies]{1998A&A...337....9R,2006NewA...11..293A}, that have characterized the frequency which vertically perturbed disks arise in the local Universe, have found that approximately 70$\%$ of them present a warped disk, typically displaying S-shaped configuration. However, evidence for more complex corrugation patterns in external galaxies, such as those hinted in this work, have been extremely scarce. Indeed, only a handful of previous studies have previously reported corrugations on external galaxies \citep[e.g.][]{2015MNRAS.454.3376S,2020MNRAS.495.3705N,2021ApJ...908...27G}. In agreement with observations, cosmological simulations of high-resolution late-type galaxies within Milky Way-sized halos, \citet{2016MNRAS.456.2779G,2017MNRAS.465.3446G} estimated that $70\%$ should show strongly vertically perturbed disks, and 35\% should present a corrugated structure. The scarcity in the detection of corrugation patterns on external galaxies so far was not that surprising since most studies attempting to characterize stellar disk vertical structures have been focused on edge-on systems in which corrugation is difficult to detect due to projection effects. Our study shows that high resolution velocity maps, obtained with techniques such as Fabry-Perot interferometer, allows us to reveal much more complex velocity structures on external galactic disks. Furthermore, due to the low inclination angles of the disks in our sample, these are likely linked with vertical flows and consistent with corrugations patterns. Eight galaxies from WiNDS show potential vertical patterns, of which 25$\%$ are in a close interaction with a massive companion, $50\%$ show nearby dwarf satellite galaxies, and 25$\%$ can be regarded as isolated.

However, the fraction of low inclination galaxies with detected velocity flow, i.e. $\approx 20 \%$, is currently significantly lower than the expected fraction of vertically perturbed disks. Several factors can be playing a role here. The first and most obvious reason is that a fraction of $\approx 30 \%$ of the galaxies in our sample did not show a wide $\mathrm {H_{\alpha}}$ coverage. Such cases either presented a patchy and poor $\mathrm {H_{\alpha}}$ emission  distribution, or it was concentrated within the inner  $\sim 0.5\ R_{\rm opt}$.  As a result, we were not able to globally explore the kinematics of their disks, especially on the outer disk regions where warps and vertical perturbations are expected to be stronger. Second, the weather conditions in $\approx 12.5 \%$ of our observations were not ideal. As a result, the observations were not sufficient to generate kinematic maps accurate enough to properly resolve their velocity fields. Considering these caveats, it is not surprising that we have been able to detect kinematic perturbation, consistent with a vertical perturbed disk, in a smaller fraction of galaxies than previously reported.

In a follow-up study deeper and better observation will be presented. Following \citet{2021ApJ...908...27G} we will compare our observations against detailed analytic kinematic models that account for the axisymmetric perturbations measured on each galaxy and thus constrain the origin of their velocity perturbations.

\section*{Acknowledgements}

FAG and CUM acknowledges financial support from FONDECYT Regular 1211370. FAG, AM and CUM gratefully acknowledge support by the ANID BASAL project FB210003 and funding from the Max Planck Society through a Partner Group grant. AM acknowledges financial support from FONDECYT Regular 1212046. FM acknowledges support through the Program `Rita Levi Montalcini' of the Italian MIUR. CUM also acknowledges financial support through the fellowship `Becas Doctorales Institucionales ULS', granted by the Vicerrector{\'i}a de Investigaci{\'o}n y Postgrado de la Universidad de La Serena. The WiNDS survey is based on observations taken at the Observatoire de Haute Provence (OHP, France), operated by the French CNRS. The authors warmly thank the OHP team for its technical assistance before and during the observations, namely the night team: Jean Balcaen, Stéphane Favard, Jean-Pierre Troncin, Didier Gravallon and the day team led by François Moreau as well as Dr. Auguste Le Van Suu. We are grateful to the whole SAM/SOAR team who assisted us in the preparation and in the execution of the observations. This work was supported by the Programme National Cosmology et Galaxies (PNCG) of CNRS/INSU with INP and IN2P3, co-funded by CEA and CNES. CUM also acknowledges all the help and insights from Jes{\'u}s G{\'o}mez-L{\'o}pez and Gustavo Morales during the writing and data processing steps. The author also acknowledges Diego Pallero, Ciria Lima-Dias and Daniel Hern{\'a}ndez for their valuable discussions and comments.

\appendix
%\section*{Appendix}
%\addcontentsline{toc}{section}{Appendices}
\renewcommand{\thesubsection}{\Alph{subsection}}

\numberwithin{equation}{section}
\renewcommand{\theequation}{A\arabic{equation}} 

\subsection{\textup{NEW OBSERVATION DATA}}
\label{subsec:new_obs}

\subsubsection{\textup{Comments for individual galaxies.}}
\label{sssec:comments_new_obs}

\textbf{NGC 1058} is a nearly normal Sc face-on spiral galaxy with an inclination angle of $6^{\circ}$. NGC 1058 does not present well-defined arms in $\mathrm {H_{\alpha}}$ emission and reveals a weaker emission in the innermost than in outer regions. The galaxy is located at a distance of 9.93 Mpc, the most recent work distance value is considered. It has an angular diameter of 3 arcmin with apparent magnitude in B band about 12 mag. NGC 1058 belongs to the group NGC 1023 being the least bright member. This galaxy was observed in $\mathrm {H_{\alpha}}$ line \citep{2015MNRAS.454.3376S} to analyze if it presents vertical flows using the long-slit technique using different P.A. and finding that velocity peaks are associated with star-forming regions. Two supernovae 1961V and 1969L have been reported in the outer disk.

\textbf{NGC 2500} is a nearby spiral galaxy with a short bar aligned with its minor kinematical axis, classified as SB(rs)d, located at 9.79 kpc and inclination angle $40^{\circ}$. This galaxy was observed using $\mathrm {H_{\alpha}}$ filter by \citet{2008MNRAS.388..500E} who find diffuse emission in their observed $\mathrm {H_{\alpha}}$ maps. NGC 2500's corrugated velocity patterns were studied using long-slits with $\mathrm {H_{\alpha}}$ observations in \citet{2015MNRAS.454.3376S}, where the vertical displacements do not seem to be related with an $\mathrm {H_{\alpha}}$ emission peak.
Like NGC 1058, this galaxy is part of the sample by \citet{2015MNRAS.454.3376S} but unlike in NGC 1058 where they find speed peaks, in NGC 2500 it was not conclusive.

\textbf{NGC 3147} is a non-barred galaxy and it is classified as SA(rs)bc. It is considered as the best type 2 Seyfert candidates using optical and X-ray observations, simultaneously \citep{2012MNRAS.426.3225B}.

\textbf{NGC 3184} nearby grand design spiral SABc type. Located at a distance of 10.05 Mpc, it has a nearly face-on orientation  with inclination angle $16.7^{\circ}$ and $184.5^{\circ}$ deviated in this article using the methodology described previously.
This galaxy belongs to the group NGC 3184 being the brightest with  $9.7\ \mathrm{mag}$. Other members are NGC 3198, NGC 3432, and NGC 3319. 

\textbf{NGC 3423} nearby spiral galaxy Sc type, absolute B-band magnitude of $12\ \mathrm{mag}$ and angular size 3.56 arcmin and is located at 15 Mp. No previous analysis studies in particular of this galaxy.

\textbf{NGC 3485} nearby spiral galaxy located at 28.17 Mpc with absolute B-band magnitude $12.7\ \mathrm{mag}$ . The kinematical inclination derived in this work is $26^{\circ}$. There are no individual works for this galaxy previously. In this work, the $\mathrm {H_{\alpha}}$ emission is faint and is it located in the galaxy arm.

\textbf{NGC 3642} Non-barred galaxy with three rings and dusty spiral arms classified SAbc-type. The galaxy shows a warped outer disk in HI line \citep{2002A&A...389..825V}. NGC 3642 belongs to a group composed of 5 galaxies. NGC 3642 is the brightest member with elliptical galaxies NGC 3610 and NGC 3613 with similar magnitude. Another member is NGC 674 and NGC 3683 are fainter by more than 2 magnitudes. It is an extended galaxy and shows around internal disk from which a spinal arm appears that forms an external disk

\textbf{NGC 4136} nearby face-on galaxy with a bar and a ring as well as a well-developed spiral structure in its outer disk with low-luminosity, classified SBc type \citep{2003AstL...29..363G}.

\textbf{NGC 4900} is a spiral barred galaxy, classified as SB(rs)c located in the Virgo cluster at a distance of 13.3 Mpc and at about $12^{\circ}$ south-east of M87. NGC 4900 presents HII nucleus using CO observations \citep{1998JKAS...31...95L}.

\subsubsection{\textup{Presentation of the $\mathrm {H_{\alpha}}$ maps of the new observation data. (Figs. 12 to 20)}} 

\label{sssec:new_obs_maps}
\begin{figure*}

  \centering
  \includegraphics[width=140mm,clip,angle=0]{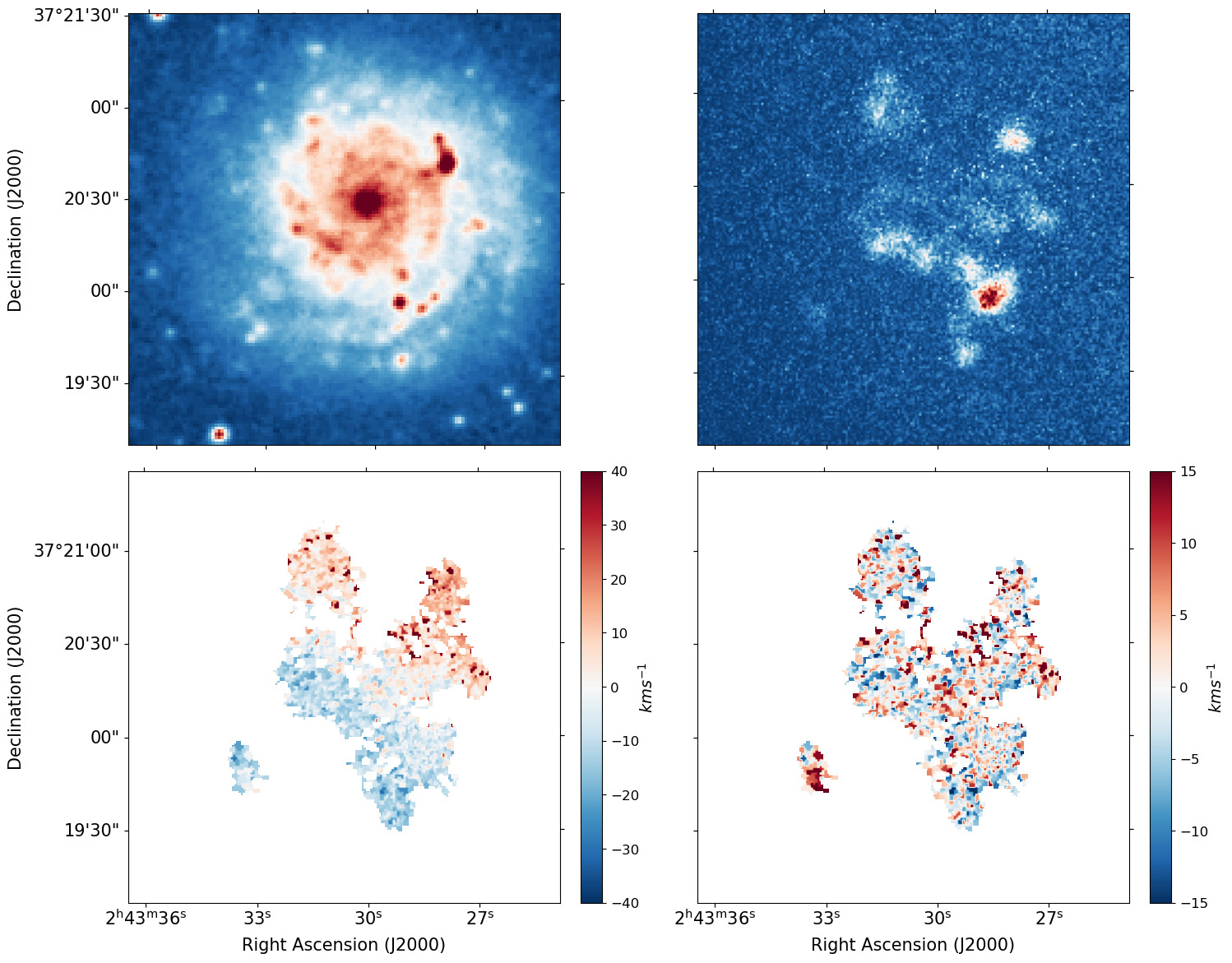}  

  \caption{NGC 1058. Top left: XDSS Blue Band image. Top right: $\mathrm {H_{\alpha}}$ monochromatic image. Bottom left: $\mathrm {H_{\alpha}}$ velocity field. Bottom right: Residual map $\mathrm {H_{\alpha}}$ field}
  \label{fig:ngc1058}
   
   \includegraphics[width=140mm,clip,angle=0]{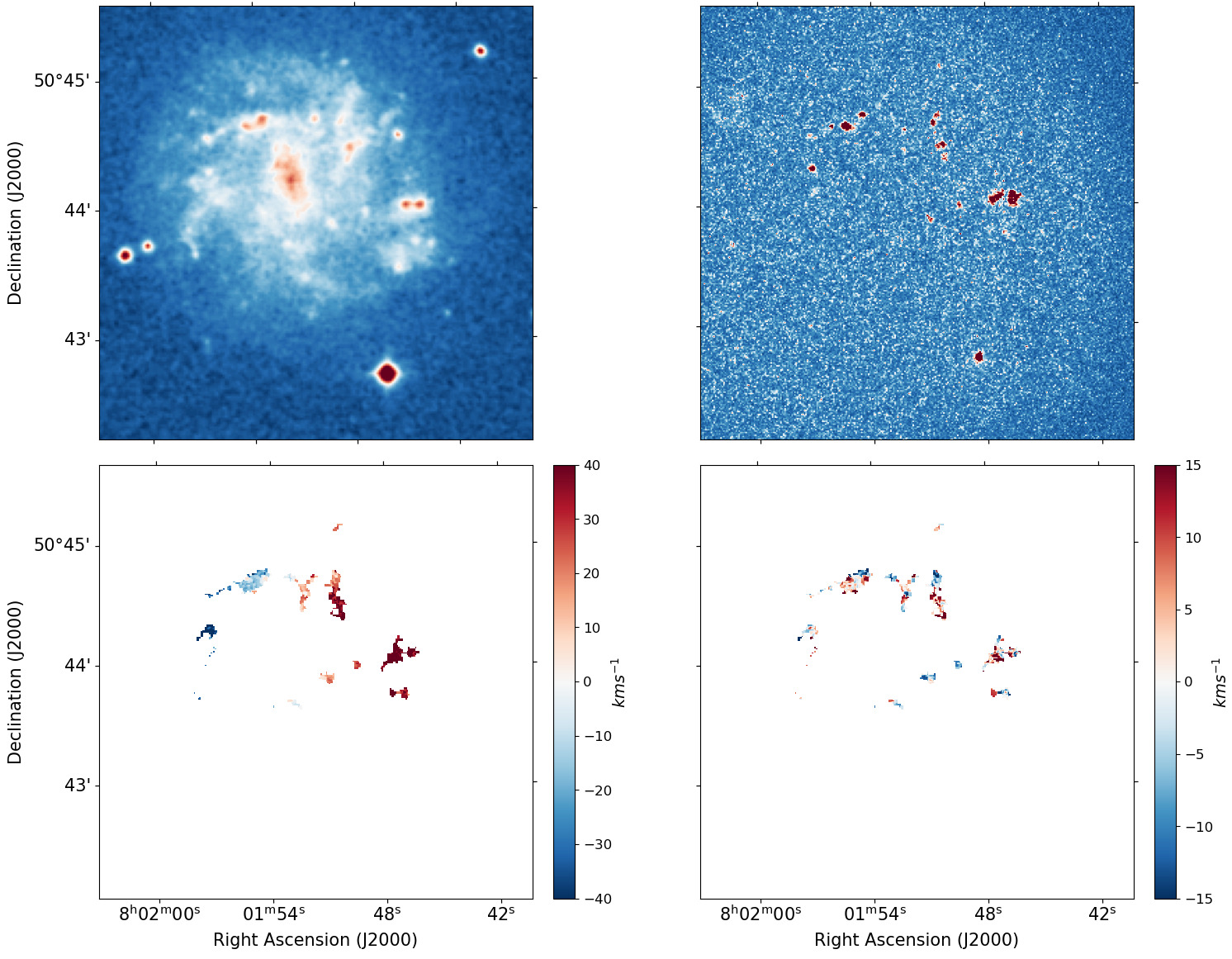}  

  \caption{NGC 2500. Top left: XDSS Blue Band image. Top right: $\mathrm {H_{\alpha}}$ monochromatic image. Bottom left: $\mathrm {H_{\alpha}}$ velocity field. Bottom right: Residual map $\mathrm {H_{\alpha}}$ field}
  \label{fig:ngc2500}

\end{figure*}
   
\begin{figure*}
  \centering
  \includegraphics[width=140mm,clip,angle=0]{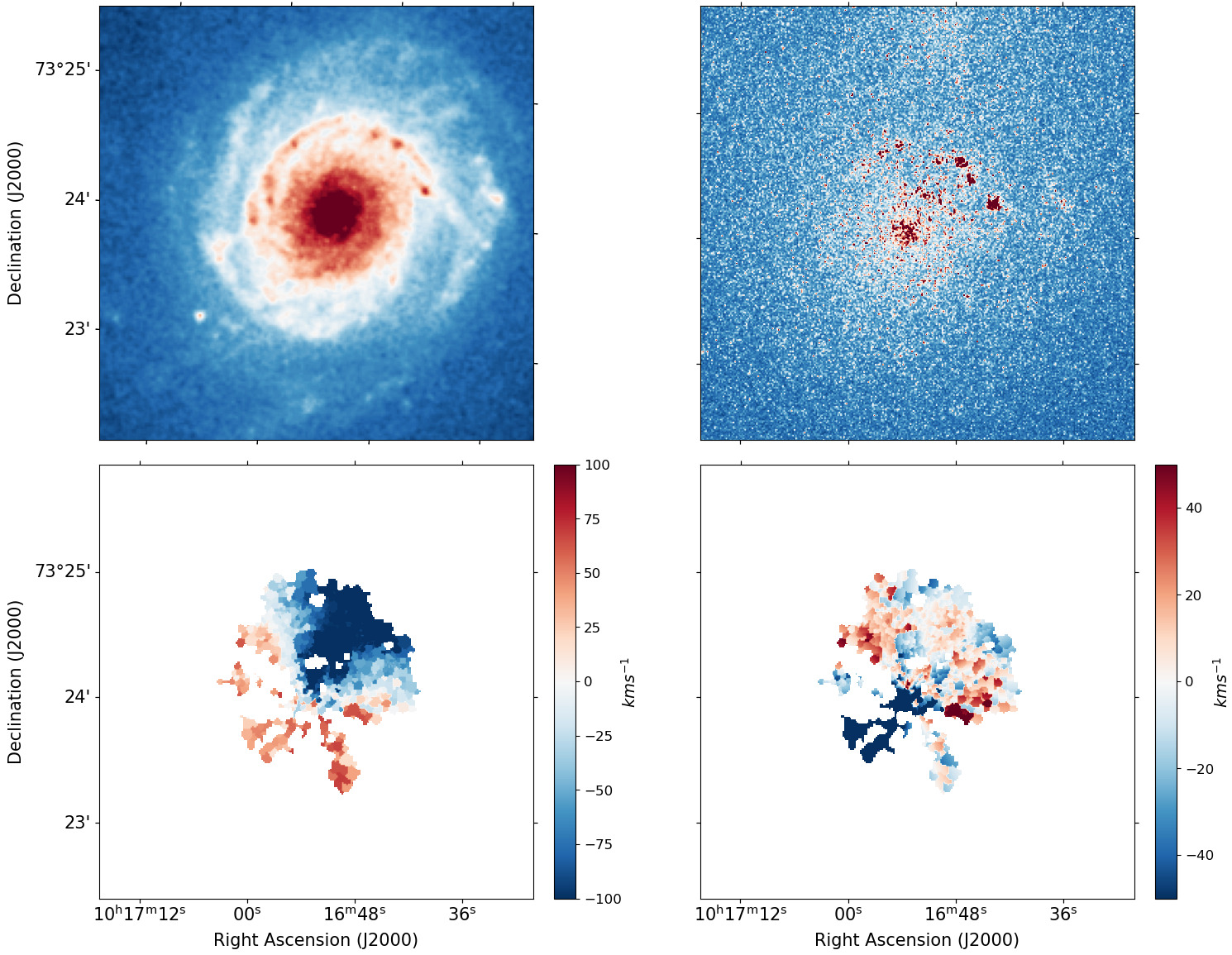}  

  \caption{NGC 3147. Top left: XDSS Blue Band image. Top right: $\mathrm {H_{\alpha}}$ monochromatic image. Bottom left: $\mathrm {H_{\alpha}}$ velocity field. Bottom right: Residual map $\mathrm {H_{\alpha}}$ field}
  \label{fig:ngc3147}
   
  \includegraphics[width=140mm,clip,angle=0]{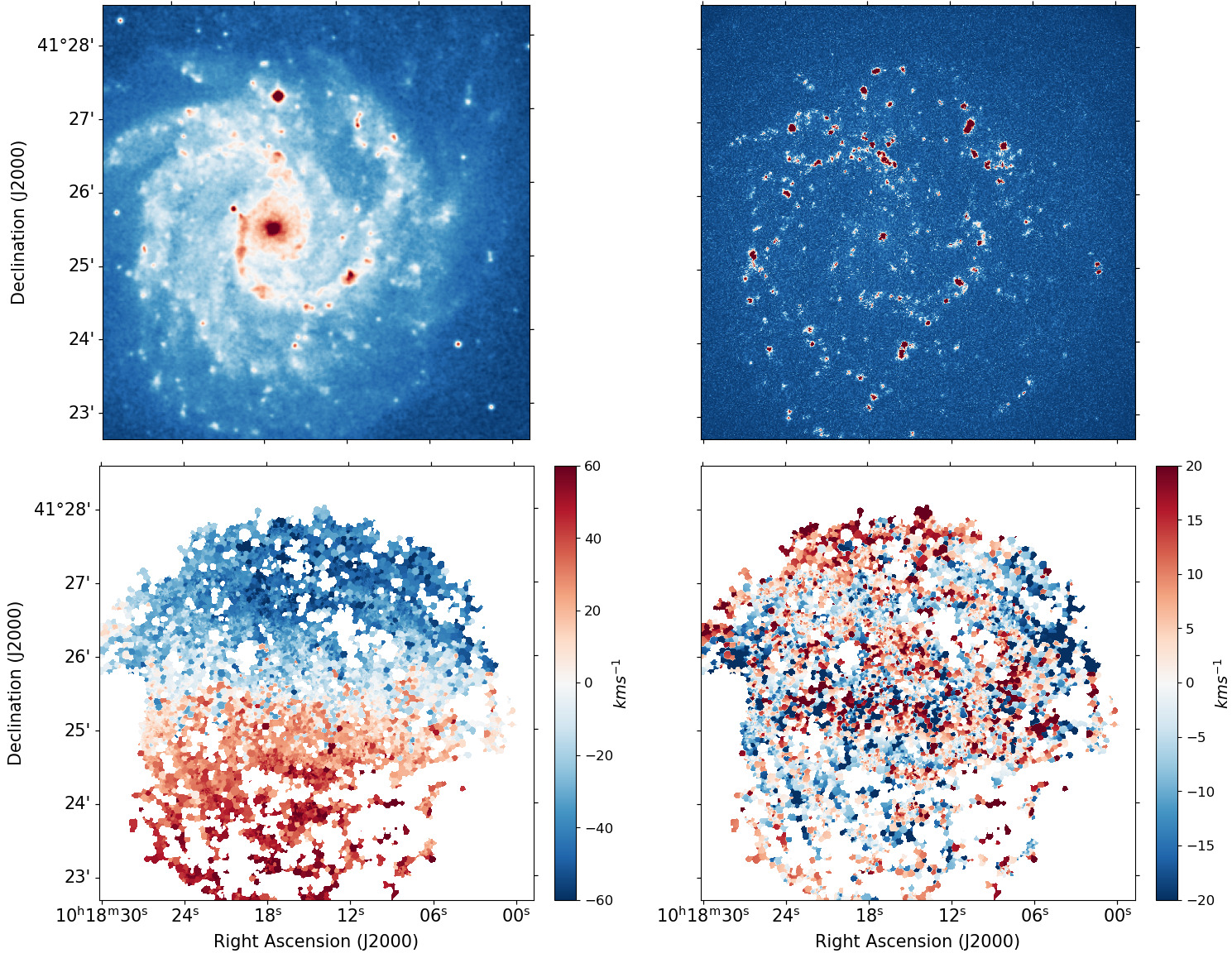}  

  \caption{NGC 3184. Top left: XDSS Blue Band image. Top right: $\mathrm {H_{\alpha}}$ monochromatic image. Bottom left: $\mathrm {H_{\alpha}}$ velocity field. Bottom right: Residual map $\mathrm {H_{\alpha}}$ field}
  \label{fig:ngc3184}
   
\end{figure*}

\begin{figure*}
  \centering
  \includegraphics[width=140mm,clip,angle=0]{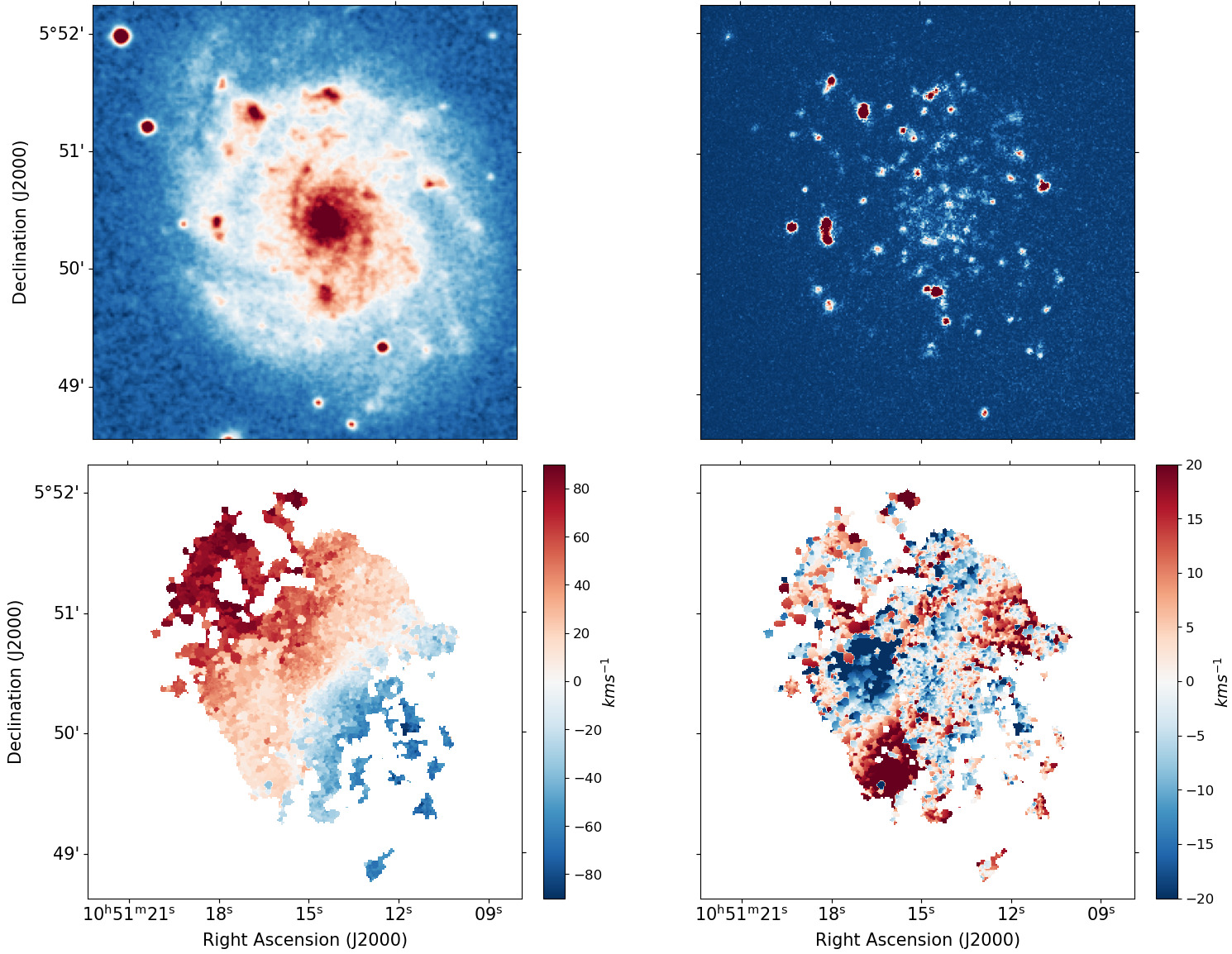}  

  \caption{NGC 3423. Top left: XDSS Blue Band image. Top right: $\mathrm {H_{\alpha}}$ monochromatic image. Bottom left: $\mathrm {H_{\alpha}}$ velocity field. Bottom right: Residual map $\mathrm {H_{\alpha}}$ field}
  \label{fig:ngc3423}
   
  \includegraphics[width=140mm,clip,angle=0]{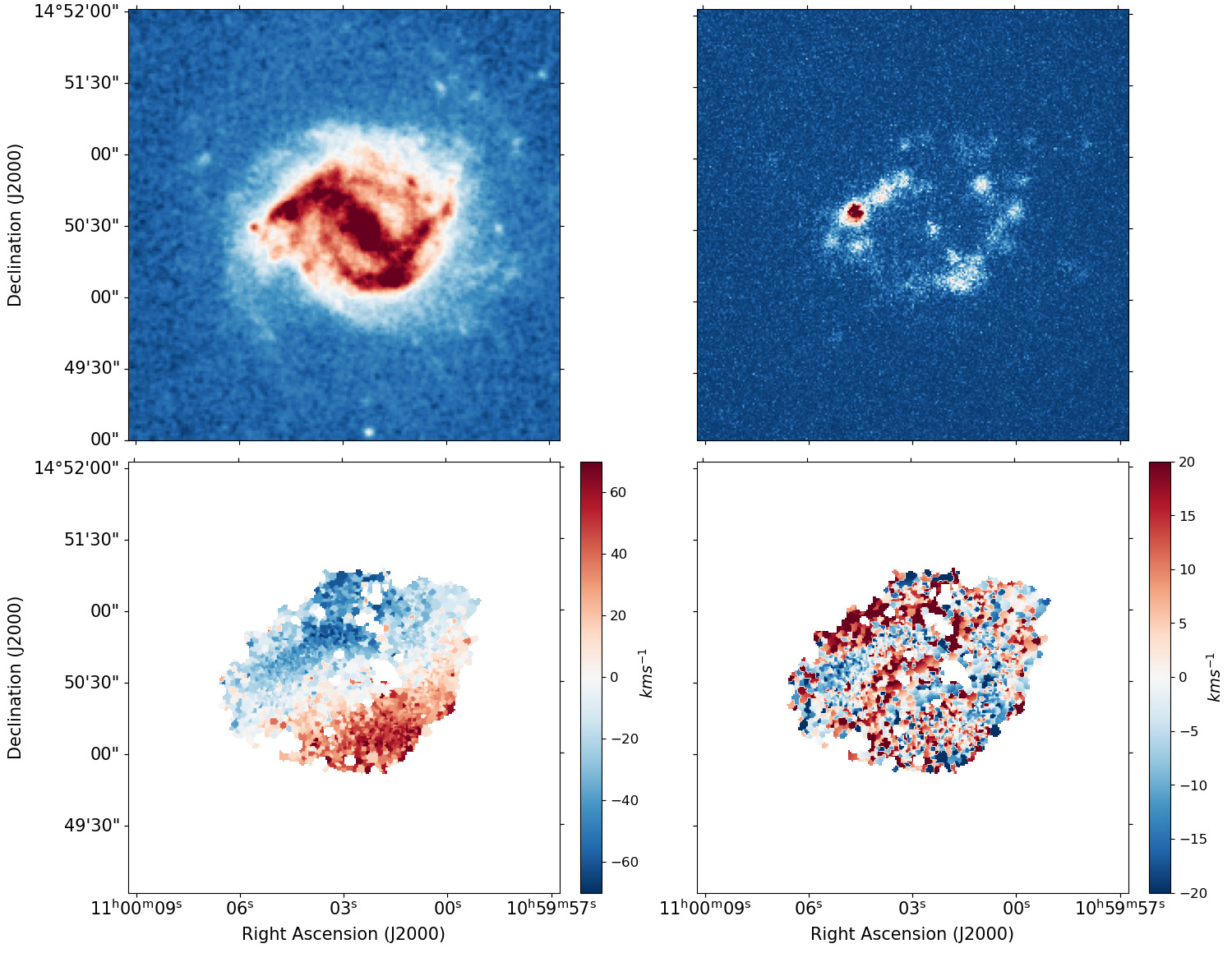}  

  \caption{NGC 3485. Top left: XDSS Blue Band image. Top right: $\mathrm {H_{\alpha}}$ monochromatic image. Bottom left: $\mathrm {H_{\alpha}}$ velocity field. Bottom right: Residual map $\mathrm {H_{\alpha}}$ field}
  \label{fig:ngc3485}
  
\end{figure*}

\begin{figure*}
  \centering
  \includegraphics[width=140mm,clip,angle=0]{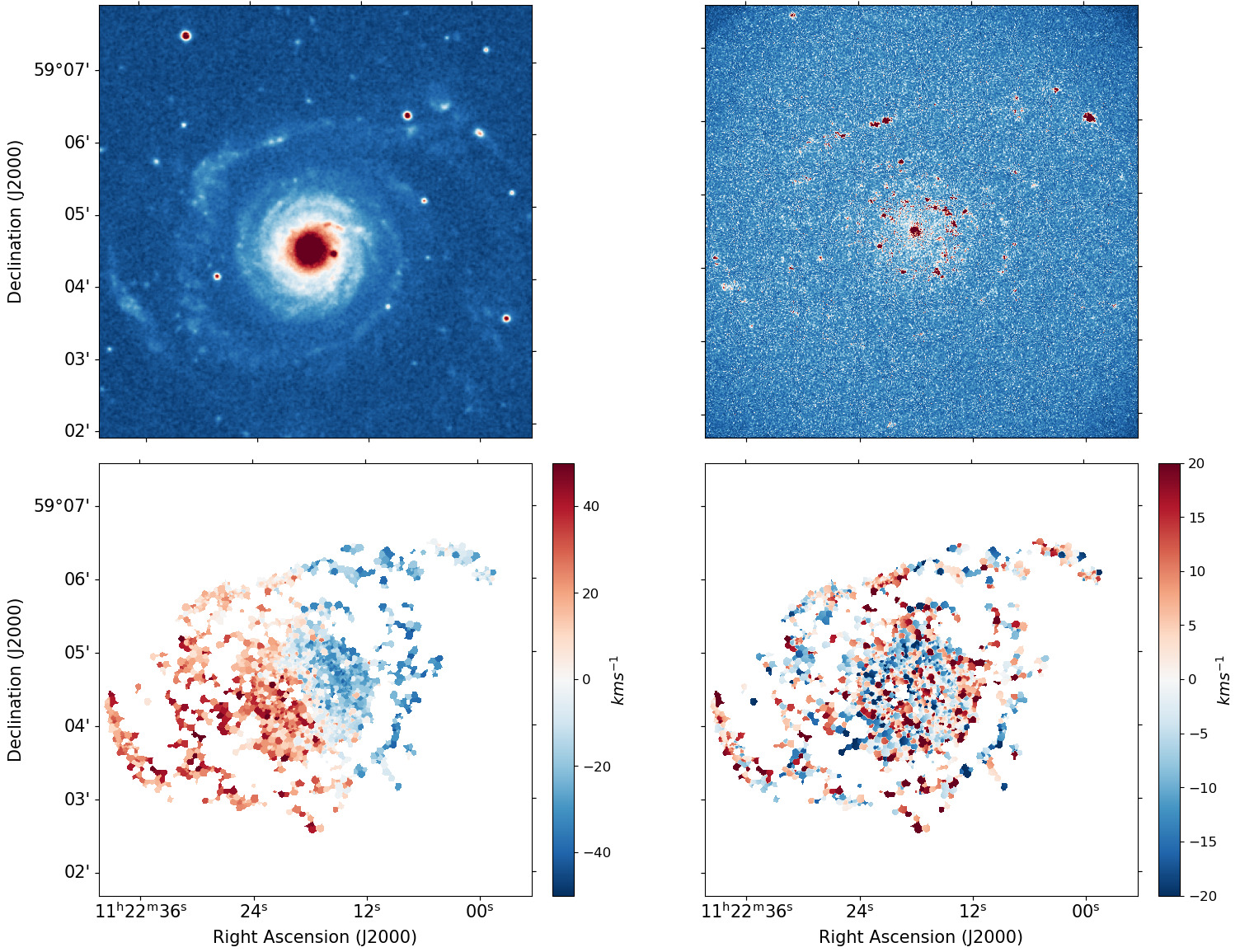}  

  \caption{NGC 3642. Top left: XDSS Blue Band image. Top right: $\mathrm {H_{\alpha}}$ monochromatic image. Bottom left: $\mathrm {H_{\alpha}}$ velocity field. Bottom right: Residual map $\mathrm {H_{\alpha}}$ field}
  \label{fig:ngc3642}
   
  \includegraphics[width=140mm,clip,angle=0]{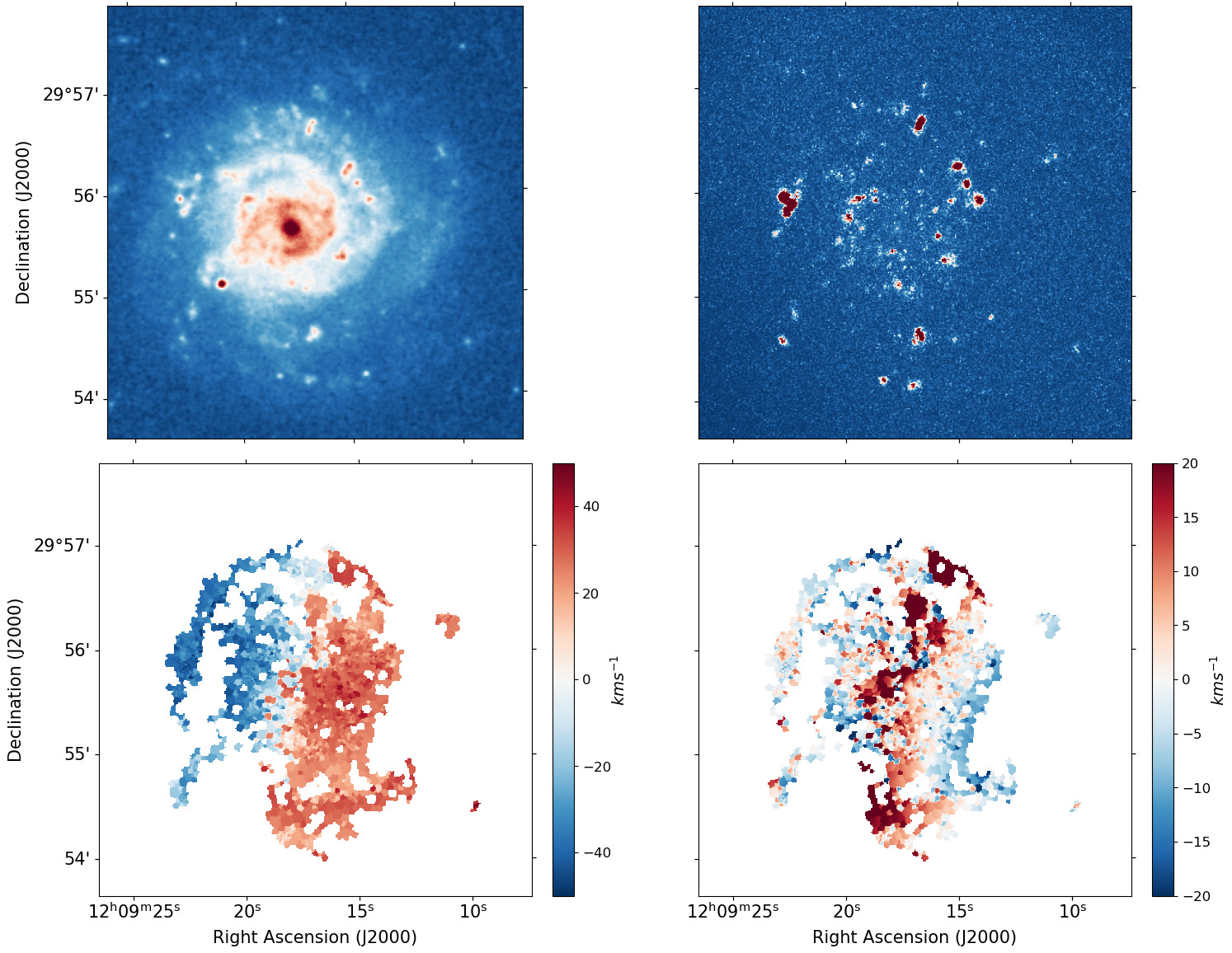}  

  \caption{NGC 4136. Top left: XDSS Blue Band image. Top right: $\mathrm {H_{\alpha}}$ monochromatic image. Bottom left: $\mathrm {H_{\alpha}}$ velocity field. Bottom right: Residual map $\mathrm {H_{\alpha}}$ field}
  \label{fig:ngc4136}
  
\end{figure*}

\begin{figure*}
  \centering
  \includegraphics[width=140mm,clip,angle=0]{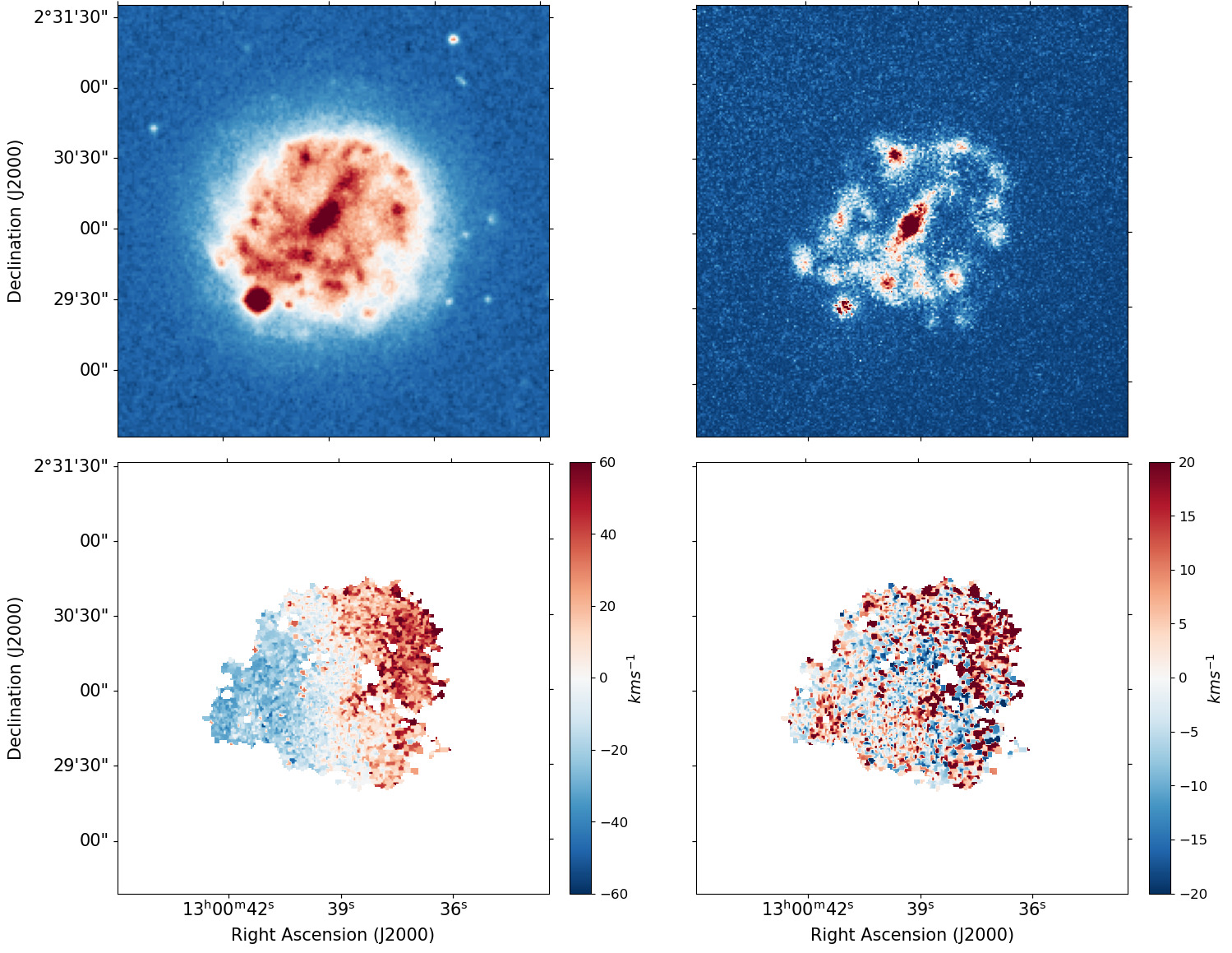}  

  \caption{NGC 4900. Top left: XDSS Blue Band image. Top right: $\mathrm {H_{\alpha}}$ monochromatic image. Bottom left: $\mathrm {H_{\alpha}}$ velocity field. Bottom right: Residual map $\mathrm {H_{\alpha}}$ field}
  \label{fig:ngc4900}
  
\end{figure*}

\subsubsection{\textup{Presentation of the $\mathrm {H_{\alpha}}$ monochromatic maps of the candidates of vertically perturbed disks. (Fig. 21)}}
\label{sssec:monochromatic_maps}
\begin{figure*}

  \centering
  \includegraphics[width=140mm,clip,angle=0]{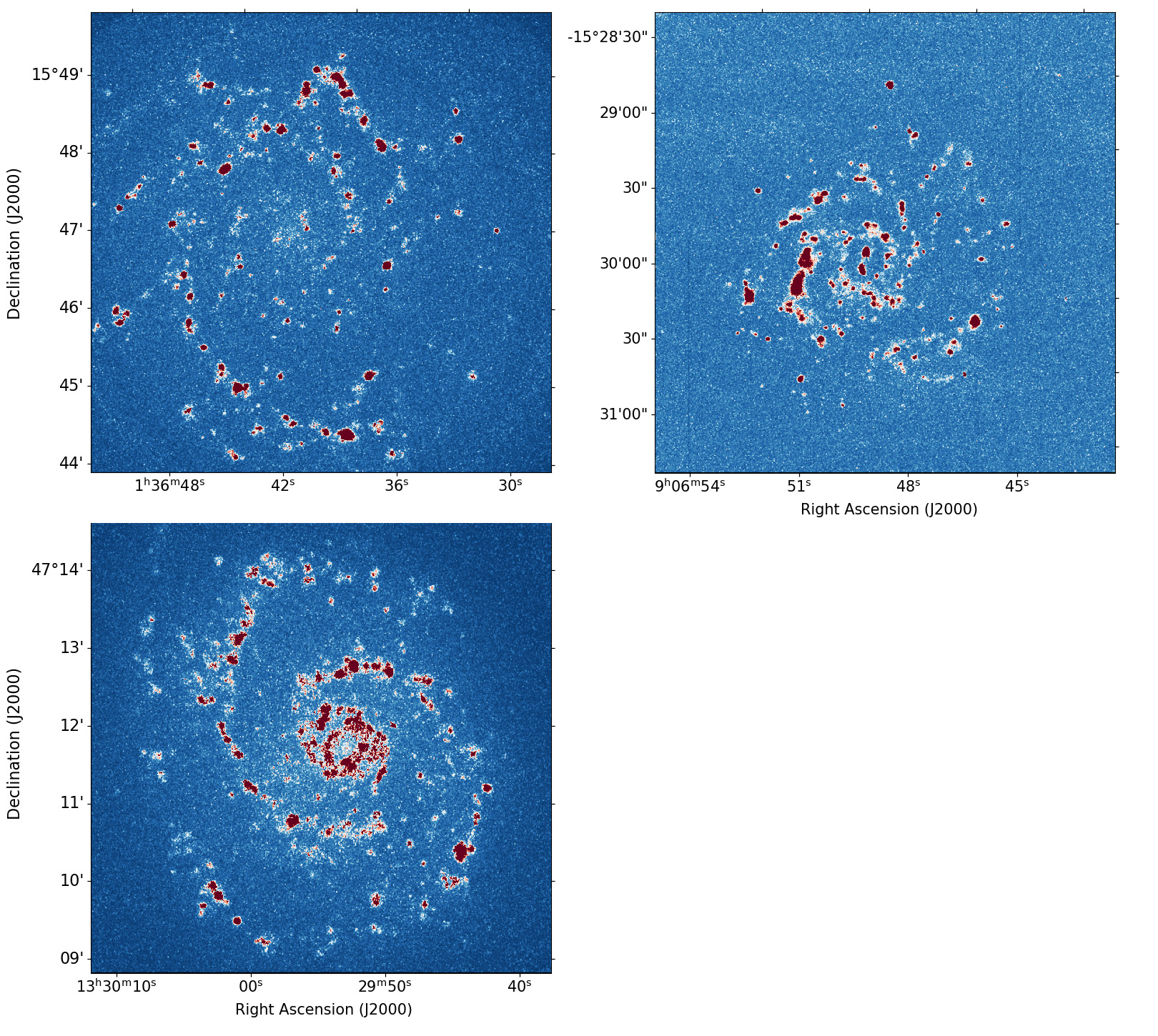}  

  \caption{$\mathrm {H_{\alpha}}$ monochromatic maps. Top left: NGC 628. Top right: NGC 2763. Bottom left: NGC 5194.}
   \label{fig:mono_maps}

\end{figure*}

\subsection{\textup{RESIDUAL MAPS OF ADDITIONAL DATA ARCHIVE}}
\subsubsection{\textup{Residual maps without signs of vertical perturbations. (Figs. 22 to 24)}}

\label{sssec:maps_without_perturbations}

\begin{figure*}
  \centering
  \includegraphics[width=140mm,clip,angle=0]{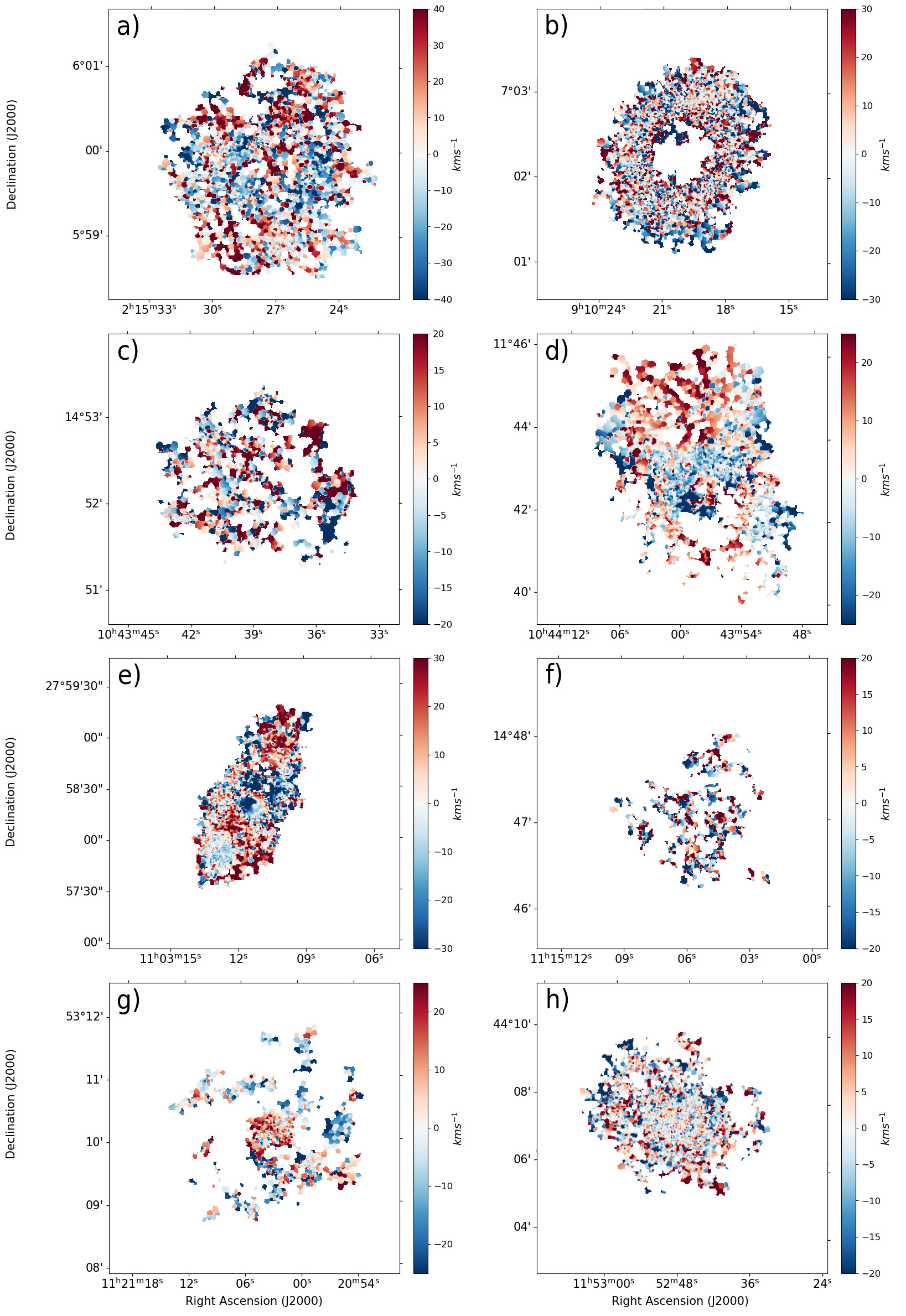}  

  \caption{$\mathrm {H_{\alpha}}$ residual maps of WiNDS data cubes. Panel a) NGC 864 (G). Panel b) NGC 2775 (G) Panel c) NGC 3346 (G). Panel d) NGC 3351 (S). Panel e) NGC 3504 (G). Panel f) NGC 3596 (G). Panel g) NGC 3631 (H). Panel h) NGC 3938 (S).}
  \label{fig:residual_maps_set01}

\end{figure*}

\begin{figure*}
  \centering
  \includegraphics[width=140mm,clip,angle=0]{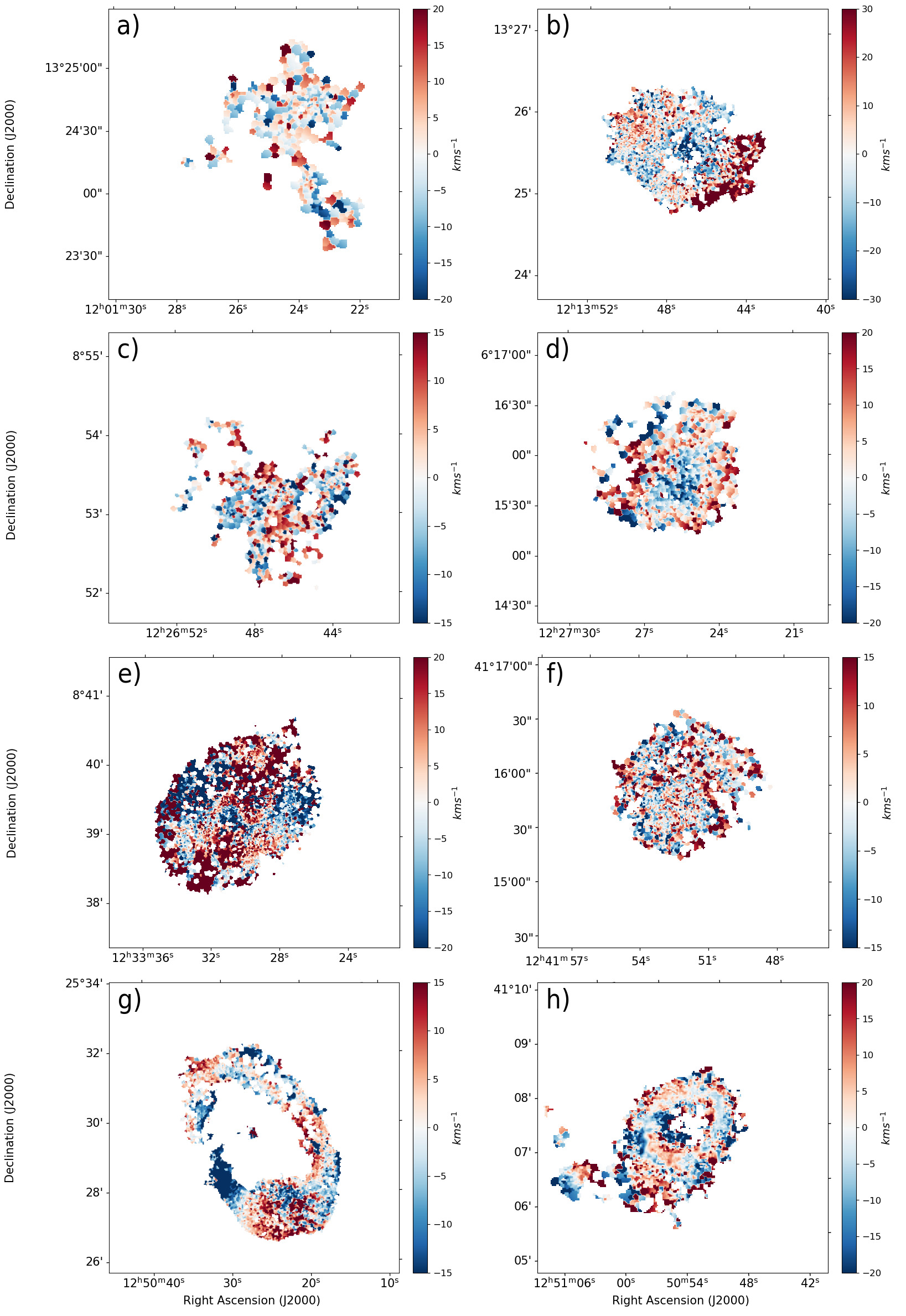}  

  \caption{$\mathrm {H_{\alpha}}$ residual maps of WiNDS data cubes. Panel a) NGC 4037 (H). Panel b) NGC 4189 (V). Panel c) NGC 4411B (G). Panel d) NGC 4430 (H). Panel e) NGC 4519 (V). Panel f) NGC 4625 (S). Panel g) NGC 4725 (S). Panel h) NGC 4736 (S).}
  \label{fig:residual_maps_set02}

\end{figure*}

\begin{figure*}
  \centering
  \includegraphics[width=140mm,clip,angle=0]{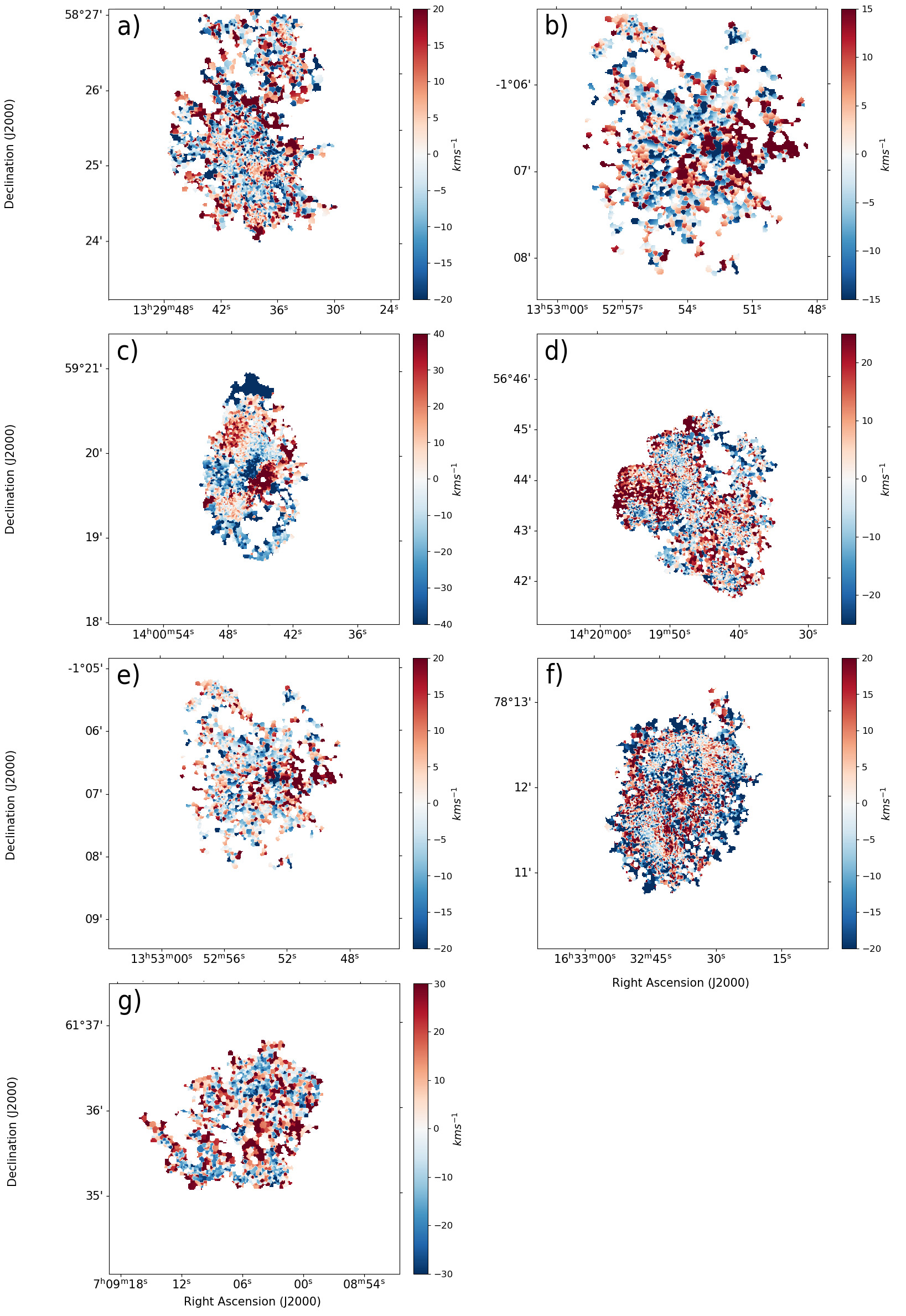}

  \caption{$\mathrm {H_{\alpha}}$ residual maps of WiNDS data cubes. Panel a) NGC 5204 (G). Panel b): NGC 5334 (H). Panel c) NGC 5430 (G). Panel d) NGC 5585 (G). Panel e) NGC 5669 (H). Panel f) NGC 6217 (G). Panel g) UGC 3685 (G).}
  \label{fig:residual_maps_set03}

\end{figure*}

\subsection{\textup{ROTATION CURVES OF WINDS}}
\subsubsection{\textup{Presentation of the rotation curves of WiNDS. (Figs. 25 to 29)}}

\label{sssec:rotation_curves}
\begin{figure*}
  \centering
  \includegraphics[width=140mm,clip,angle=0]{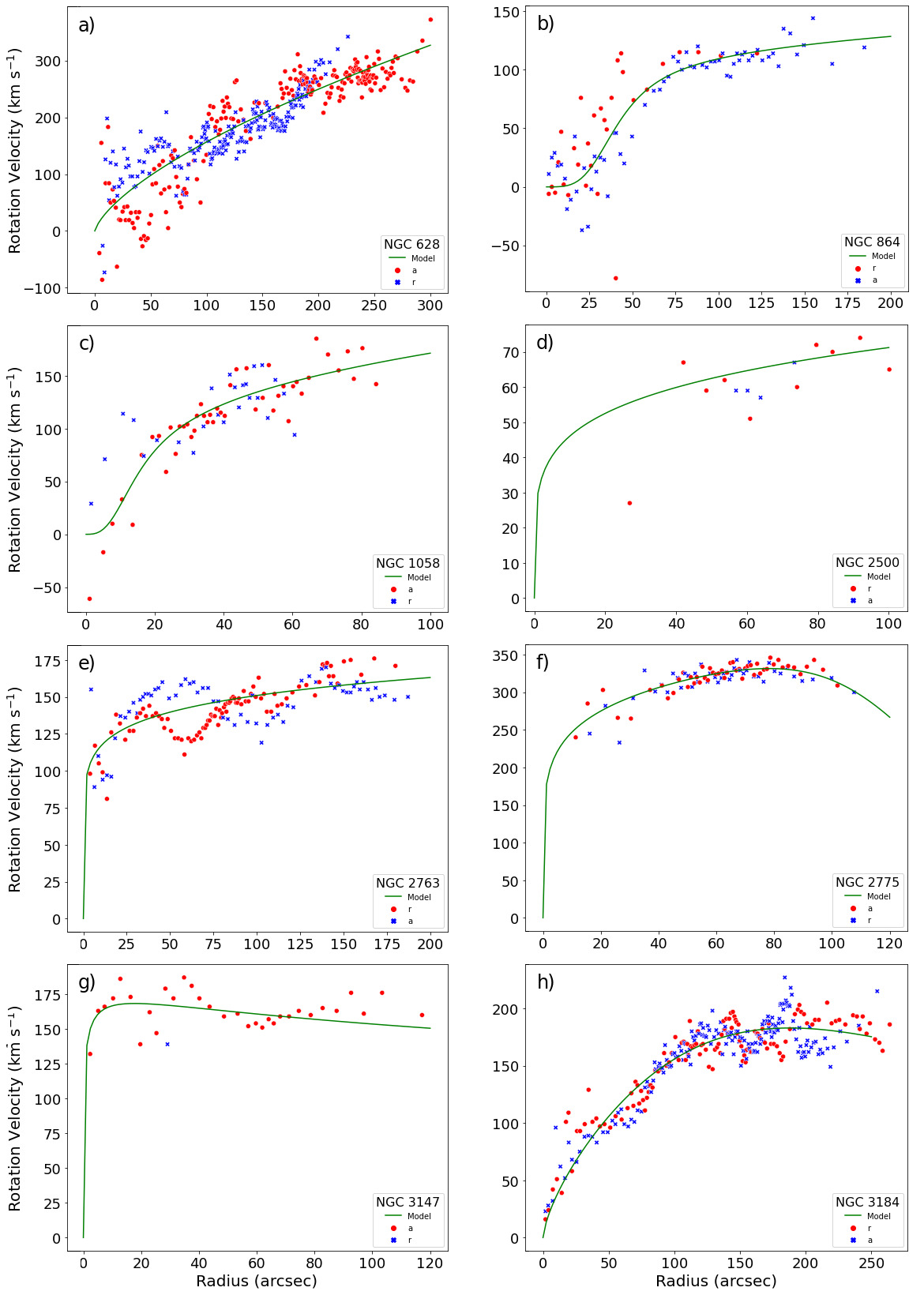}  

  \caption{Rotation curves of WiNDS.  Panel a) NGC 628. Panel b) NGC 864. Panel c) NGC 1058. Panel d) NGC 2500. Panel e) NGC 2763. Panel f) NGC 2775. Panel g) NGC 3147. Panel h) NGC 3184. The symbols represent the receding (dots) and approaching (crosses) side (with respect to the center).}
   \label{fig:rc_set01}

\end{figure*}

\begin{figure*}
  \centering
  \includegraphics[width=140mm,clip,angle=0]{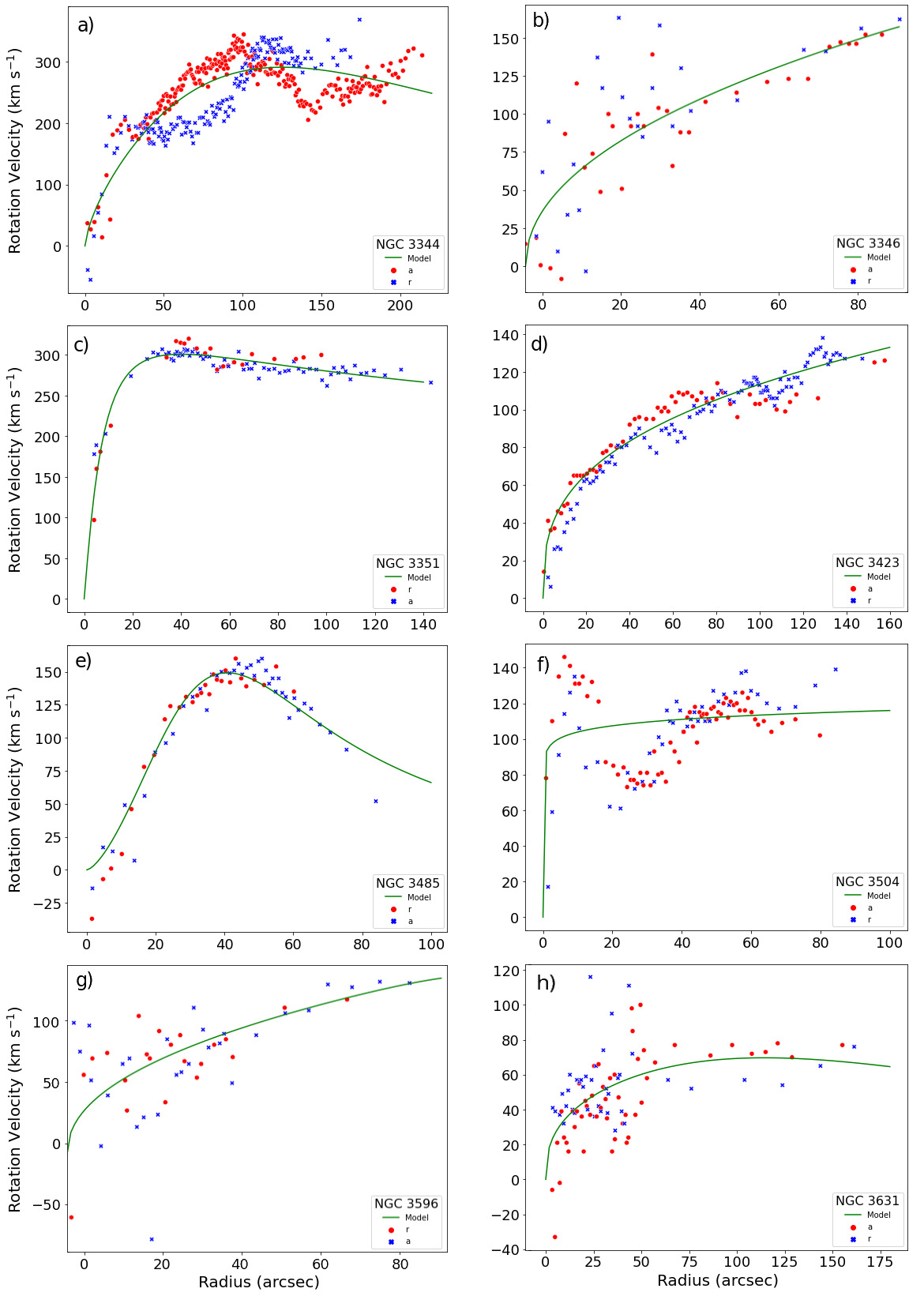}  

  \caption{Rotation curves of WiNDS. Panel a) NGC 3344. Panel b) NGC 3346. Panel c) NGC 3351. Panel d) NGC 3423. Panel e) NGC 3485. Panel f) NGC 3504. Panel g) NGC 3596. Panel h) NGC 3631. The symbols represent the receding (dots) and approaching (crosses) side (with respect to the center).}
   \label{fig:rc_set02}

\end{figure*}

\begin{figure*}
  \centering
  \includegraphics[width=140mm,clip,angle=0]{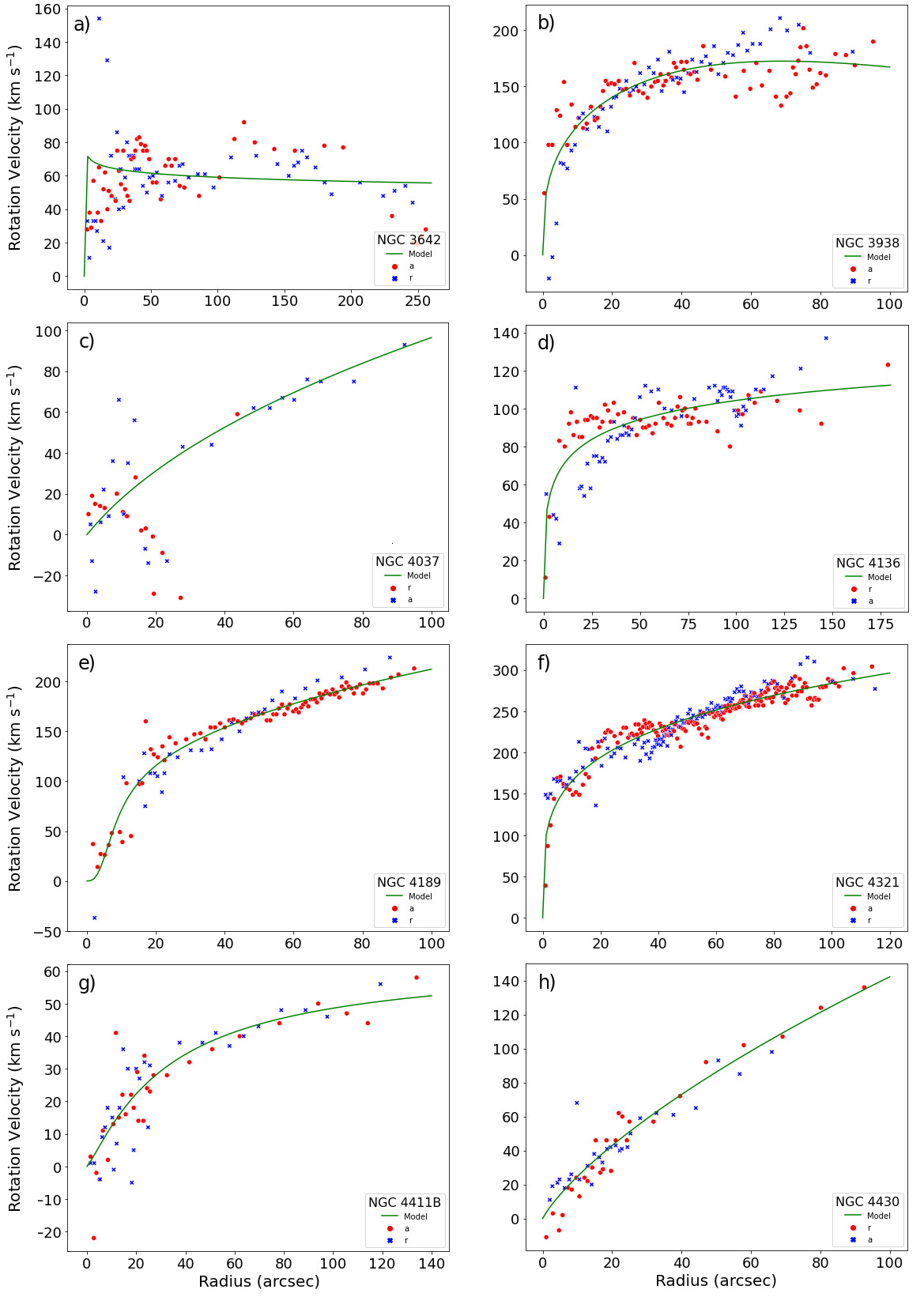}  

  \caption{Rotation curves of WiNDS. Panel a) NGC 3642. Panel b) NGC 3938. Panel c) NGC 4037. Panel d) NGC 4136. Panel e) NGC 4189. Panel f) NGC 4321. Panel g) NGC 4411B. Panel h) NGC 4430. The symbols represent the receding (dots) and approaching (crosses) side (with respect to the center).}
   \label{fig:rc_set03}

\end{figure*}

\begin{figure*}
  \centering
  \includegraphics[width=140mm,clip,angle=0]{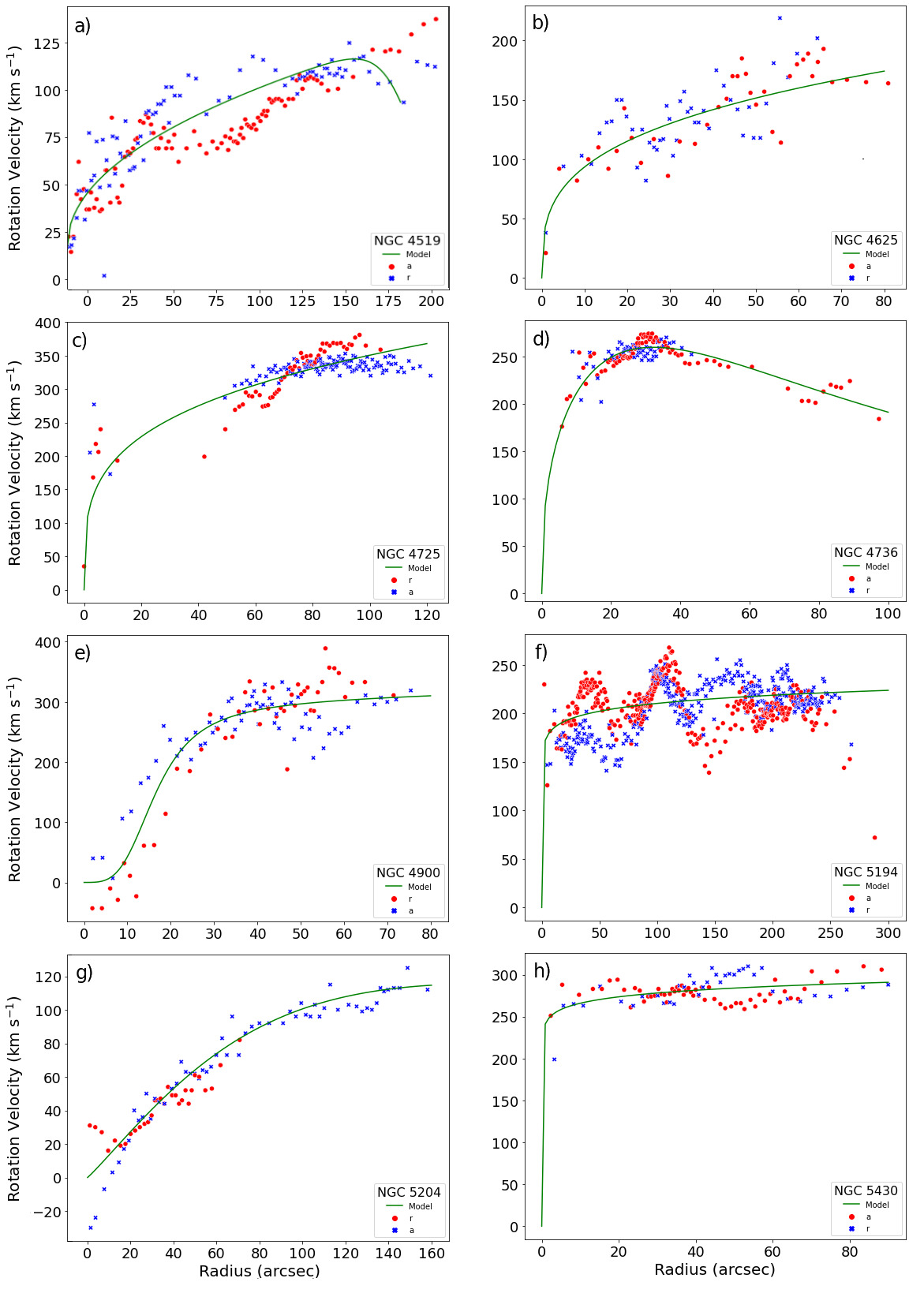}  

  \caption{Rotation curves of WiNDS. Panel a) NGC 4519. Panel b) NGC 4625. Panel c) NGC 4725. Panel d) NGC 4736. Panel e) NGC 4900. Panel f) NGC 5194. Panel g) NGC 5204. Panel h) NGC 5430. The symbols represent the receding (dots) and approaching (crosses) side (with respect to the center).}
  \label{fig:rc_set04}

\end{figure*}

\begin{figure*}
  \centering
  \includegraphics[width=140mm,clip,angle=0]{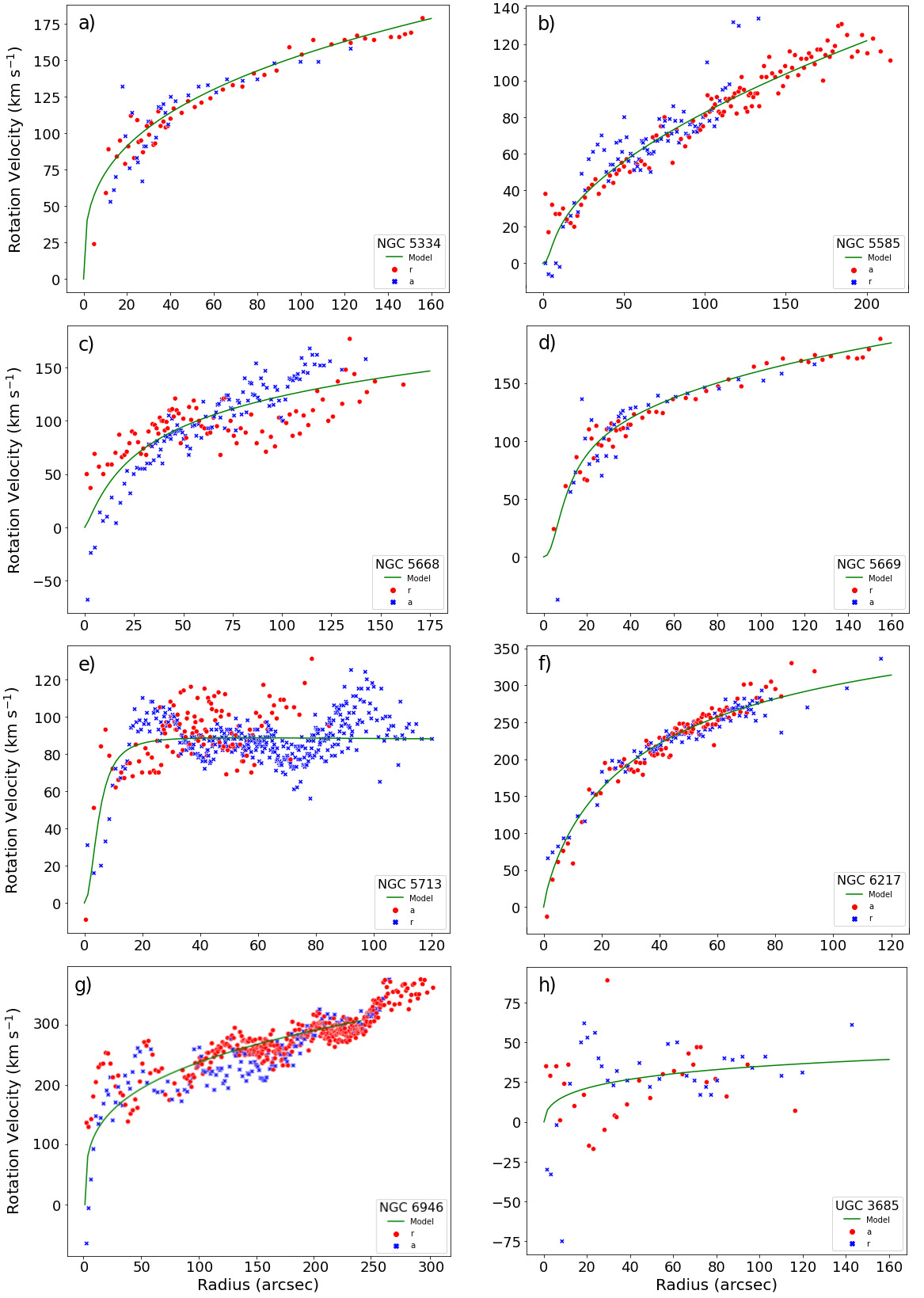}  

  \caption{Rotation curves of WiNDS. Panel a) NGC 5334. Panel b) NGC 5585. Panel c) NGC 5668. Panel d) NGC 5669. Panel e) NGC 5713. Panel f) NGC 6217. Panel g) NGC 6946. Panel h) UGC 3685. The symbols represent the receding (dots) and approaching (crosses) side (with respect to the center).}
  \label{fig:rc_set05}

\end{figure*}

\subsection{\textup{IMAGE PROCESSING}}

\subsubsection{\textup{Imaging process for the quantification of vertical perturbations on $V_{\rm res}$ field.}}
\label{sssec:low_pass_filter}

To highlight the larger velocity amplitudes in the $V_{\rm res}$ map, we use a filter on the residual field through the Fourier Transform and Gaussian low-pass filter.
The image filtering process is carried out following the process described in Fig. 30.

Let be a spatial image \textit{f(x,y)} with dimension \textit{$N\times N$}, the Discrete Fourier Transform (DFT) of \textit{f} called \textit{F(u,v)} is defined as 

\begin{equation}
\label{eq:fft}
F(u,v)=\frac{1}{N}\sum_{x=0}^{N-1} \sum_{y=0}^{N-1} f(x,y) e^{-2\pi i (ux+vy)/N}
\end{equation}

with $u = 0,1, 2, ..., N-1$ and $v = 0,1, 2, ..., N-1$, coordinates in Fourier space.

The Inverse Discrete Fourier Transform is defined as 
\begin{equation}
\label{eq:ifft}
f(x,y)=  \frac{1}{N} \sum_{x=0}^{N-1} \sum_{y=0}^{N-1} F(u,v) e^{2\pi i (ux+vy)/N}
\end{equation}

To remove or attenuate high frequencies in the Fourier domain, related to image noise, the Gaussian low-pass filter is used.
\begin{equation}
\label{eq:lpf}
H(u,v)= e^{-D^{2} (u,v)/2D_{0}^{2}}
\end{equation}

where \textit{D(u,v)} corresponds to the Euclidean distance from \textit{(u,v)} to the origin of the frequency plane and $D_{0}$ is cutoff frequency in pixel in Fourier space.

\begin{figure*}
  \centering
  \includegraphics[width=150mm,clip,angle=0]{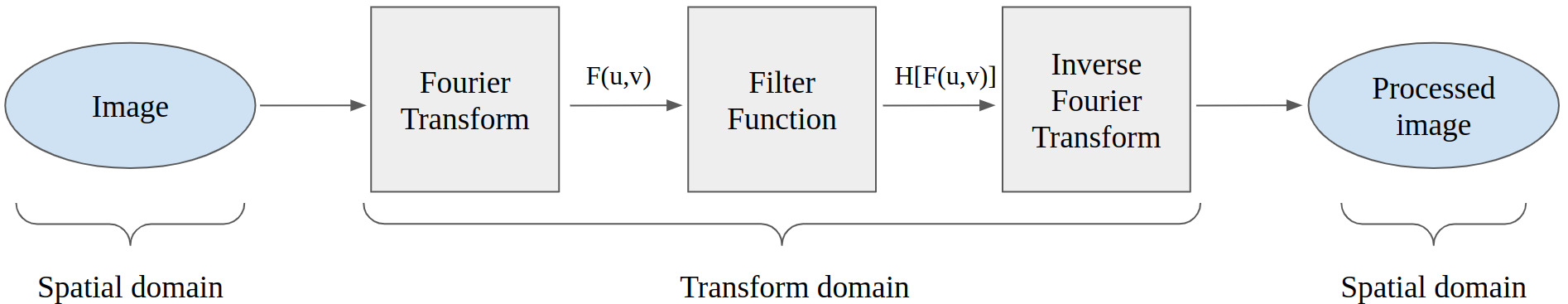}  

   \caption{Sketch of the image processing pipeline adopted in this work.}
   \label{fig:image_processed}
\end{figure*}

\bibliographystyle{apj}
\bibliography{apj-jour,winds}

\end{document}